%% file: DiracMat_AdvPhys.tex
\documentclass[]{tADP2e}

\usepackage{graphicx,amsmath,amssymb}
\usepackage{epstopdf}
\usepackage{amsbsy}
\usepackage[german,english]{babel}
\usepackage{color}
\usepackage{hyperref}

\newcommand{\hc}{\ensuremath{\text{H.c.}}}
\newcommand{\ang}{\ensuremath{\text{\AA}}}
\newcommand{\diff}{\ensuremath{\text{d}}}
\newcommand{\sgn}{\ensuremath{\text{sgn}}}
\newcommand{\bra}[1]{\ensuremath{\langle #1|}}
\newcommand{\ket}[1]{\ensuremath{|#1\rangle }}

\newcommand{\vF}{\ensuremath{v_{\rm F}}}

\newcommand{\beq}{\begin{eqnarray}}
\newcommand{\eeq}{\end{eqnarray}}
\newcommand{\beqa}{\begin{equation}}
\newcommand{\eeqa}{\end{equation}}

\newcommand{\ba}{\mathbf{a}}

\newcommand{\bh}{\mathbf{h}}
\newcommand{\bk}{\mathbf{k}}
\newcommand{\bl}{\mathbf{l}}
\newcommand{\br}{\mathbf{r}}
\newcommand{\bp}{\mathbf{p}}
\newcommand{\bq}{\mathbf{q}}
\newcommand{\bA}{\mathbf{A}}
\newcommand{\bG}{\mathbf{G}}

\renewcommand{\Re}{{\rm Re}\,}
\renewcommand{\Im}{{\rm Im}\,}
\newcommand{\up}{\uparrow}
\newcommand{\dn}{\downarrow}
\newcommand{\bbb}{\mathbf{b}}
\newcommand{\bu}{\mathbf{u}}

\begin{document}
\articletype{REVIEW ARTICLE}
\title{Dirac materials}

\author{T.~O.~Wehling$^{a,b}$
, A.~M.~Black-Schaffer$^{c}$, and A.~V.~Balatsky$^{d,e}$$^{\ast}$\thanks{$^\ast$Corresponding author. Email: avb@nordita.org, avb@lanl.gov \vspace{6pt}}
\\\vspace{6pt}
$^{a}${\em{Institut f{\"u}r Theoretische Physik, Universit{\"a}t Bremen, Otto-Hahn-Allee 1, 28359 Bremen, Germany}};
$^{b}${\em{Bremen Center for Computational Materials Science, Universit{\"a}t Bremen, Am Fallturm 1a, 28359 Bremen, Germany}};
$^{c}${\em{Department of Physics and Astronomy, Uppsala University, Box 516, S-751 20 Uppsala, Sweden}};
$^{d}${\em{Nordic Institute for Theoretical Physics (NORDITA), Roslagstullsbacken 23, S-106 91 Stockholm, Sweden}};
$^{e}${\em{Institute for Materials Science, Los Alamos National Laboratory, Los Alamos, New Mexico 87545,USA}}
\\ \received{\today}}

\maketitle

\begin{abstract}
A wide  range of materials, like $d$-wave superconductors, graphene, and topological insulators, share a fundamental similarity: their low-energy fermionic excitations behave as massless Dirac particles rather than fermions obeying the usual Schr{\"o}dinger Hamiltonian. This emergent behavior of Dirac fermions in condensed matter systems defines the unifying framework for a class of materials we call ``Dirac materials''. 
In order to establish this class of materials, we illustrate how Dirac fermions emerge in multiple entirely different condensed matter systems and we discuss how Dirac fermions have been identified experimentally using electron spectroscopy techniques (angle-resolved photoemission spectroscopy and scanning tunneling spectroscopy). 
As a consequence of their common low-energy excitations, this diverse set of materials shares a significant number of universal properties in the low-energy (infrared) limit. We review these common properties including nodal points in the excitation spectrum, density of states, specific heat, transport, thermodynamic properties, impurity resonances, and magnetic field responses, as well as discuss many-body interaction effects. 
We further review how the emergence of Dirac excitations is controlled by specific symmetries of the material, such as time-reversal, gauge, and spin-orbit symmetries, and how by breaking these symmetries a finite Dirac mass is generated. 
We give examples of how the interaction of Dirac fermions with their distinct real material background leads to rich novel physics with common fingerprints such as the suppression of back scattering and impurity-induced resonant states.
\end{abstract}
\begin{classcode}
73.20.-r	Electron states at surfaces and interfaces
73.25.+i	Surface conductivity and carrier phenomena
73.50.-h	Electronic transport phenomena in thin films
74.20.-z	Theories and models of superconducting state
75.76.+j	Spin transport effects
\end{classcode}


\noindent
{\bf Table of contents}

\noindent
1. Introduction\\
2. Microscopic origin of Dirac excitations\\
2.1 Graphene\\
2.2 Topological insulators\\
2.3 $d$-wave superconducors\\
2.4 Weyl semimetals\\
3. Stability of Dirac materials\\
3.1 Symmetry protection\\
3.2 Mass generation\\
3.3 Dirac cone annihilation\\
5. Experimental confirmation of Dirac energy spectrum\\
4.1 Angle-resolved photoemission spectroscopy\\
4.2 Scanning tunneling spectroscopy\\
5. Many-body interactions\\
5.1 Electronic excitations\\
5.2 Electric screening\\
5.3 Velocity renormalization\\
5.4 Mass gap\\
5.5 Superconductivity\\
6. Universal properties of Dirac materials\\
6.1 Thermodynamic properties\\
6.2 Magnetic field dependence\\
6.3 Transport in Dirac materials\\
6.4 Creation of low-energy impurity states\\
6.5 Real space properties of impurity resonances\\
7. Conclusions\\
8. Acknowledgements \\

\section{Introduction}
Condensed matter physics is witnessing a rapid expansion of materials with Dirac fermion low-energy excitations \cite{Dirac1928}, with examples ranging from superfluid phases of $^3$He \cite{Volovikbook,VollhardtWoelfle}, high-temperature $d$-wave superconductors \cite{Balatsky_RMP}, graphene \cite{AHC_RMP,Katsnelson_book} to topological insulators \cite{Kane_RMP2010,Qi11RMP}. 
These seemingly diverse materials possess properties that are a direct consequence of the Dirac spectrum of the quasiparticles and are universal. For example, the presence of nodes in the excitation spectrum controls low-energy energy properties, such as the fermionic specific heat of these materials. Thus, neutral superfluids like $^3$He, high-temperature $d$-wave superconductors and graphene, despite being vastly different materials, all exhibit the same power-law temperature dependence of the fermionic specific heat, controlled only by the dimensionality of the excitation phase space. Other universal features include the response to impurities and magnetic fields, suppressed backscattering, transport properties and optical conductivity \cite{DasSarmaRMP}.

The observed similarity of the low-energy spectrum points to a powerful organizing principle. There are symmetries  that control the formation of the Dirac points (or possibly lines) in the excitation spectrum.  These symmetries are different for different materials, yet their net effect is to preserve Dirac nodes. Examples include time-reversal symmetry in topological insulators and sublattice symmetry in graphene. Once Dirac nodes are established in the spectrum, the universality of response functions and susceptibilities naturally follows. The materials that have Dirac nodes in the spectrum, regardless of their origin, we call {\em Dirac materials}. They form a distinct class that is different from conventional metals and (doped) semiconductors.

This view of defining a material class is based on the most basic approach for characterizing condensed matter systems, i.e. using their low-energy excitations, which generally determine the response of the system to external probes. Depending on the low-energy excitations materials can be metallic or insulating. Metals have a finite phase-space for low-energy electronic  excitations, are electric conductors and their specific heat increases to leading order linearly with the temperature.  Insulators, on the other hand,  require a finite energy  gap  to be overcome for  electronic  excitations  and  thermally  excited  electron-hole pairs are therefore exponentially  suppressed  at low temperatures. For many metals, as well as doped semiconductors, the concept of nearly free quasiparticles obeying the Schr{\"o}dinger equation with the Hamiltonian $H_S=\bp^2/2m^*$, where $m^*$  is the  effective mass, provides an extremely successful description of the low-energy excitations. These excitations are often simply referred to as ``Schr{\"o}dinger fermions". 
In contrast, in Dirac materials the low-energy fermionic excitations or quasiparticles do not obey the Schr{\"o}dinger Hamiltonian  $H_S$ but rather a Dirac Hamiltonian \cite{Dirac1928} with the effective ``speed of light" $c$ being given by the Fermi velocity $\vF$. In two spatial dimensions, this Hamiltonian has the form
\beq H_D=c{\bm\sigma}\cdot\bp+mc^2\sigma_z, \label{eqn:Dirac}\eeq
where $\bm\sigma=(\sigma_x,\sigma_y)$ and $\sigma_z$ are the usual Pauli matrices. Quasiparticles described by the Hamiltonian $H_D$ are frequently called "Dirac fermions".

In the limit of vanishing Dirac mass $m\to 0$, there is no gap in the spectrum of $H_D$ and the quasiparticle dispersion is linear, which is qualitatively different from the parabolic dispersion of conventional metals or semiconductors. Moreover, even for non-zero mass, positive and negative energy eigenstates of the Dirac Hamiltonian are made from the same space of spinor wave functions. Thus, particles and holes are interconnected and have the same effective mass $m$, which is directly related to the spectral gap $\Delta=2mc^2$ \cite{Katsnelson_QED}. This is very different from systems like conventional metals and semiconductors, where electrons and holes obey separate Schrödinger equations with different effective masses and no unique relation between gap and mass. Therefore, Dirac fermions with non-zero mass still are qualitatively different from Schr{\"o}dinger fermions, as soon as experimentally probed energies are on the order of $mc^2$ or higher. It is thus natural to include both systems with massless and massive Dirac fermion excitations in the unifying framework of Dirac materials.

A variety of Dirac materials has been discovered to date ranging from ``normal state'' crystalline materials to exotic quantum fluids (c.f.~Table \ref{tab:DM}). 
In the superfluid $^3$He-A phase \cite{Volovikbook,VollhardtWoelfle},
 for example, the low-energy fermionic excitations near the north and south pole of the Fermi surface form two nodal points where the Bogoliubov quasiparticles are described by a Dirac Hamiltonian (\ref{eqn:Dirac}). A related example is the case of the cuprate superconductors, a class of superconductors with an order parameter with $d$-wave symmetry, $\Delta_\bk = \Delta_0(\cos k_xa - \cos k_ya)$, and low-energy fermionic excitations being described by the Dirac Hamiltonian (\ref{eqn:Dirac}) \cite{Volovikbook,Anderson_98,Khveshchenko_PRL01,Tesanovic_PRL01,Balatsky_RMP}. The rise of graphene \cite{Novoselov_science2004,Geim2005,Zhang2005} --- a layer of carbon atoms arranged in a
 honeycomb lattice --- draws attention to the fact that the same Dirac-like spectrum, Eq.~(\ref{eqn:Dirac}), as in the superconducting or superfluid materials can be an inherent property of the band structure of a material, ultimately stemming from the crystalline order \cite{Wallace-1947,Semenoff-1984,AHC_RMP,Katsnelson_book}. The same crystalline order produces Dirac fermions also in silicene and germanene \cite{Silicene_CiraciPRL09}, the Si and Ge equivalents of graphene, as well as in ``artificial" graphene \cite{Wunsch_NJP08,Park_NanoLett08}.
  In a more recent development, a new kind of insulators has been discovered \cite{Kane_PRL05,Moore_PRB07,Koenig_Science07,Hasan_09,SCZhang_09,Zhang_PhysToday2010,Kane_RMP2010}, the so-called topological insulators, which have a fully gapped energy spectrum in the bulk but Dirac fermions on the surface. Furthermore, ultra-cold atoms in optical lattices provide another realizations of Dirac fermions in condensed matter systems \cite{Sengstock_2011, Esslinger_2012}. 

The possibility of finding materials with three-dimensional Dirac-like spectrum has recently also gained a lot of attention. In three dimensions, all three Pauli matrices are used in the momentum dependent term: $H_D = c{\bm\sigma}\cdot\bp$, and a mass term is thus per definition absent. This Hamiltonian enters the Weyl equation in particle physics, and materials with this low-energy spectrum have subsequently been coined Weyl semimetals \cite{WanTurner11}. If there is a band degeneracy present at the Dirac nodal point (but not causing a finite gap) these materials are instead called three-dimensional Dirac semimetals \cite{Young12}. Recent ARPES measurements on Na$_3$Bi \cite{Liu13Na3Bi} and Cd$_3$As$_2$ \cite{Borisenko13,Neupane13} have found evidence for a three-dimensional Dirac semimetal state in these materials.

At first (microscopic) sight, a material like graphene does hardly display any similarity with typical $d$-wave superconductors or superfluids. There are important materials specific properties making all these materials distinct: some are superconductors and some are (bulk) insulators; some are crystalline with honeycomb lattice (graphene or silicene), others have a more  complicated Perovskite crystal structure (cuprate superconductors) or do not exhibit any crystalline order ($^3$He-A phase). Also the physical realizations of the Dirac pseudospin differ between these materials (c.f.~Table \ref{tab:DM}) and the list of differences can be continued. But again, it is the universal properties related to the existence of the low-energy Dirac excitations that justify the concept of Dirac materials. 
As a unifying principle, the presence of nodes leads to a sharp reduction of the phase space for low-energy excitations in Dirac materials. More precisely, the dimensionality of the set of points in momentum space where we have zero-energy excitations is reduced in Dirac materials as compared to normal metals. For example, nodes for a three-dimensional Dirac material mean the effective Fermi surface is shrunk from a two-dimensional object to a point. Lines of Dirac nodes in three dimensions would mean that the Fermi surface has shrunk from a two-dimensional surface to a one-dimensional line. In either case there is a reduction of dimensionality for the zero-energy states. This reduction of phase space controlled by additional symmetry in the system is an indicator for Dirac materials. Reduced phase space and controlling symmetries are important for applications. First of all, it is possible to lift the protected symmetry of the Dirac node and therefore destroy the nodes and open an energy gap. This modification of the spectrum of quasiparticles drastically changes the response of the Dirac material, as for example is the case for topological insulator in a magnetic field \cite{SCZhangimp}. Second, Dirac nodes and the resultant reduction of phase space  suppress dissipation and are thus attractive for applications exploiting the coherence of low-energy states in the nodes.

\begin{table}%
\begin{tabular}{lccccc}
Material & Pseudospin & Energy scale (eV) & References\\
\hline \hline
Graphene, Silicene, Germanene & Sublattice & $1\mbox{--}3$~eV & \cite{Wallace-1947,AHC_RMP,Katsnelson_book,Takeda94, GuzmanVerri07,Silicene_CiraciPRL09} \\
Artificial Graphenes & Sublattice & $10^{-8} \mbox{--} 0.1$~eV & \cite{Gibertini_2DEG_09,Manoharan_Nature2012,Sengstock_2011,Esslinger_2012,Polini_DM13} \\
Hexagonal layered heterostructures & Emergent & $0.01\mbox{--} 0.1$~eV & \cite{Park_Louie_PRL08,Guinea_PTRSA_2010,vandenBrink_PRB2012,Falko_PRB2013,LeRoy_NatPhys2012,Ponomarenko_Nature2013,Jarillo-Herrero_Science2013} \\
Hofstadter butterfly systems & Energent & $0.01$~eV & \cite{Ponomarenko_Nature2013} \\
\multicolumn{4}{l}{\; Graphene-hBN heterostructures in high magnetic fields}\\
\hline
Band inversion interfaces & Spin-orbit ang. mom. & $0.3$~eV & \cite{Pankratov_JETP85,Fradkin_PRL86,Pankratov_SSC87} \\
\multicolumn{4}{l}{\; SnTe/PbTe, CdTe/HgTe, PbTe}\\
2D Topological Insulators & Spin-orbit ang. mom.  & $< 0.1$eV & \cite{Kane_PRL05,Bernevig_2006,Koenig_Science07,Kane_RMP2010,Qi11RMP,YAndo_JPSJ13} \\
\multicolumn{4}{l}{\; HgTe/CdTe, InAs/GaSb, Bi bilayer, ...}\\
3D Topological Insulators & Spin-orbit ang. mom. & $\lesssim 0.3$eV & \cite{Kane_PRL07,Moore_PRB07,RoyPRB09,Kane_RMP2010,Qi11RMP,Felser_review12,YAndo_JPSJ13}\\
\multicolumn{4}{l}{\; Bi$_{1-x}$Sb$_x$, Bi$_2$Se$_3$, strained HgTe, Heusler alloys, ...}\\
Topological crystalline insulators & orbital & $\lesssim 0.3$eV & \cite{Fu11crystTI,HsiehTim11,Tanaka12,Dziawa12} \\
\multicolumn{4}{l}{\; SnTe, Pb$_{1-x}$Sn$_{x}$Se}\\
\hline
$d$-wave cuprate superconductors & Nambu pseudospin & $\lesssim 0.05$eV & \cite{Tsuei00RMP, Thompson12} \\
\hline
$^3$He & Nambu pseudospin & $0.3\,\mu$eV & \cite{Volovikbook,VollhardtWoelfle}\\
\hline
3D Weyl and Dirac semimetals & Energy bands & Unclear & \cite{Liu13Na3Bi, Borisenko13,Neupane13}\\
Cd$_3$As$_2$, Na$_3$Bi \\ 
\hline \\
\end{tabular}
\caption{Table of Dirac materials indicated by material family, pseudospin realization in the Dirac Hamiltonian, and the energy scale for which the Dirac spectrum is present without any other states.
}
\label{tab:DM}
\end{table}

Historically, the Dirac equation
\begin{equation}
i\hbar\frac{\partial}{\partial t}\psi=(c{\bm\alpha}\cdot\bp+\beta mc^2)\psi
\label{eq:Dirac_hep}
\end{equation}
was introduced to formulate a quantum theory that was compatible with special relativity and which explained the fine structure of atomic spectra \cite{Dirac1928}. Naturally, the Dirac equation was first formulated in (3+1)-dimensional space-time, where  ${\bm\alpha}$ and $\beta$ form an algebra of anticommuting $4\times 4$ matrices and $\psi$ is a four component spinor. In (2+1) or (1+1) dimensions the right-hand-side of the Dirac equation (\ref{eq:Dirac_hep}) reduces to the $2\times 2$ Hamiltonian in (\ref{eqn:Dirac}). The original Dirac equation is Lorentz covariant, i.e.~it does not change its form under transformation between two inertial frames, and it can be rewritten in a manifestly covariant form as $(\gamma^\mu p_\mu-mc)\psi=0$ \cite{BjorkenDrell}.
For condensed matter systems, the Dirac Hamiltonian contains the Fermi velocity instead of the speed of light and is tied to the rest frame of the material. It is thus not Lorentz covariant. A Schr{\"o}dinger equation with a Dirac Hamiltonian $H_D$ nevertheless equals the form of the Dirac equation (with $c\to v_F$) and is therefore often simply called a ``Dirac equation". 

\begin{figure}
\includegraphics[width=\columnwidth]{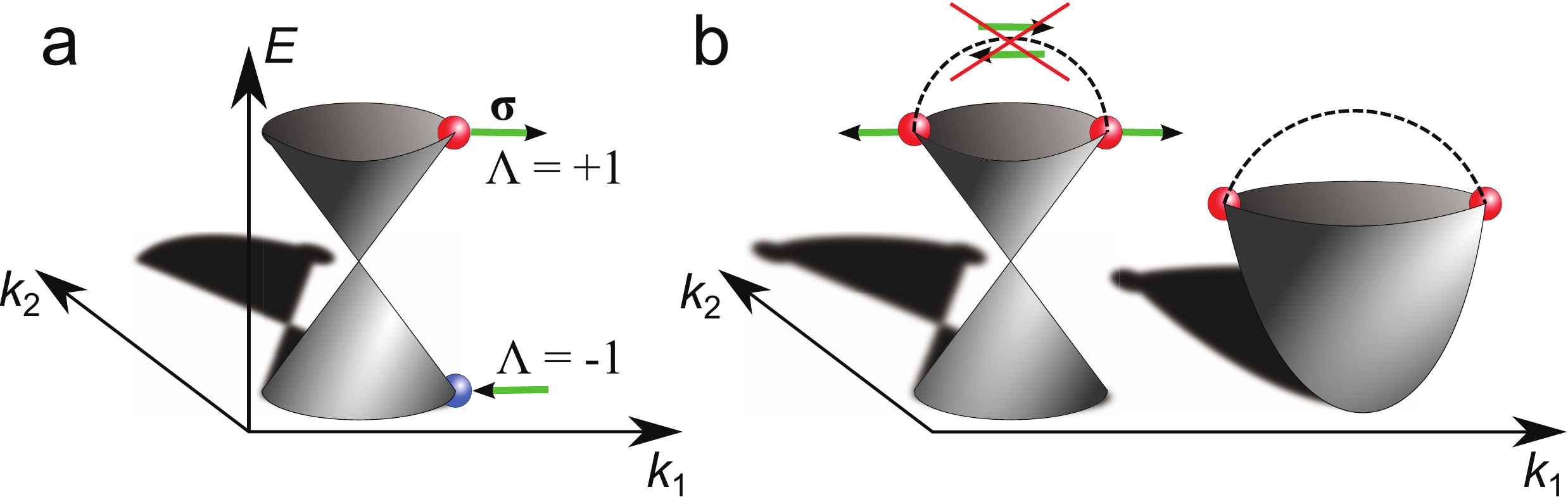}%
\caption{Dispersion of Dirac particles and suppression of back scattering in two dimensions. (a) The energy $E$ as a function of momentum $\bk=(k_1,k_2)$ obeying a conical dispersion relation. In Dirac materials, the momentum is coupled to a (pseudo)spin $\bm{\sigma}$ (green arrows). The helicity (or chirality) $\Lambda=\bm{\sigma}\cdot\bk/|\bk|=\pm 1$ is well defined and conserved, with particles and holes having opposite chirality. (b) Back scattering of Dirac fermions ($\bk\to-\bk$) requires a simultaneous flip of the the (pseudo)spin $\bm{\sigma}\to-\bm{\sigma}$. If an impurity potential preserves chirality, back scattering is thus forbidden, in contrast to usual metals with Schr{\"o}dinger-like quasiparticles (right) which, in general, allow for back scattering. (Dashed lines schematically depict the impurity potential.)}
\label{fig:Dirac_chiral}
\end{figure}

Naturally, a variety of novel phenomena can be directly translated from relativistic quantum mechanics to Dirac materials. 
For any particle, the ratio $cp/mc^2$ of the momentum $p=|\bp|$ to its rest mass $m$, or energy gap in Dirac materials, determines whether the particle behaves quasi non-relativistically $(cp\ll mc^2)$ or whether relativistic effects take over and eventually completely change the dynamics. Importantly, a new conserved quantity, the chirality $\Lambda={\bm\sigma}\cdot\bp/p$, emerges when the kinetic energy largely exceeds the rest mass $(cp\gg mc^2)$ in the ultrarelativistic limit. The chirality takes the values $\Lambda=\pm 1$ and distinguishes between right- and left-handed particles, i.e.~particles where spin and momentum are parallel or antiparallel, respectively, as displayed in Figure \ref{fig:Dirac_chiral}(a). The spin in a Dirac material does not have to be the electron spin, but is more often derived from another degree of freedom.
Chirality strongly affects the particle dynamics. Back scattering $\bp\to-\bp$ has to be accompanied by a spin-flip ${\bm\sigma}\to-{\bm\sigma}$ when the chirality is conserved or is otherwise forbidden, see Figure \ref{fig:Dirac_chiral}(b). Therefore, ultrarelativistic particles can tunnel through arbitrarily high and wide potential barriers. This so-called Klein paradox is a manifestation of particle-antiparticle conversion at a sharp potential step. Quite dramatically, the Dirac equation in presence of a sufficiently strong Coulomb potential leads to the so-called relativistic collapse of electrons into superheavy nuclei and a rearrangement of the vacuum with the production of particle-antiparticle pairs. Another directly transferable phenomenon is the enhanced sensitivity to magnetic fields in two dimensions. The spacing between the quantized energy levels of the electrons in a magnetic field $B$ changes from being linear in $B$ for Schr\"odinger fermions to a $\sqrt{B}$ dependence in massless Dirac systems.
The numerous distinctions between relativistic and non-relativistic behavior which are often introduced in the context of elementary particles are very closely related, and arguably most directly observable, in condensed matter systems, where analogues of virtually all above mentioned ultrarelativistic effects have been observed in table top experiments. 

This review is dedicated to discuss multiple classes of Dirac materials and their common features. In Section 2 we start with showing how Dirac fermions appear in many different materials, ranging from graphene, topological insulators, to high-temperature cuprate superconductors and Weyl semimetals. We then discuss the stability and symmetry protection of Dirac nodes in Section 3. In Section 4 we proceed by reviewing how a low-energy Dirac spectrum has experimentally been established in numerous Dirac materials by presenting results from angle-resolved photoemission spectroscopy (ARPES) and scanning tunneling spectroscopy (STS).
Section 5 is dedicated to the effects of universal many-body interactions within the low-energy Dirac spectrum.  In Section 6 we discuss the universality of the various properties of Dirac materials. Finally, we end the review with a brief concluding statement in Section 7.
 
\section[]{Microscopic origin of Dirac excitations} \label{sec:MicroscopicOrigin}
We will start by explicitly illustrating how, in the low-energy limit, material specific microscopic Hamiltionians can lead to a massless Dirac equation. We will do so for the two-dimensional Dirac materials graphene, topological insulators, and $d$-wave superconductors, as well as  for the proposed and very recently discovered three-dimensional Dirac materials: Weyl semimetals and Dirac semimetals, respectively. These are all distinctly different systems, yet they share a common low-energy spectrum. It will here become clear that the occurrence of massless Dirac fermions materials generally traces back to a material specific symmetry, which enforces degeneracy and a vanishing mass term ($m=0$) in the Dirac Hamiltonian (\ref{eqn:Dirac}).

\input{Subsection_graphene}

\input{Subsection_TI}

\input{Subsection_dwave}

\input{Subsection_Weyl}

\input{Section_Stability}

\section{Experimental confirmation of Dirac energy spectrum}
Having established the microscopic origin of the Dirac spectrum in several very different Dirac materials we will now  explicitly review experimental results confirming the existence of Dirac materials. 
The defining linear spectrum of a Dirac material can be directly probed using experimental techniques measuring the $\bk$-resolved band structure.
Angle-resolved photoemission spectroscopy (ARPES) is the most direct and also the leading experimental probe for obtaining the band structure of a material. As long as the material in question has a clean enough surface which exposes either the relevant surface states or allows probing of the bulk bands, ARPES offers an invaluable tool for experimental band structure measurements. Scanning tunneling spectroscopy (STS) has also emerged as an indispensable and complementary surface characteristics tool. Although it captures real space data, careful Fourier transform analysis can provide valuable insight into reciprocal space, both in terms of dispersion relations but also pseudo spin information. In this section we will review experimental results from both of these techniques aimed at verifying and demonstrating the Dirac spectrum in Dirac materials.
%
\input{Subsection_ARPES}

\input{Subsection_QPI}

\input{Section_ManyBodyEffects}

\section{Universal properties of Dirac materials}
As seen in the previous section, many-body effects are often shared between different Dirac materials. We now turn to review even more properties that are shared between Dirac materials, despite the often widely different microscopic origin of the Dirac spectrum.
Some of these are straightforward consequences of having the same low-energy spectrum and its accompanied reduced phase space. These properties are then common to all Dirac materials, irrespective of the nature of the quasiparticle excitations. This includes  thermodynamic properties as well as some transport properties.
Others, such as scattering and impurity effects of Dirac materials are also universal and intimately tied to the (pseudo)spin-momentum locking in the Dirac spectrum. 
While there exist a multitude of different properties possible to review, we have here chosen to focus mainly on thermodynamic properties, magnetic field dependence, transport, and impurity responses, as these are both important for potential applications of Dirac materials and very representative in illustrating how Dirac materials often behave fundamentally different from normal metals.

\subsection{Thermodynamic properties}
A  direct consequence of the Dirac Hamiltonian in Eq.~(\ref{eqn:Dirac}) is the low-energy form of the density of states (DOS), $N_0(E)$, for $d$-dimensional Dirac materials around the Dirac point:
\beq
N_0(E)\sim E^{d-1}.
\eeq
Therefore, thermodynamic response functions of Dirac materials can be characterized by a universal set of exponents that describe their power-law dependence on temperature. For example, the specific heat of Dirac materials is $C(T)_{T \rightarrow 0}\sim T^d$ which is notably different from $C(T)_{T \rightarrow 0}\sim T$ encountered for normal metals. This universal and characteristic behavior of Dirac materials is a direct consequence of the low-energy fermionic excitations being created around the nodal points. Notably, thermodynamical properties are not sensitive to the charge characteristics of the quasiparticles, and are shared between both superconducting and normal state Dirac materials.
Experimentally, the low temperature specific heat in several superconducting cuprates has been shown to display a $T^2$ dependence, although both linear (possibly extrinsic) and cubic (phonon) terms are also often present \cite{Moler94PRL, Wright99PRL, Tsuei00RMP, Hussey02}.
It is, however, also important to note that the universal behavior for clean Dirac materials only accounts for the band structure effects, and it is possible that electron-electron interactions also influence the temperature dependence. For example, in undoped graphene, Coulomb interactions have been proposed to logarithmically suppress the temperature dependence of the specific heat due to poor screening at the Dirac point \cite{Vafek07PRL}. In doped Dirac materials, normal metal behavior can be recovered as a more metal-like Fermi surface develops, as illustrated for graphene in Ref.~\cite{Ramezanali09}.

\input{Subsection_magneticfield}

\input{Subsection_chiralanomaly}

\subsection{Transport in Dirac materials}
A lot of attention has also been directed towards Dirac materials for their special electronic transport properties. Effects like suppressed back scattering \cite{Ando_98a}, Klein tunneling \cite{Katsnelson_book}, creation of midgap states \cite{Balatsky_RMP}, universal minimum conductivity, weak antilocalization, suppression and restoring of Anderson localization \cite{Mirlin_Evers_RMP}, and scattering off random (pseudo)magnetic fields \cite{Katsnelson_book} have all been investigated theoretically. Many of these effects have also been realized in experiments on graphene and are the subject of several reviews \cite{AHC_RMP, Peres_RMP10, DasSarma_RMP11, Katsnelson_book}. In the case of three-dimensional topological insulators like Bi$_2$Se$_3$, the situation is more complex due to large bulk contributions to the electron transport \cite{Culcer_Review12} but sample fabrication techniques for three-dimensional topological insulators are currently approaching a point where studies of two-dimensional Dirac electron transport are becoming possible \cite{YAndo_JPSJ13}.

In this section we chose to mainly focus the discussion on the very characteristic suppression of back scattering in Dirac materials. In the next sections we will also review the creation of low-energy impurity resonances or so-called midgap states,  and their manifestations in electron transport and local probe experiments, which are both universal and very distinctive for Dirac materials.

\subsubsection{Suppression of back scattering}
One of the most profound features that controls the DC transport of Dirac materials is the suppressed phase space for back scattering. This feature has been discussed within the context of graphene \cite{Ando_98a,Ando_98b,Katsnelson_KP06}, $d$-wave superconductors \cite{Hirschfeld05} and more recently in topological insulators \cite{Yazdani_Nature09}, and is thus a very universal phenomenon for two-dimensional Dirac materials, largely independent on the charge of the Dirac fermions.

In the Dirac Hamiltonian  in Eq.~(\ref{eqn:Dirac}) the momentum is coupled to a (pseudo)spin described by the Pauli matrices $\sigma_i$, which results in helicity $\Lambda=\bk\cdot\bm{\sigma}/|\bk|$ being a good quantum number, pictorially illustrated in Figure \ref{fig:Dirac_chiral}(a). In two dimensions, the corresponding eigenfunctions are spinors and read as 
\begin{align}
\Psi_\bk (r)=e^{i\bk\cdot\br}\chi_{\bk,\Lambda} \, \, \, {\rm with} \, \, \, \chi_{\bk,\Lambda}=e^{i\sigma_z\theta_{\bf k}/2}\left(\begin{array}{c}1\\ \Lambda \end{array}\right) \, \, \, {\rm and} \, \, \, \theta_{\bf k} = \arctan\frac{k_y}{k_x}.
\end{align}
For back scattering $\bk\to-\bk$, i.e. $\theta_{\bf k}\to \theta_{\bf k}+\pi$ while conserving the helicity, the pseudospin parts of the wave functions become orthogonal: $\langle\chi_{\bk,\Lambda},\chi_{-\bk,\Lambda}\rangle={\rm Tr} \, e^{i\sigma_z\pi/2}=0$. Hence, back scattering off potentials which are diagonal in pseudospin space is suppressed, as also illustrated in Figure \ref{fig:Dirac_chiral}(b).
This statement holds also for multiple scattering events due to the Berry phase intrinsic to the two-dimensional Dirac Hamiltonian: If a particle is propagated around a closed loop in momentum space that encloses the origin, its wave function will acquire a
phase $\pm \pi$ after each $\theta_{\bf k} \rightarrow \theta_{\bf k} \pm
2\pi$. As a consequence, scattering in momentum space along time reversed paths, i.e. $\theta_{\bf k}\to \theta_{\bf k}+\pi$ as compared to $\theta_{\bf k}\to \theta_{\bf k}-\pi$, results in final state wave functions which differ by a phase $\pi$ and thus interfer destructively \cite{Ando_98b}. Hence, the back scattering amplitude is suppressed also for multiple scattering events off potentials which are diagonal in pseudospin space. This destructive interference is most effective in zero magnetic field, weakened in finite magnetic field and leads to weak antilocalization. Weak antilocalization related with Dirac fermions has been most unambiguously been measured in graphene \cite{Peres_RMP10}. In topological insulators weak antilocalization can be expected in the topologically protected surface states but also in bulk or quantum well states \cite{YAndo_JPSJ13}. The assignment of weak antilocalization to the topological protected surface states is thus more difficult but might be achieved in combination with further sample characterizations \cite{Fu-Chun_PRL2011,Bao_SciRep2012,Juhn-Jong_PRB13}.

In the case of graphene, a potential diagonal in pseudospin space is generated by any electric field which is smooth on the scale of the lattice constant. Hence, transport in  graphene is insensitive to such perturbations \cite{Ando_98a,Ando_98b} and graphene p-n junctions display 100$\%$ transmissions for charge carriers at normal incidence, which is often referred to as Klein tunneling \cite{Katsnelson_KP06}. Experimentally suppressed back scattering has been demonstrated both by magnetotransport measurements in graphene showing weak antilocalization \cite{WAL_Gr_Savchenko_09} and characteristic phase shifts in Fabry-Perot oscillations in graphene n-p-n junctions \cite{Young_Kim_NPhys09}. Potentials which vary significantly on the scale of the lattice and couple either the two valleys in graphene or break the sublattice symmetry can restore back scattering. In contrast, in topological insulators any electrostatic potential enters the low-energy Dirac Hamiltonian as a scalar potential and thus cannot cause back scattering --- at least if there is no interaction of the topological surface states with bulk states or surface quantum well states at the same energy.

For both, graphene \cite{Brihuega_PRL08} and topological insulators \cite{Yazdani_Nature09,Kapitulnik10}, STS experiments reported the absence of back scattering signatures in quasiparticle interference maps and in FT-STS. However, the interpretation of these STS experiments can be difficult due to the issue of pseudospin averaging which might hinder the detection of $2k_F$ oscillations \cite{Peres_NJP09,Mallet_PRB12}, as explained in section 4.2.

\subsection{Creation of low-energy impurity states}
\begin{figure}
\begin{center}
\includegraphics[width=0.8\columnwidth]{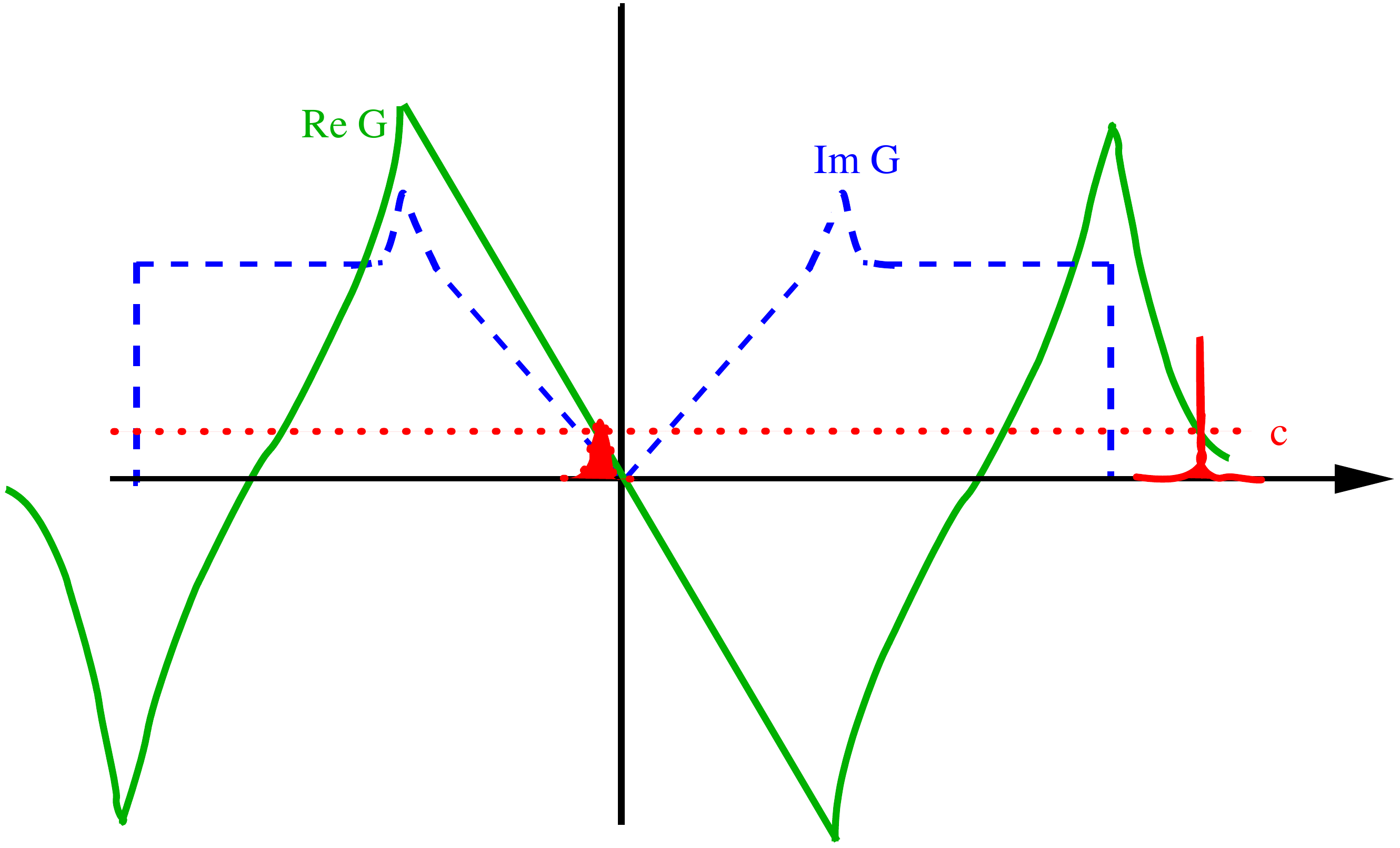}
\caption{\label{fig:G_Eimp} Universal local electronic properties of Dirac materials. Real (solid green line) and imaginary (dashed blue line) parts of the local Green function.
and the graphic solution of Eq.~(\ref{eqn:G_Eimp}) for $V = 1/c >0$ (red dotted line). We here show a
physically relevant solution, where the width of the resonance $\Gamma\ll
E_{\rm imp}$ is small. If $\Im G(E_{\rm imp})\gg \Re G(E_{\rm imp})$ the
resonance is broadened and merges with the continuum. The virtual bound
state (red) inside the gap is well resolved for large $V$, i.e.~small $c=1/V$. (Reprinted figure with permission from A.~V.~Balatsky {\it et al.}, Rev. Mod. Phys. 78, 373 (2006) \cite{Balatsky_RMP}. \href{http://link.aps.org/doi/10.1103/RevModPhys.78.373}{Copyright \copyright~(2006) by the American Physical Society.})}
\end{center}
\end{figure}

We now turn to the opposite disorder limit, that is impurities localized on an atomic scale. Here multiple scattering becomes important and two-dimensional Dirac materials respond again in a universal way: For any material with linear
density of states $N_0(E)=\frac{|E|}{D^2}\cdot\Theta(D-|E|)$, where $D$ is a bandwidth parameter, the local Green function has the form
\begin{align}
G^0(E)\approx\frac{E}{D^2}\ln\left|\frac{E^2}{D^2}\right|-i\pi N_0(E),
\end{align}
and illustrated in Figure~\ref{fig:G_Eimp}.

The change in the Green function, $\hat G=\hat G^0+\hat G^0\hat T\hat G^0$, encoding the local electronic properties as well as the contribution of an impurity to the quasiparticle scattering rate $\tau^{-1}\sim |T(E_F)|^2 N_0(E_F)$, is determined by the $T$-matrix: $\hat T=(1-\hat V \hat G^0)^{-1}\hat V$. An
atomically sharp impurity, described by the $\delta$-function potential
$\bra{x}\hat V\ket{x}=V\delta(x)$ causes a resonance at energy $E_{\rm imp}$ in the $T$-matrix for
\begin{equation}
\label{eqn:G_Eimp}1/V=\Re G^0(E_{\rm imp}),
\end{equation}
given that $\Im G(E_{\rm imp})$ is sufficiently small, see Refs.~\cite{Balatsky_RMP,Wehling_CPL09} for reviews.
In practice, impurity potentials will have a finite range. As long as the impurity generated perturbation has a short range component on the scale of the lattice constant, the $\delta$-function potential approximation employed in the analysis is adequate and provides at least a qualitative guidance on the response of materials to atomic defects, like interstitial, substitutional or adatom impurities.
In this cases, Eq.~(\ref{eqn:G_Eimp}) can be solved graphically, as illustrated in Figure \ref{fig:G_Eimp}, as well as analytically:
\begin{equation}
\label{eqn:modelDOS_Eimp}
 E_{\rm imp}\approx\frac{D^2}{2V\ln\left|\frac{D}{2V}\right|}.
\end{equation}
Hence, if the potential strength is on the order of the bandwidth or
stronger a resonance close to the nodal point is created. This is in striking contrast to normal two-dimensional semiconductors, where arbitrarily weak perturbations can create sharp resonances at the gap edges. Two-dimensional Dirac materials feature a linearly vanishing density of states around the Dirac point which is frequently called ``pseudogap". Hence, the low-energy impurity states discussed here are often termed ``intragap" or ``midgap" states.

\begin{figure}[tb]
\begin{minipage}{.49\linewidth}
\vfill
\includegraphics[width=.98\linewidth]{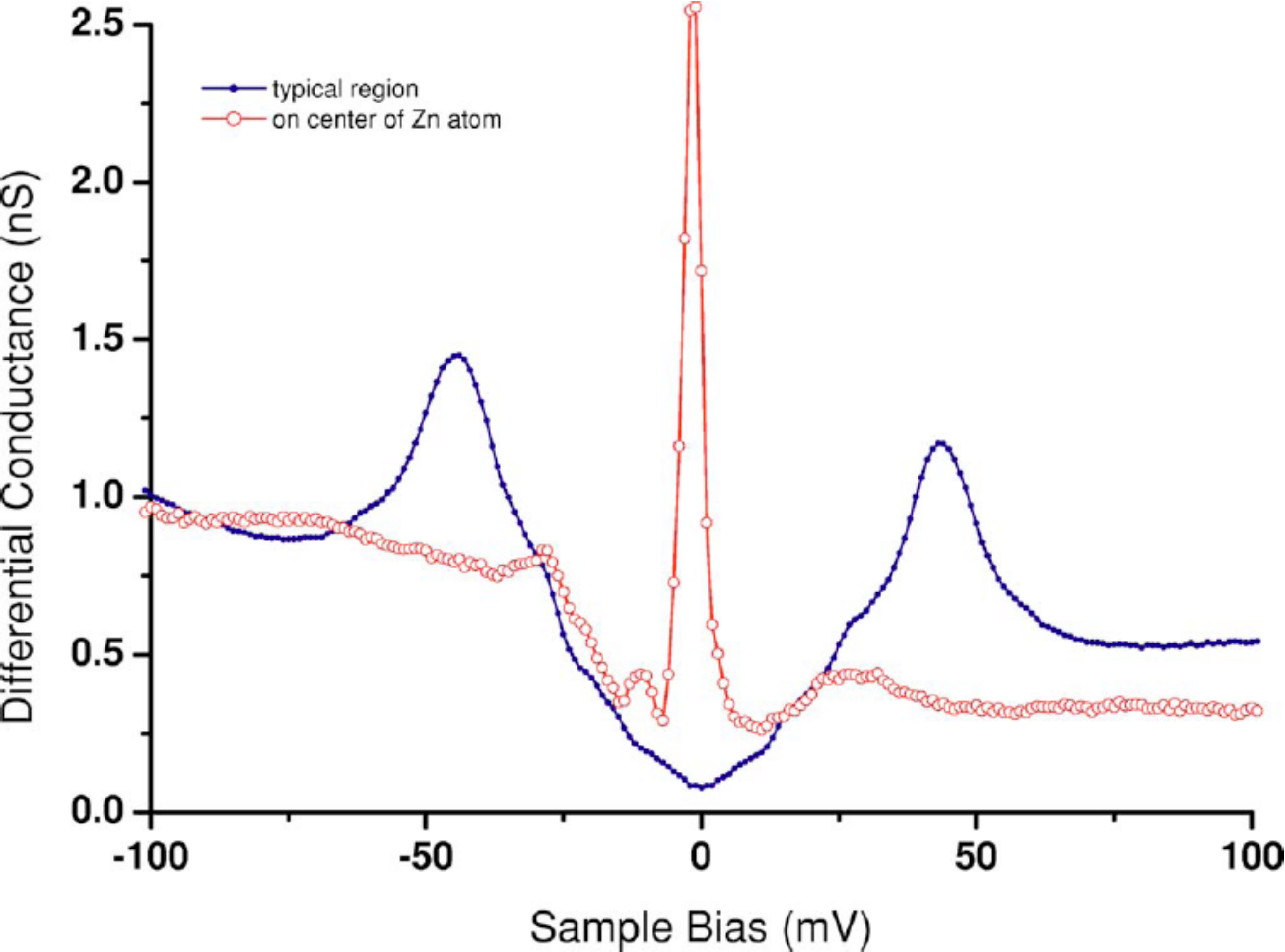}
\end{minipage}
\begin{minipage}{.49\linewidth}
\centering
\includegraphics[width=.8\linewidth]{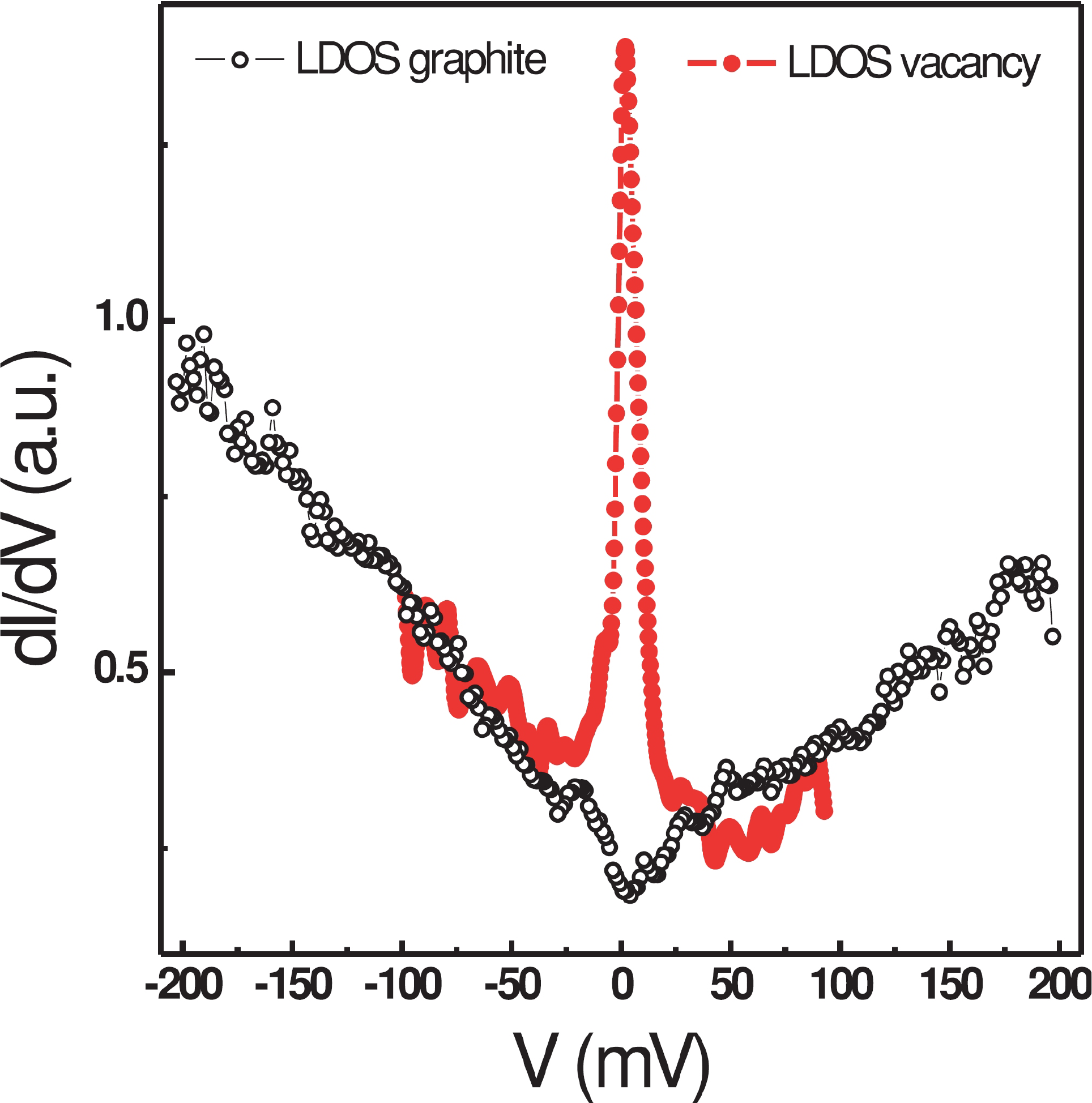}
\end{minipage}
\caption{\label{fig:STM_imp} Low-energy impurity resonances in $d$-wave cuprate superconductors and in graphite. Left: STM studies on single crystals of the high-temperature cuprate superconductor Bi$_2$Sr$_2$Ca(Cu$_{1-x}$Zn$_x$)$_2$O$_{8+\delta}$ with $x=0.6\%$. The differential tunneling spectra taken at the Zn-atom site (red open
circles) and far away from the impurity (blue filled circles). 
(Reprinted by permission from Macmillan Publishers Ltd: Nature, S.~H.~Pan {\it et al.}, Nature 403, 746 (2000) \cite{Hudson00}, copyright \copyright ~(2000).)
Right: STM d$I/$d$V$ spectra of graphite near a single vacancy (red filled circles) and in a defect-free area (black open circles). As in the case of the high-temperature cuprate superconductor, the defect causes a sharp resonance at the Fermi level. 
(Reprinted figure with permission from M.~M.~Ugeda {\it et al.}, Phys.~Rev.~Lett.~104, 096804 (2010) \cite{Gomez-Rodriguez_PRL10}. \href{http://link.aps.org/abstract/PRL/v104/p096804}{Copyright \copyright~(2010) by the American Physical Society.})
}
\end{figure}

For nodal superconductors the characteristic energy scale $D$ is set by the gap $\Delta_0\approx 30$\,meV \cite{Balatsky_RMP} below which the DOS depends linearly on the energy. Hence, any local impurity exceeding this minimum potential strength can create intragap resonances. An experimental demonstration of these resonances has been reported in STM experiments \cite{Hudson00} on the high temperature cuprate superconductor
Bi$_2$Sr$_2$Ca(Cu$_{1-x}$Zn$_x$)$_2$O$_{8+\delta}$. At $x=0.6\%$ Zn doping, a characteristic resonance close to the nodal point was found in the vicinity of Zn impurities, as seen in Figure \ref{fig:STM_imp} (left).
Zn is believed to substitute Cu in the Cu-O plane without changing the doping. Then, Zn$^{2+}$ is in the $d^{10}$ configuration, and the third ionization energy yields a rough measure of the impurity potential $V$. By comparing these energies for a Cu atom $E_{{\rm Cu}^{2+}}=-36.8$\,eV and a Zn atom $E_{{\rm Zn}^{2+}}=-39.7$\,eV, we estimate $V=-2.9$\,eV. Since $|V|/D\sim 10^2$, we therefore expect the Zn to act as a strong potential scatterer for holes in the CuO$_2$ plane and a low-energy resonance is created according to Eq.~(\ref{eqn:modelDOS_Eimp}).

It is very instructive to apply same considerations to graphene to explicitly show how impurity resonances are universal for two-dimensional Dirac materials. Graphene is a Dirac material with entirely different real material background which manifests in the characteristic energy scale of $D\approx 6$\,eV \cite{Wehling_PRB07} being two orders of magnitude larger than for the $d$-wave cuprate superconductors. Nevertheless, intragap states close to the Dirac point exist in graphene and related materials. Scanning tunneling spectra in the vicinity of impurities in graphite have been shown to exhibit sharp low-energy resonances \cite{Fukuyama06,Gomez-Rodriguez_PRL10}, see e.g.~Figure \ref{fig:STM_imp}(right). Here, monovalent impurities like H adatoms or various organic groups present natural sources for the generation of intragap states as they crack graphene's $sp^2$ network and form strong covalent bonds with the carbon atoms \cite{casolo-2008,boukhvalov:08,Wehling_Midgap_PRB09,Wehling_ResScatt10}. Consequently, there is hopping $V_t\sim D$ from carbon to the additional orbital brought along by each of these impurities. With the impurity orbital energy being typically in the range $|\epsilon_d|\lesssim D/20$, this frequently leads to \textit{effective} local potentials on the order of $V=V_t^2/\epsilon_d\gtrsim 20D$, which act on the graphene Dirac fermions \cite{Wehling_ResScatt10}. Hence, intragap states within some $30$\,meV around the Dirac point are created by various adsorbates. Recent experiments have shown that these midgap states are very likely the dominant scattering mechanism in current high-mobility graphene samples \cite{Nietal10,Katoch_PRB10,Peres_RMP10}.

Topolgoical insulators also appear to have a very similar response to local defects. For either magnetic or nonmagnetic impurities one expects to see essentially the same impurity induced resonances as in the case of graphene and $d$-wave cuprate superconductors \cite{SCZhangimp,Biswasimp,Black-Schafferimp,Black-Schafferimp2}. Indeed, impurity-induced resonances are seen in scanning tunneling spectroscopy measurements \cite{Alpichshev_PRL12, Honolka_PRL12, Teague12} but a full comparison between theoretical predictions and observations is still an outstanding question. Possible explanations of the measured low-energy resonances include Dirac electron midgap states but also donor or acceptor levels derived from the bulk valence and conduction bands.

The above examples stress the point that the real material background is important in determining the response of a specific Dirac material to specific impurities. However, while there are specific paths for different Dirac materials to enter the regime of strong local impurities, midgap states appear as a hallmark for all two-dimensional Dirac materials.
In graphene (and also silicene) the Dirac electron bandwidth is on the order of several eV. Thus, energies associated with the formation of midgap states are also in this range and can even affect adsorbate chemistry. Midgap states are created by shifting spectral weight from far below and far above the Dirac point to approximately zero energy. Thus, half-filled midgap states are energetically most unfavorable. Therefore, adsorption of chemical species which create midgap states are expected to be more favorable in charge doped graphene rather in in charge neutral graphene. This expectation has been confirmed by first-principles calculations \cite{Huang_JCP}. Several instabilities to (partly) release the energy in half-filled midgap states have been suggested to occur at finite impurity concentrations. First, the high density of states at the Fermi level might cause a Stoner instability and render the system magnetic \cite{Yazyev_RepProgPhys10}. Moreover, structural instabilities could occur. Indeed, midgap states originating from different impurity sites hybridize and mediate an interaction between the adsorbates. This interaction has been shown to turns out attractive on average and thus favors the formation of impurity clusters \cite{Levitov_PRL09}.

\subsection{Real space properties of impurity resonances}
In the last section we reviewed how strong impurities create universal low-energy resonances in Dirac materials. Here we will now focus on the real space properties of the impurity states as they both alter the local density of states (LDOS), $N(r,E)=-\frac{1}{\pi}\Im\bra{r}\hat G(E)\ket{r}$, and cause oscillations in $N(r,E)$ which provide peculiar fingerprints in Dirac materials.
First, oscillations in $N(r,E)$ show a particular power-law decay which is characteristic of Dirac
materials. Second, the impurity resonances exhibit characteristic pseudospin structures which directly relate to the pseudospin inherent to the Dirac Hamiltonian. We illustrate the theoretical background and experimental observations of these Dirac material characteristics in the following.

The Green function $G(r,0,E)=\bra{r}\hat G(E)\ket{0}$ gives the information on how LDOS modulations extend away from a defect localized at $r=0$. Specifically, we can write $N(r,E)=N_0(E)+\Delta N(r,E)$ where then $\Delta N(r,E)=G(\br,0,E)T(E)G(0,\br,E)$. From the Dirac Hamiltonian we then find
\begin{equation}
 G(\br,0,E)\sim \frac{E}{4}\left(f_0(\rho)\sigma_0+f_1(\rho)\hat{\br}\cdot\bm{\sigma}\right),
\end{equation}
where $\rho=|\br| E/\hbar v_F$, $\hat{\br}=\br/|\br|$, $f_0(\rho)=Y_0(|\rho|)-i\sgn(\rho)J_0(|\rho|)$, and $f_1=i\sgn(\rho)Y_1(|\rho|)+J_1(|\rho|)$. $J_{1,2}$ and $Y_{1,2}$ denote the Bessel functions of first and second kind, respectively. The Green function thus decays as $G(\br,0,E)\sim 1/\sqrt{r}$ for $E\neq0$ and $G(\br,0,E)\sim 1/r$ for $E=0$ \cite{Balatsky_RMP,Bena05,Biswasimp}.
Therefore, any impurity state at the Dirac nodal point $E=0$ has to decay as $\Delta N(\br,E)\sim 1/r^2$ or faster. Explicit proof of this decay has been given by solving the spherical two-dimensional Dirac equation \cite{Lin06}, as well as specifically for graphene \cite{pereira_06}.

\begin{figure}%
\includegraphics[width=\columnwidth]{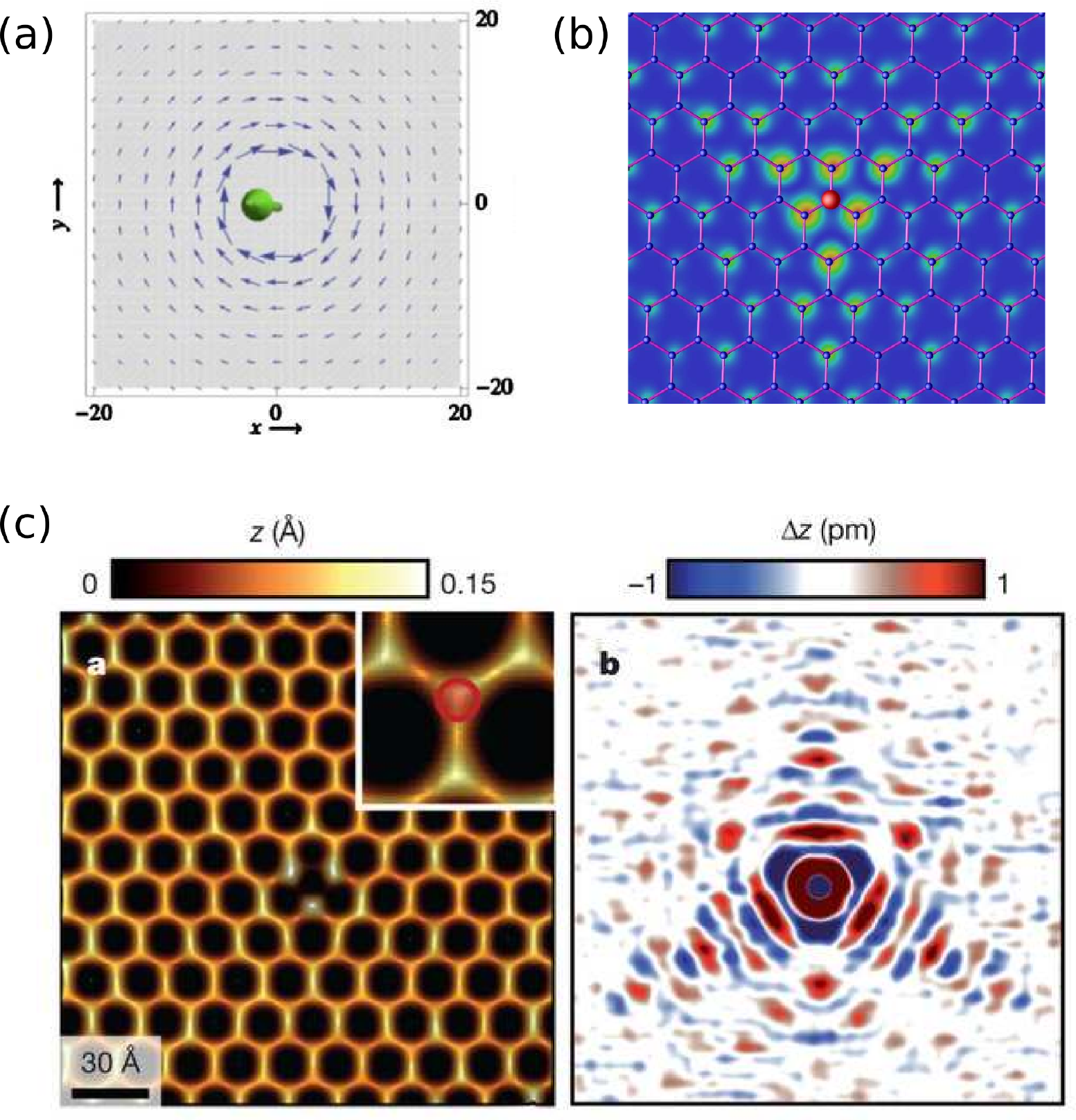}
\caption{Pseudospin structure of impurities in Dirac materials. (a) Vortex patterns in the spin-polarized LDOS from magnetic impurities on the surface of three-dimensional topological insulators when the impurity spin is $z$-polarized (solid green arrow). The topological surface state spin components in the $xy$-plane are denoted by small vectors (blue). 
(Reprinted figure with permission from R.~R.~Biswas and A.~V.~Balatsky, Phys.~Rev.~B~81, 233405 (2010) \cite{Biswasimp}. \href{http://link.aps.org/abstract/PRB/v81/p233405}{Copyright \copyright~(2010) by the American Physical Society.})
(b,c) Sublattice polarization of a midgap state in graphene induced by an impurity potential acting on sublattice A. (b) Theoretical prediction of the LDOS variation in the vicinity of the impurity. (Reprinted figure with permission from T.~O.~Wehling {\it et al.}, Phys.~Rev.~B~75, 125425 (2007) \cite{Wehling_PRB07}. \href{http://link.aps.org/abstract/PRB/v75/p125425}{Copyright \copyright~(2007) by the American Physical Society.})
(c) Experimental realization of a strong local impurity in artificial graphene. Left: STM topography of quasi-neutral artificial graphene with an additional CO molecule at the sublattice A site (location indicated in inset). Right: Impurity scattering quasi-particle interference mapped through the subtraction of two topographs measured in identical fields of view and distinguished only by the presence of the extra CO molecule at the top site. 
(Adapted by permission from Macmillan Publishers Ltd: Nature Physics, K.~K.~Gomes {\it et al.}, Nature 483, 306 (2012) \cite{Manoharan_Nature2012}, copyright \copyright ~(2012).)
}%
\label{fig:pseudospin}%
\end{figure}

For $E>0$ and  $|r|\gg \hbar v_f/E$ we instead obtain $G(\br,0,E)\to c \sqrt{\frac{1}{\rho}}e^{i\rho}(1+\hat{\br}\cdot\bm{\sigma})$. Hence, an impurity with the $T$-matrix, $T\sim (\sigma_0+\bm{\sigma}\cdot\bu)$, where $\bu=(u_1,u_2,u_3)$, leads to LDOS modulations of the kind
\begin{equation}
\Delta N(\br,E)\sim (1/r) \left(\bm{\sigma}\cdot\bu-(\hat{\br}\cdot\bu)(\bm{\sigma}\cdot\hat{\br})-i\bm{\sigma}(\bu\times \hat{\br})\right).
\label{eq:Nimp_r}
\end{equation}
Thus, any local impurity which breaks pseudospin symmetry ($\bu\neq0$) leaves a characteristic psuedospin polarization pattern in the LDOS. For topological insulators vortex patterns are expected around magnetic impurities in the spin polarized LDOS \cite{Biswasimp}, as displayed in Figure \ref{fig:pseudospin}(a). In graphene, the pseudospin instead corresponds to the sublattice degree of freedom and local impurities are then generally associated with a characteristic sublattice pattern in the LDOS. For instance, a strong local impurity acting on sublattice A leads to a low-energy impurity state which is mainly localized in sublattice B \cite{pereira_06,Wehling_PRB07}, as shown in Figure \ref{fig:pseudospin}(b). Recently, this sublattice pseudospin polarization has be demonstrated experimentally by means of STM on artificial graphene \cite{Manoharan_Nature2012}, see Figures \ref{fig:pseudospin}(c).

Also the real space decay of the impurity-induced LDOS variations shows a specific behavior in Dirac materials. The leading $1/r$ term to the LDOS modulation is suppressed if the impurity is isotropic in pseudospin space $\bu=0$, as seen in Eq.~(\ref{eq:Nimp_r}), or if only ${\rm Tr}\,\Delta N(\br,E)$ is measured. For $E\neq 0$ it thus depends on the type of local probe experiment, i.e.~whether $\Delta N(\br,E)$ or ${\rm Tr}\,\Delta N(\br,E)$ is measured, and on the type of impurity, i.e.~on $\vec u$ but also whether there is scattering between different Dirac nodes, if the LDOS oscillations has a $1/r$ or $1/r^2$ decay\cite{Bena_PRB09}. But we can still make the following universal argument: Whenever a $1/r^2$ decay of LDOS modulations is encountered, it strongly hints at a midgap state in a Dirac material. This is because impurity states in normal two-dimensional metals always decay  as $\Delta N(r,E)\sim 1/r$. Experimentally, $1/r^2$ decay is consistent with STM characterizations of intragap impurity states in the high-temperature cuprate superconductor Bi$_2$Sr$_2$CaCu$_2$O$_{8+\delta}$ \cite{Hudson_Science99}, in graphite \cite{Gomez-Rodriguez_PRL10}, and possibly also in topological insulators \cite{Alpichshev_PRL12}.

\section{Conclusions}
The classification of materials according to their common properties and functionalities is crucial for gaining unified and
 structured understanding of diverse sets of solids. In physics, classification
 often refers to broken symmetries or low-energy excitations that control the response of a particular state to external probes. Broken symmetry can be a powerful predictor of the materials properties: superconductors, for example, have broken gauge-symmetry and the resulting low-energy Goldstone mode is responsible for their ability to carry currents without dissipation.  
Another fundamental and widely employed classification of materials is whether they are metallic or insulating. Metals have non-zero phase space for low-energy electronic excitations, can conduct an electric current and their specific heat increases to leading order linearly with temperature. Insulators, on the other hand,  require a finite energy gap to be overcome by electronic excitations and thermally excited electron-hole pairs are exponentially suppressed at low temperatures. At the boundary between metals and insulators are the gapless semiconductors \cite{Tsildilkovski_book}, where both the bottom of the conduction band and the top of the valence band coincide with the Fermi level. Superconductors, metals, and insulators all form classes of states in materials, where general characteristics can be provided without full knowledge of the microscopic details.

In this work we have reviewed a class of materials which we call {\it Dirac materials}, unified by their low-energy excitation spectrum containing Dirac nodes with an accompanying linear Dirac dispersion relation. In absence of mass terms, these Dirac materials are all gapless. However, not all gapless semiconductors are Dirac materials, as most gapless semiconductors have a parabolic dispersion, as expected from the non-relativistic Schr\"{o}dinger equation. Dirac materials, on the other hand, exhibit excitations described by the Dirac Hamiltonian, where particles and holes are always interconnected.

Very generally, to achieve a Dirac spectrum and the associated band crossing Dirac nodal point, specific symmetries needs to be fulfilled. These symmetries vary depending on the material, for example in graphene the sublattice symmetry of the honeycomb lattice protects the Dirac spectrum, whereas in topological insulators it is time-reversal symmetry that gives the Dirac spectrum. The suppression of the symmetry universally results in a gapped state. 
We have in this review systematically discussed the protective symmetries for many different Dirac materials and also how they can be lifted, resulting in the generation of a finite Dirac mass.

The presence of Dirac nodes in the energy spectrum leads to the sharp reduction of the phase space for nodal excitations. The dimensionality of the set of points in momentum space where we have zero energy excitations is reduced in Dirac materials with the Fermi level aligned with the Dirac point(s). For example, Dirac nodes for a three-dimensional Dirac material mean that the effective Fermi surface has shrunk from two-dimensional object to a point. For a two-dimensional Dirac material the effective Fermi surface is likewise reduced from the expected one-dimensional line to a point. In either case there is a reduction of dimensionality of the zero-energy states. The reduction of phase space  and the controlling symmetries are important for the applications of Dirac materials. First one can lift the protected symmetry of the Dirac node and therefore destroy the nodes and open an energy gap. This modification of the spectrum of quasiparticles drastically changes the response of Dirac material, as for example is the case for topological insulator in a magnetic field \cite{Kane_RMP2010, Qi11RMP}. Second, Dirac nodes and the resultant reduction of phase space suppress dissipation and are thus attractive for applications utilizing the coherence of low-energy states in the nodes.

Beyond the shared low-energy band structure and the phase-space reduction Dirac materials have many more common features which all ultimately derive from the low-energy Dirac spectrum. These include:
\\
\\
 i) the density of states $N_0(E) \sim E^{d-1}$ for $d$-dimensional Dirac materials. This phase space restriction affect thermodynamic properties, for example by reducing the specific heat to $C(T) \sim T^{d}$;
\\
\\
ii) Landau level quantization with a square-root field-dependence $E_n(B)  \sim \sqrt {nB}$ for two-dimensional Dirac materials;
\\
\\
iii) Suppressed back scattering due to the Berry phase inherent to the Dirac matrix structure of the Hamiltonian. This suppression is responsible for the unique transport properties of graphene, $d$-wave cuprate superconductors and topological insulators, alike.
\\
\\
iv) Local response to impurities, where strong impurity scattering produces low-energy impurity resonances, so-called ''intragap states''.
\\
\\

Furthermore, the shared low-energy dispersion results in similar sensitivities to many-body effects, such as a lack of electric screening and velocity renormalization, although the effects of electron-electron interactions can also be sensitive to the material host. One example of the latter is that different ordered states are possible in the limit of strong electron-electron interactions, dependent on both the interactions and material. Another example are the $d$-wave superconductors where the superconducting condensate effectively screen the Coulomb repulsion.

 Looking forward, we notice that the field is moving fast in discovering new states and materials, which might well turn out to be additions to the Dirac materials class. Recent verified examples include materials like silicene \cite{Vogt12}, with its close cousin, germanene, still awaiting experimental confirmation. Likewise, the three-dimensional Dirac semimetals were just very recently experimentally discovered \cite{Liu13Na3Bi,Borisenko13,Neupane13}, with their close relative, the Weyl semimetals, still not yet experimentally found.
In addition, recent synthesis capabilities have reached the stage where we can grow films of materials on demand. This tunability offers another venue where we can perhaps synthesize Dirac materials either by atom deposition \cite{Gibertini_2DEG_09, Manoharan_Nature2012} or by synthesis of films using molecular epitaxy. All these  possibilities suggest that the field of Dirac materials is still only in its infancy and will grow significantly in the future.

Most importantly, Dirac materials as a general framework offers us a guidance in predicting the features that we have established to be universal. One example of this predictive power is the expectation that Dirac material is susceptible to gap opening once we lift the symmetry that protects its Dirac nodes. Another important prediction is that Dirac materials generates intragap impurity resonances, be it $d$-wave cuprate superconductors, graphene, or topological insulators. We expect a similar sensitivity from any, yet to be discovered, Dirac material.  At the same time, there is a natural energy range for any Dirac material above which the specific details of the band structure and lattice structure come into play and material differences begin to matter. It is this interplay of universality and material specifics which makes Dirac material physics so rich, and just like the currently known Dirac materials, we expect that new Dirac materials will play a key role in modern condensed matter physics and materials physics.

\section*{Acknowledgements}
We are grateful to D.~Arovas,D.~Abergel, R.~Biswas, A.~H.~Castro Neto, H.~Dahal, V.~Fal'ko, M.~Fogelstr\"{o}m, J.~Fransson, M.~Graf, Z.~Huang,  M.~I.~Katsnelson, A.~I.~Lichtenstein,  J.~Linder, F.~Lombardi, H.~Manoharan, J.~Moore, N.~Nagaosa, K.~Scharnberg, Z.~X.~Shen, Y.~Tanaka, O.~Tjernberg,  A.~Yazdani, S.~C.~Zhang, J.~X.~Zhu for discussions.
This work has been supported by US DoE BES, LDRD,  University of California UCOP-09-027, SFB-668 and SPP-1459 (Germany), Dirac Materials ERC-DM-321031, and the Swedish Research Council (VR). TOW thanks KITP Santa Barbara for hospitality during a visit where parts of this work were written.

For figures with copyright from the American Physical Society: Readers may view, browse, and/or download material for temporary copying purposes only, provided these uses are for noncommercial personal purposes. Except as provided by law, this material may not be further reproduced, distributed, transmitted, modified, adapted, performed, displayed, published, or sold in whole or part, without prior written permission from the American Physical Society.

\bibliographystyle{tADP}

\end{document}

%% file: Subsection_graphene.tex
\subsection{Graphene}
\begin{figure}
\begin{center}
\includegraphics[width=0.59\columnwidth]{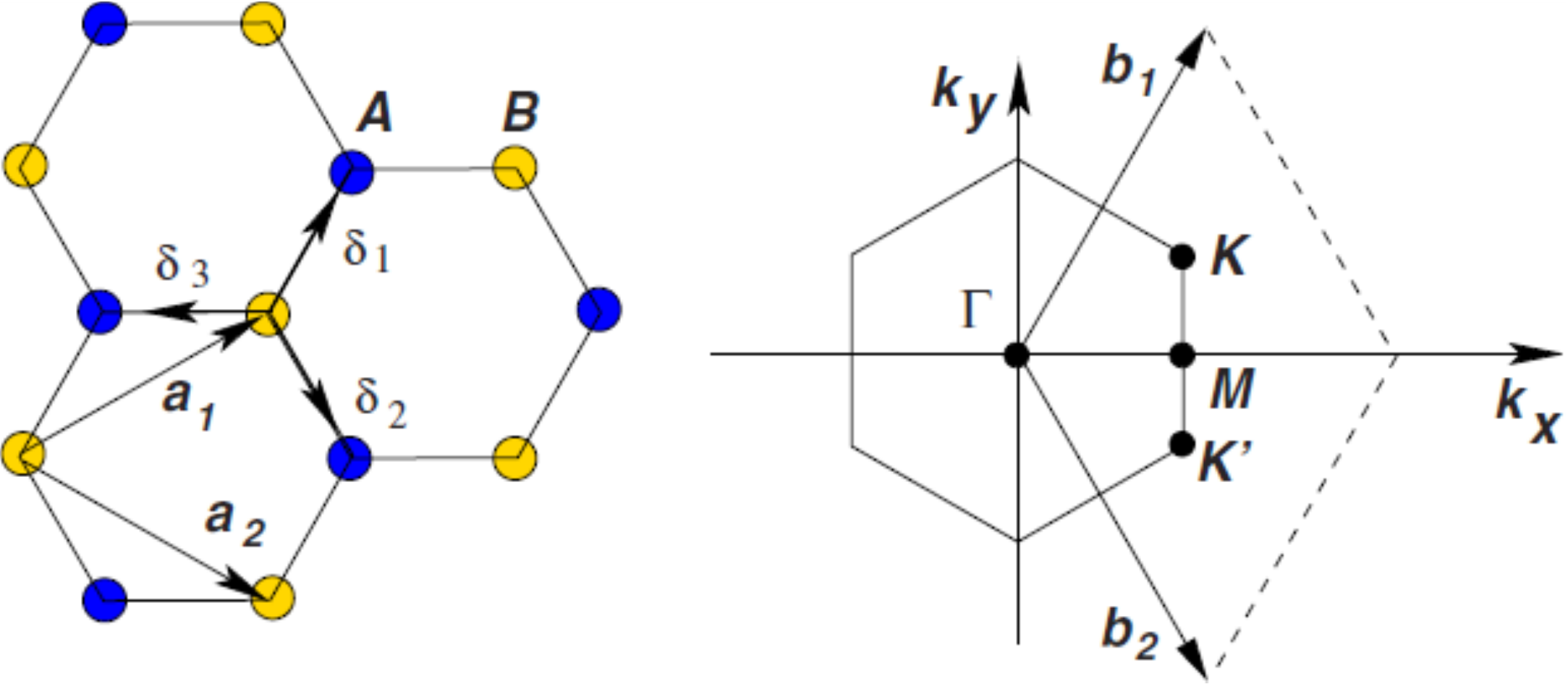}%
\includegraphics[width=0.39\columnwidth]{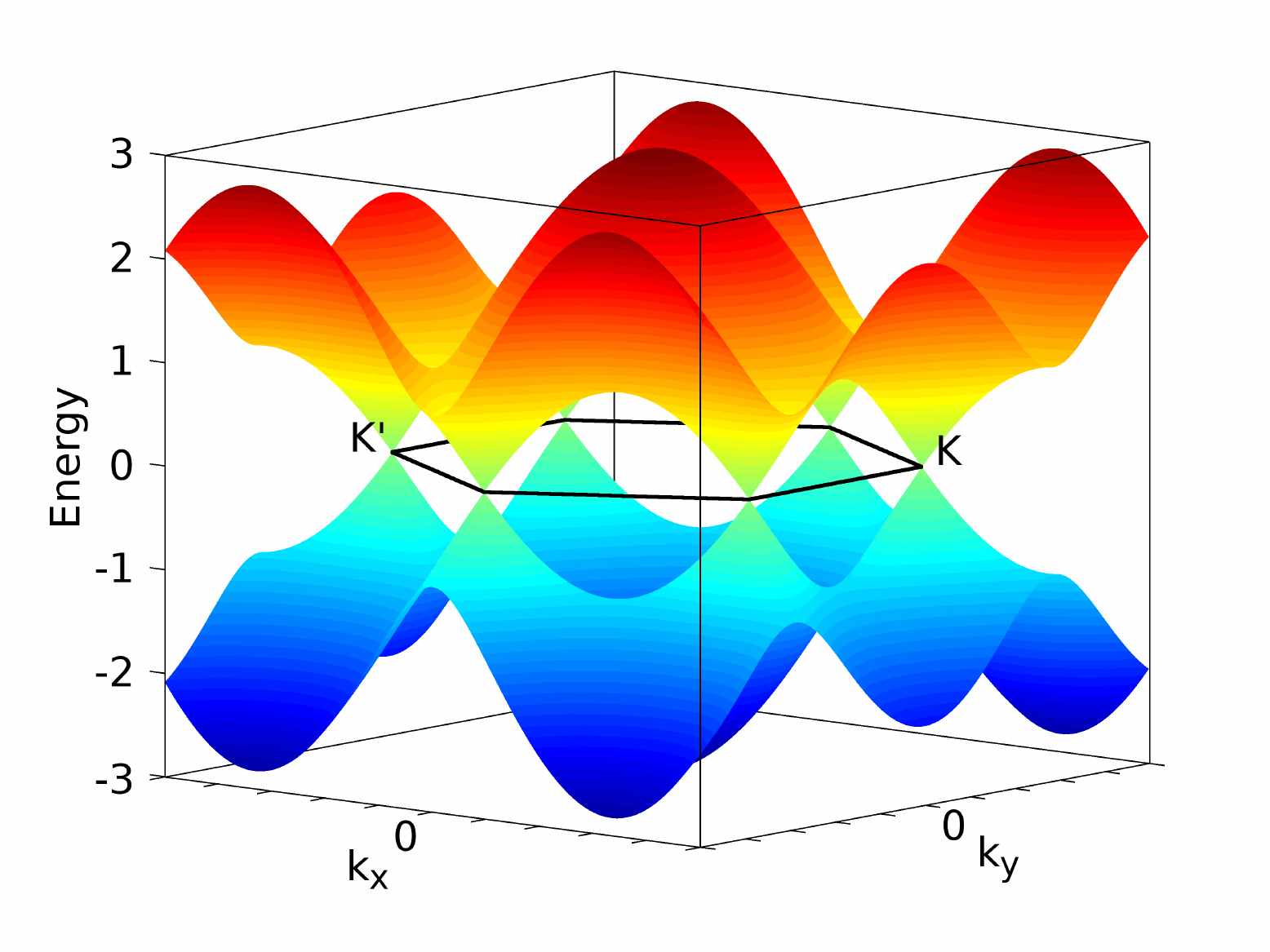}%
\caption{(a) Crystal structure of graphene consisting of sublattices A (blue) and B (yellow). The lattice vectors $\ba_{1,2}$ as well as the vectors $\bm\delta_{1,2,3}$ connecting nearest neighbour atoms are indicated. (b) Brillouin zone of graphene with the reciprocal lattice vectors $\bbb_{1,2}$. (Reprinted figures by permission from A.~H.~ Castro Neto et al., Rev. Mod. Phys. 81, 109 (2009)  \cite{AHC_RMP}. \href{http://link.aps.org/abstract/RMP/v81/p109}{Copyright \copyright~(2009) by the American Physical Society.}). Dispersion of the graphene $\pi$-bands in the nearest-neighbor tight binding model, equations~(\ref{eqn:TB-real-sp}) and (\ref{eqn:TB_graphene_2}).}
\label{fig:gr_struct}
\end{center}
\end{figure}
Graphene might arguably be the most famous Dirac material. It consists of a layer of carbon atoms arranged in a honeycomb lattice as shown in Figure \ref{fig:gr_struct}(a). Each unit cell of the hexagonal Bravais lattice contains two carbon atoms, which give rise to two sublattices, A and B. Obviously, atoms in sublattice A are surrounded by three nearest neighbors in sublattice B and vice versa. The lattice is thus bipartite, if only nearest neighbor coupling is included.

In the vicinity of the Fermi level, all electronic states consist of the out-of-plane carbon $p_z$ orbitals. These form $\pi$-bonds with neighboring atoms and the resulting $\pi$-bands, as displayed in Figure \ref{fig:gr_struct}(c), can be easily understood from a tight-binding Hamiltonian 
\begin{equation}
\label{eqn:TB-real-sp}
\hat{H}=-t\sum_{\langle i,j\rangle}a^\dagger_i b_j+a^\dagger_j b_i,
\end{equation}
which consists of a hopping term $t\approx 2.7$~eV between nearest neighbor atoms. Here, $a_i$ and $b_i$ annihilate electrons in the $p_z$ carbon atomic orbitals in unit cell $i$ in sublattice A and B, respectively \cite{Wallace-1947}. As there are two atoms per unit cell, the Hamiltonian (\ref{eqn:TB-real-sp}) leads to a $2\times 2$ matrix in momentum space representation: 
\begin{equation}
\label{eqn:TB_graphene_2}
H(\bk)  =  \left( \begin{array}{cc}
0 & \xi(\bk)\\
\xi^\ast(\bk)&  0\end{array} \right),
 \end{equation}
with $\xi(\bk)=-t(e^{i{\bm\delta}_1\cdot \bk}+e^{i{\bm\delta}_2\cdot \bk}+e^{i{\bm\delta}_3\cdot \bk})$. The three partial hopping amplitudes entering in $\xi(\bk)$ stem from hopping processes connecting each atom in sublattice A with its three nearest neighbors in sublattice B through the vectors ${\bm \delta}_i$ and vice versa. The resulting energy bands are $\epsilon(\bk)=\pm|\xi(\bk)|$, Figure \ref{fig:gr_struct}(c). It is only at the two inequivalent corners  of the Brillouin zone, $K$ and $K'=-K$, that the dispersive term vanishes $\xi(\bk)=0$ and there the two bands are degenerate. Here, the three hopping amplitudes add up with phase factors $e^{i(\bm\delta_{1,2,3}) {\bm K}}$, which reduce to $1$, $e^{+2\pi i/3}$, and $e^{-2\pi i/3}$ (up to a common prefactor) and thus interfere destructively to zero.

As there are two atoms per unit cell, which contribute one $p_z$ electron each, the lower $\pi$ band is completely filled while the upper $\pi^*$ band is empty in intrinsic graphene. The Fermi level thus lies at the energy $\epsilon(\pm K)=0$ of the band degeneracy points at $K$ and $K'$. An expansion in the vicinity of $\pm K$ yields up to a constant phase factor
\begin{equation}
\label{eqn:TB_graphene_Dirac}
H(\pm K+\bq)  = \hbar v_F \left( \begin{array}{cc}
0 & q_x\pm iq_y \\
q_x\mp iq_y&  0\end{array} \right),
 \end{equation}
where the vector $\bq=(q_x,q_y)$ is given in the coordinate frame indicated in Figure \ref{fig:gr_struct}(b) \cite{Semenoff-1984}. This is the two-dimensional Dirac Hamiltonian (\ref{eqn:Dirac}), with the speed of light replaced by the Fermi velocity $\hbar v_{\rm F} \approx 5.8$\,eV$\ang$. Graphene thus hosts four flavors of Dirac fermions: two (real electron) spin degenerate cones in each of the two \textit{valleys} near $K$ and $K'$. The pseudospin in the graphene Dirac Hamiltonian (\ref{eqn:TB_graphene_Dirac}) corresponds to the sublattice degree of freedom and is thus fundamentally different from the real electron spin. Importantly, the Dirac particles in graphene are derived from the electronic band structure and are thus charged (electron or hole) quasiparticles.

The above derivation shows that the emergence of Dirac fermions in graphene requires two ingredients: first, the destructive interference of the three partial hopping amplitudes at $\xi(\bk=K)$ connecting each atom in sublattice A with its three nearest neighbors in sublattice B and vice versa; and second, the absence of a term $\sim \sigma_z$, which opens an energy gap, or equivalently produces a finite Dirac mass. The latter can be understood as follows:
In perfect freestanding graphene, the two sublattices transform into each other under real space inversion and are therefore equivalent. Thus, inversion $I$ means $\bk\to-\bk$ and exchanging the sublattices by $\sigma_x$:
\[I:H(\bk)=\sigma_x H(-\bk)\sigma_x.\]
In absence of a magnetic field, the system is also naturally time-reversal invariant:
\[T:H(\bk)=H^\ast(-\bk).\]
The combination of both symmetries requires that
\[TI:H(\bk)=\sigma_x H^\ast(\bk)\sigma_x,\]
which forces mass terms proportional to $\sigma_z$ in (\ref{eqn:Dirac}) to vanish \cite{Manes_07}. We note that the real electron spin has been disregarded in this discussion, which is appropriate in the absence of spin-orbit coupling, which is negligible in graphene.

Recently, silicene, the graphene equivalent of silicon, has been synthesized and analyzed using STM \cite{Vogt12}. Despite a slightly buckled structure, band structure calculations \cite{Takeda94, GuzmanVerri07,Silicene_CiraciPRL09} predict that suspended silicene and its close relative, germanene, will have Dirac cones just like graphene, and therefore, both belong to the class of Dirac materials. A significant spin-orbit coupling in silicene might in fact induce a topological state \cite{Liu_siliceneQSH_PRL11} which has an insulating bulk with a small energy gap and low-energy one-dimensional Dirac edge excitations. 

\subsubsection{Artificial graphene, secondary Dirac points, and organic Dirac materials}
Very generally, two-dimensional crystals with trigonal symmetry ($C_{3v}$) host Dirac fermion excitations in the corners, $K$ and $K'$, of their hexagonal Brillouin zone \cite{Wunsch_NJP08,Park_NanoLett08}. This is well understood from a group theoretical point of view. The $K$ and $K'$ points of the Brillouin zone are invariant under $C_{3v}$, up to reciprocal lattice vectors. Since there exists a two-dimensional irreducible representation of $C_{3v}$, two-fold degenerate bands lead, in general, to a Dirac-like dispersion at $K$ and $K'$.
Therefore, trigonally symmetric structures can be used to design artificial Dirac materials similar to graphene. Successful experimental realizations have been reported for lithographically patterned two-dimensional electron gases in semiconductors \cite{Gibertini_2DEG_09}, metal surfaces with hexagonal assemblies of CO molecules \cite{Manoharan_Nature2012} and ultracold atoms trapped in honeycomb optical lattices \cite{Sengstock_2011,Esslinger_2012}. Reference \cite{Polini_DM13} provides an excellent review of recent experimental advances in this field.

Recently, large efforts have also been made to fabricate heterostructures of graphene and other layered hexagonal crystals such as h-BN or MoS$_2$. These naturally lead to long wavelength hexagonal moir{\'e} structures superimposed on the graphene lattice due to lattice mismatch or rotational misalignment. In these structures, secondary Dirac points can emerge due to the moir{\'e} superlattice potential \cite{Park_Louie_PRL08,Guinea_PTRSA_2010,vandenBrink_PRB2012,Falko_PRB2013}. First, the above symmetry argument leads to the emergence of new Dirac points in the superlattice Brillouin zone corners as long as an overall $C_{3v}$ symmetry prevails \cite{Guinea_PTRSA_2010,vandenBrink_PRB2012,Falko_PRB2013}. Moreover, in the absence of \textit{local} sublattice symmetry breaking, additional Dirac points emerge at the middle points M of the superlattice Brillouin zone edges \cite{Park_Louie_PRL08,Guinea_PTRSA_2010,Falko_PRB2013}. Depending on microscopic interface details, the additional Dirac points might or might not be masked by additional bands. Signatures of these {\it secondary} Dirac points emerging in graphene on h-BN have been detected by scanning tunneling spectroscopy \cite{LeRoy_NatPhys2012} and in magnetotransport experiments \cite{Ponomarenko_Nature2013,Jarillo-Herrero_Science2013}. For graphene on h-BN, the moir{\'e} lattice period can exceed 10~nm. The correspondingly large moir{\'e} unit cell area turns out to facilitate experiments on electrons in a magnetic field, where the magnetic flux per unit cell can be on the order of one flux quantum. In this normally extremely hard to realize regime, the electronic spectrum rearranges to a so-called Hofstadter butterfly pattern \cite{Ponomarenko_Nature2013, PKim_Nature2013, Jarillo-Herrero_Science2013}. In this ultrahigh magnetic field regime \textit{ternary} Dirac points emerge, whenever a unit fraction of the magnetic flux quantum permeates the moir{\'e} unit cell. The system becomes translation invariant again with respect to an integer multiple of moir{\'e} unit cells, and the {\it ternary} Dirac points emerge at the corners of the new (magnetic) moir{\'e} Brillouin zone \cite{Ponomarenko_Nature2013}.
 
Another class of materials which can be relevant as Dirac Materials are organic conductors. The rich variety of molecular arrangements observed in organic materials allows for the formation of different lattices, including effective quasi-two-dimensional hexagonal structures, which can host Dirac nodes. One example is the organic conductor $\alpha$-(BEDT-TTF)$_2$I$_3$ under high pressure \cite{Tajima00, Katayama06, Kobayashi07, Tajima09}. The van der Waals nature of the stacking also makes it relatively straightforward to control the electronic parameters such as the bandwidth, which makes these materials attractive candidates for the studies of correlation effects and frustration in low dimensions, see e.g.~\cite{Kanoda11, Lebed08}.
Another reason some organic materials belong to the class of Dirac Materials is the fact that they develop unconventional nodal superconductivity, like in the case of the $d$-wave superconducting state in $\kappa$-(BEDT-TTF)$_2$Cu(NCS)$_2$ \cite{Arai01} and $\kappa$-(BEDT-TTF)$_2$Cu[N(CN)$_2$]Br \cite{Ichimura08}. We review the general nature of Dirac nodal superconducting excitations in more detail when we discuss $d$-wave superconductors later in this chapter.

The idea of exploiting symmetry protected band degeneracies to create Dirac materials in low dimensions can be even further generalized. Recently, it has been proposed that nanowires cut from crystals in a way that a non-symmorphic symmetry is preserved can give rise to Dirac bands in one dimension \cite{Deak2013}. In this way, e.g.~anatase TiO$_2$ nanowires would become a Dirac material.

%% file: Subsection_TI.tex
\subsection{Topological insulators}
Topological insulators belong to recently discovered new state of matter which has received a lot of attention, both because of their topological nature but also because of them being Dirac materials \cite{Kane_RMP2010,Qi11RMP}.
They are strongly spin-orbit coupled materials with an insulating bulk but conducting surface states. Like normal insulators, valence and conduction bands of a topological insulator are separated by a finite energy gap in the entire Brillouin zone of the bulk material. The surface of a topological insulators or any boundary to a normal insulator (or the vacuum) hosts, however, \textit{always} Dirac surface states, which close the gap unless time-reversal symmetry is broken. 
As the name suggests, the Dirac surface states are related to a topological property of the material.

Two insulators are defined to be topologically equivalent, if the Hamiltonians describing their band structure can be smoothly deformed into each other without closing the energy gap. Accordingly, insulators can be grouped into topological equivalence classes which are labelled by a topological invariant $\nu$, where $\nu$ is a property of the bulk material. This particular classification leads to an invariant, $\nu \in \mathbb{Z}_2$, which takes the values 0 and 1, see e.g.~\cite{Kane_PRL05, Moore_PRB07,Qi08}. Those insulators which are non-equivalent to the vacuum are termed ``topological insulators", while all others are called ``normal insulators". By the so-called bulk-boundary correspondence, the change of the invariant, $\Delta \nu$, across an interface of two materials is intimately tied to the occurrence of surface states, which close the energy gap at the interface \cite{Kane_RMP2010}. The number $N_k$ of such surface states fulfills
\begin{equation}
N_k \mod 2=\Delta \nu \mod 2.
\label{eq:bulk_boundary}
\end{equation}
These topologically protected surface states have a massless Dirac spectrum, where the momentum is locked to the spin, and thus results in a spin-helical metal. Generally, $n$-dimensional topological insulators host $n-1$-dimensional gapless surface or edge states.

\begin{figure}%
\centering
\includegraphics[width=0.6\columnwidth]{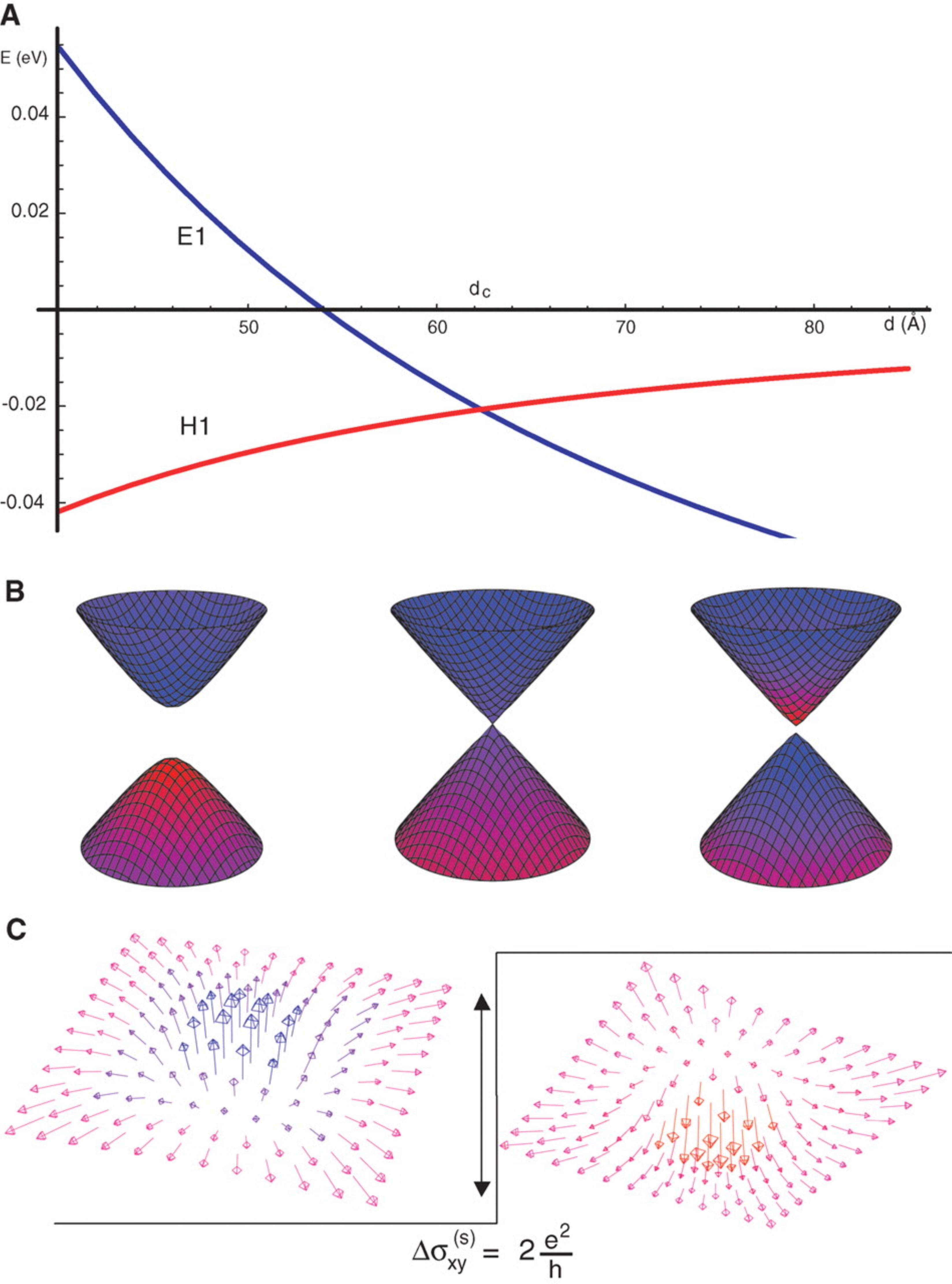}%
\caption{(A) Thickness dependent band inversion in HgTe / CdTe quantum wells. The character of the valence band maximum and the conduction band minimum interchange at a critical thickness $d= 63.5$~nm. $E1$ refers to $s$-orbtial states, whereas $H1$ refers to spin-orbit coupled p-orbital states. (B) Dirac dispersions with positive (left), vanishing (middle) and negative (right) mass terms as realized in HgTe / CdTe quantum wells with thicknesses $d=40$~nm, 63.5~nm, and 70~nm, respectively. (C) Schematic representation of the massive Dirac Hamiltonian through the vector $\bh(\bk)$ for positive (left) and negative mass (right). (Reprinted from B. A. Bernevig et al., Science 314, 5806 (2006) \cite{Bernevig_2006}. Reprinted with permission from AAAS.)}
\label{fig:BHZ_Hamiltonian_Hedgehog}%
\end{figure}

The concept of topologically distinct insulators and emergent gapless edge states can be illustrated with the two-dimensional Dirac Hamiltonian for particles with mass $m$ 
\begin{equation}
H(\bk)=v_F\bk\cdot {\bm \sigma}+m\sigma_z,
\label{eq:Dirac_massive}
\end{equation} 
where ${\bm \sigma}=(\sigma_x,\sigma_y)$. This Hamiltonian can be represented in terms of a fictitious magnetic field $\bh(\bk)=(v_F k_x, v_F k_y, m)$ acting on the pseudospin ${\bm\sigma}=(\sigma_x,\sigma_y,\sigma_z)$. For $m\neq 0$, the Hamiltonian (\ref{eq:Dirac_massive}) describes an insulator with gap $2|m|$, see Figure  \ref{fig:BHZ_Hamiltonian_Hedgehog} (b). The fictitious magnetic field textures, as displayed in Figure \ref{fig:BHZ_Hamiltonian_Hedgehog} (c), change with the sign of $m$: $\bh(\bk=0)$ points upwards (downwards) for $m>0$ ($m<0$). Importantly, the winding of $\bh(\bk)$ around the normal direction at $\bk=0$ changes from left to right handed upon reversing the sign of $m$. A smooth deformation of the $m>0$ case into the $m<0$ case is thus not possible without closing the gap and the two situations are therefore topologically inequivalent. 

\begin{figure}%
\centering
\includegraphics[width=0.5\columnwidth]{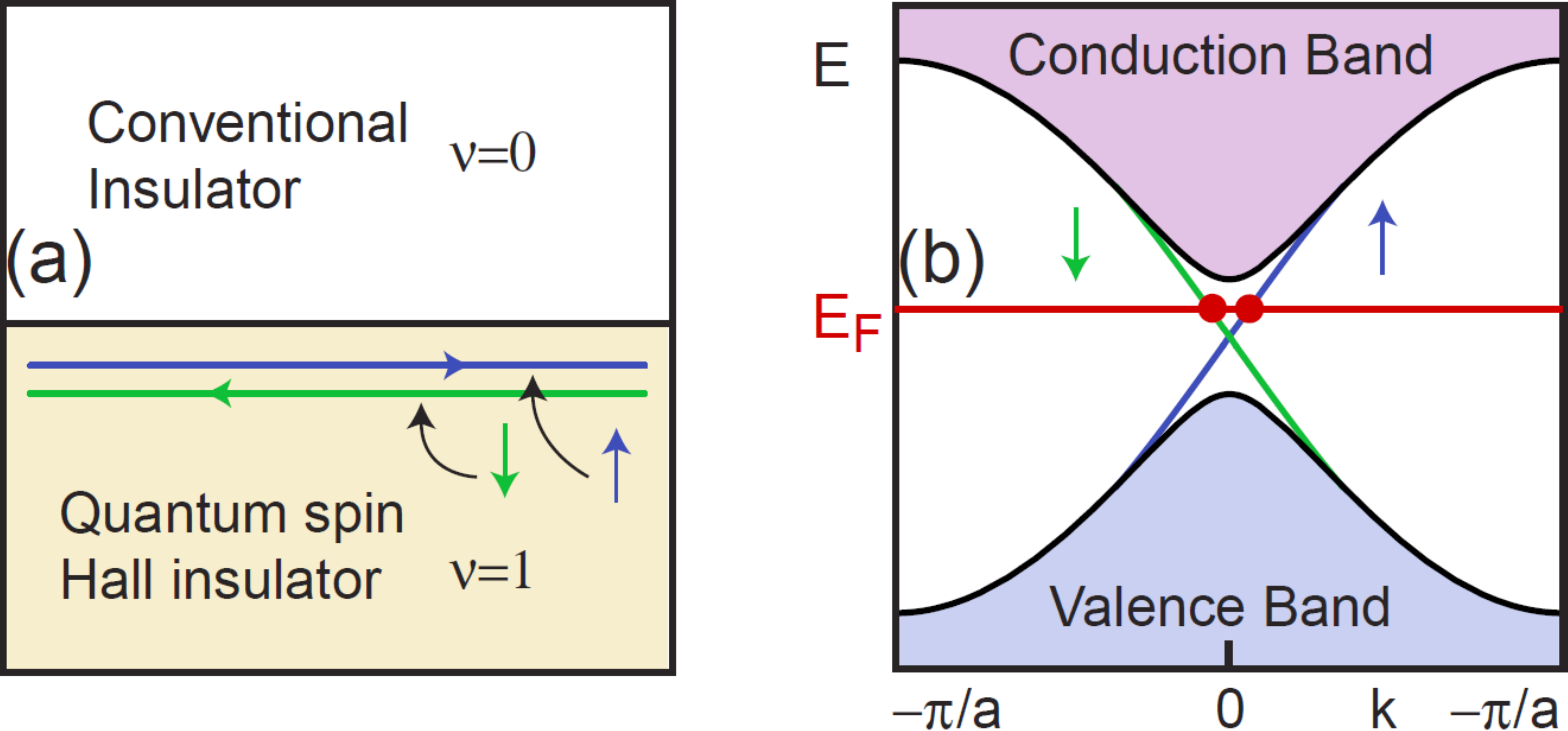}%
\caption{Counterpropagating edge states at the interace of a two-dimensional quantum spin Hall insulator and a normal insulator (a), giving rise to a one-dimensional massless Dirac dispersion with right moving spin-up and left-moving spin-down particles (b). (Reprinted figures by permission from M. Z. Hasan and C. L. Kane, Rev. Mod. Phys. 82, 3045 (2010)  \cite{Kane_RMP2010}. \href{http://link.aps.org/abstract/RMP/v82/p3045}{Copyright \copyright~(2009) by the American Physical Society.}
}%
\label{fig:Dirac_QSHI}%
\end{figure}

The emergence of gapless states at the interface of a topological insulator and a normal insulator can be understood straightforwardly for a two- dimensional system. An interface at $y=0$ corresponds to a mass term $m=m(y)$ which changes sign as function of $y$: $m<0$ for $y<0$ and $m>0$ for $y>0$. The momentum $k_x$ is still a good quantum number in this situation. The Schr\"odinger equation resulting from the replacement $\bk\to -i\nabla$ is
\begin{equation}
(-i v_F\nabla\cdot\vec\sigma+m(y)\sigma_z)\Psi(x,y)=\epsilon(k_x)\Psi(x,y)
\label{eq:TI_Dirac_explicit_I}
\end{equation}
and has an elegant exact solution \cite{Kane_RMP2010} with wave function
\begin{equation}
\Psi(x,y)=e^{ik_x x}\exp\left(-\int_0^y m(y')\diff y'\right)\left(\begin{array}{c}1\\1\end{array}\right).
\label{eq:TI_edge}
\end{equation}
and linear dispersion $\epsilon(k_x)=v_F k_x$. There is, thus, \textit{one} non-degenerate edge state with a positive group velocity and a Dirac pseudospin oriented in the positive $x$-direction. That is, this model gives rise to half a Dirac spectrum in one dimension. Such chiral modes are also encountered at the edges of quantum Hall systems.
However, in contrast to the quantum Hall effect, which occurs in strong magnetic fields, topological insulators exist in the absence of a magnetic field. In topological insulators spin-orbit coupling takes over the role of the magnetic field and time-reversal symmetry is preserved at interfaces between a topological insulator and a normal insulator unless further external perturbations, such as magnetic impurities, are present. Thus, Kramer's theorem is applicable and implies that the edge state from Eq.~(\ref{eq:TI_edge}) must have a time-reversed counter part (opposite spin and momentum) at equal energies at topological insulators interfaces. Then, the edge states give rise to a full one-dimensional Dirac spectrum, as displayed in Figure \ref{fig:Dirac_QSHI}.

In the simplest model, a topological insulator implements two copies of the Hamiltonian (\ref{eq:Dirac_massive}) to describe pairs of counter-propagating time-reversed states:
\begin{equation}
H_{\rm TI}(\bk)=\left(\begin{array}{cc} H(\bk) & 0 \\
0 & H^*(-\bk)
\end{array}\right).
\label{eq:2d_TI}
\end{equation}
After a unitary transformation $\sigma_y H^*(-\bk)\sigma_y=v_F\bk\cdot\vec\sigma-m\sigma_z$, the lower right block is equivalent to $H(\bk)$ with a reversed Dirac mass.
In 2005, Kane and Mele suggested that graphene realizes such a situation \cite{Kane_PRL05}. The spin-orbit interaction, which was neglected in the preceding section on graphene, does allow for mass terms which are compatible with all lattice symmetries of graphene. Within the graphene $p_z$-manifold, the leading order term of the spin-orbit interaction commutes with the $z$-component of the real electron spin $S_z$ and contributes opposite and valley dependent mass terms for spin-up and spin-down. The two decoupled blocks of $H_{\rm TI}(\bk)$ thus correspond to spin-up and down. The resulting edge states are eigenstates of $S_z$ and spin-up and spin-down states will have opposite group velocities $\pm v$: $H_{\rm edge}\sim v k_x S_z$, for an edge in the $x$-direction. $H_{\rm edge}$ has the form of a one-dimensional Dirac Hamiltonian, where the real electron spin coincides with the Dirac fermion spin. This is generally the case for topological insulators.

The spin-orbit induced energy bulk gap in graphene is of the order of $25-50\,\mu$eV \cite{Fabian_PRB2010}, which makes topological insulators physics in graphene very difficult to observe. Heavy elements provide larger spin-orbit interaction. Indeed, HgTe/CdTe quantum well structures were predicted to realize essentially the same topological insulator state (Eq.~(\ref{eq:2d_TI})), but at a bulk energy gap on the order of $10$~meV \cite{Bernevig_2006}. Beyond a certain width $d_c$, the HgTe/CdTe quantum wells realize an inverted band structure, see Figure \ref{fig:BHZ_Hamiltonian_Hedgehog} (A). Quickly after the prediction, these quantum well structures could be realized experimentally and electron transport through the topological insulators edge states was demonstrated \cite{Koenig_Science07}. This kind of two-dimensional topological insulators is often also referred to as quantum spin Hall insulators.

The quantum Hall effect is a purely two (or even) dimensional phenomenon. In contrast, the concept of topological insulators can be generalized to three-dimensional materials. There are, in particular, so-called strong three-dimensional topological insulators, which host an odd number of two-dimensional Dirac fermion flavors on each surface \cite{Kane_PRB07, Kane_PRL07}. The occurrence of surface states in strong three-dimensional topological insulators can be understood analogously to the two-dimensional quantum spin Hall insulator case discussed above.
Examples of three-dimensional topological insulators include Bi$_2$Se$_3$, Bi$_2$Te$_3$ or Sb$_2$Te$_3$ \cite{Hasan_09,SCZhang_09}, where the emergence of Dirac-like surface states from the bulk band structure has been illustrated in Ref.~\cite{SCZhang_09}. For example, in the bulk of Bi$_2$Se$_3$ there are four bands in the vicinity of the Fermi level. These are mostly derived from Bi and Se $p_z$ orbitals and can be described by a Hamiltonian
\begin{equation}
H(\bk)=\epsilon_k+\left(\begin{array}{cccc}
	M(k) & A_1 k_z & 0 & A_2 k_- \\
	A_1 k_z & -M(k) & A_2 k_- & 0 \\
	0 & A_2 k_+ & M(k) & -A_1 k_z \\
	A_2 k_+ & 0 &-A_1 k_z &-M(k)
\end{array}\right),
\label{eqn:Hk_Bi2Se3_bulk}
\end{equation}
where $\bk=(k_x,k_y,k_z)$ and $k_\pm=k_x\pm ik_y$ denote the crystal momentum, $M(k)=M-B_1k_z^2-B_2k_+k_-$. $A_{1,2}$, $B_{1,2}$, and $M$ are material specific constants. This Hamiltonian is written in a basis $\ket{P1^+_z,\up},\ket{P2^-_z,\up},\ket{P1^+_z,\dn},\ket{P2^-_z,\dn}$, where $\up,\dn$ refers to the electron spin and $P1^+_z$, $P2^-_z$ stand for Bi and Se $p_z$ orbitals, respectively. For $M,B_1,B_2>0$, the bands are inverted at $\bk=0$ as compared to large $\bk$ and the band structure is topologically non-equivalent to the vacuum. This is the case for Bi$_2$Se$_3$, Bi$_2$Te$_3$ or Sb$_2$Te$_3$ \cite{SCZhang_09}. For the surface perpendicular to the $z$-direction, there are two normalizable surface states, $\ket{\psi_0,\up}$ and $\ket{\psi_0,\dn}$, in the vicinity of $k_x,k_y=0$ and projecting (\ref{eqn:Hk_Bi2Se3_bulk}) onto these surface states yields
\begin{equation}
H_s(k_x,k_y)=\left(\begin{array}{cc}
0 & A_2 k_-\\
A_2 k_+ & 0
\end{array}\right).
\label{eq:Hk_Bi2Se3_surf}
\end{equation}
This is equivalent to the Dirac Hamiltonian in Eq.~(\ref{eqn:Dirac}) with $v_f=A_2$. 

The list of topological insulators has been growing rapidly in the last few years thanks to extensive research efforts particularly based on ab-initio density functional theory calculations. Topological insulators based on various compounds in the HgTe system, the Bi$_2$Se$_3$ system, and strongly correlated topological insulators (see \cite{Kane_RMP2010, Qi_SCZhang_RMP11, Felser_review12} for reviews), as well as artificially engineered layered materials have been suggested \cite{DasBalatsky_artificialTI13,Liangzhi_NanoLett2013}. The topologically nontrivial band structure leads to a wide range of interesting properties of the low-energy states at the surface of topological insulators \cite{Kane_RMP2010}. Yet, as will be shown in the following, the common form of the quasiparticle dispersions lets vastly different Dirac material respond very similar to external perturbations.

Very recently crystalline topological insulators, where not time-reversal symmetry but crystal point group symmetries such as mirror planes give rise to non-trivial topological states, have also been discovered \cite{Fu11crystTI,HsiehTim11}. On symmetry protected surfaces these can also host (an even number of) Dirac cones as recently experimentally verified in SnTe \cite{Tanaka12} and Pb$_{1-x}$Sn$_{x}$Se \cite{Dziawa12}.

%% file: Subsection_dwave.tex
\subsection{$d$-wave superconductors}
\label{subsec:d-wave}
While the Dirac spectrum in graphene and topological insulators are ultimately consequences of the band structure, the Dirac spectrum of the quasiparticle excitations in $d$-wave superconductors have a distinctly different origin. Still, the same low-energy Hamiltonian appears.
The high-temperature cuprate superconductors are the most prominent examples of $d$-wave superconductors \cite{Harlingen95, Tsuei00RMP}. Many heavy fermion superconductors are likely also $d$-wave superconductors, such as the first discovered heavy fermion superconductor CeCu$_2$Si$_2$ \cite{Steglich79} and the Ce115 family, which includes for example CeCoIn$_5$ \cite{Thompson12}. In addition, organic superconductors can also have $d$-wave symmetry, as exemplified by, e.g.~ $\kappa$-(BEDT-TTF)$_2$Cu(NCS)$_2$ \cite{Arai01} and $\kappa$-(BEDT-TTF)$_2$Cu[N(CN)$_2$]Br \cite{Ichimura08}.
Since the cuprate superconductors have significantly higher transition temperatures we will focus on their properties, but the Dirac spectrum and its properties are common to all $d$-wave superconductors.

Many cuprate superconductors have a tetragonal crystal symmetry with a $D_{4h}$ point-group symmetry. These include La$_{2-x}$Sr$_x$CuO$_4$ (LSCO), Tl$_{2}$Ba$_2$CaCu$_2$O$_8$ (Tl-2212), and HgBa$_2$CaCu$_2$O$_6$ (Hg-1212), as well as some YBa$_2$Cu$_3$O$_7$ (YBCO-like) compounds with significant cation substitution. Others have an orthorhombic crystal structure, such as YBa$_2$Cu$_3$O$_{7-x}$ (YBCO) and Bi$_2$Sr$_2$CaCu$_2$O$_{8+x}$ (Bi-2212), where the point-group symmetry is $D_{2h}$. Common to all of these crystal structures is the CuO$_2$ plane with either a square or rectangular lattice, belonging to the $C_{4v}$ and $C_{2v}$ point-groups, respectively. This classification ignores any CuO$_2$ plane buckling or CuO$_6$ tilt. These are known to affect the transition temperature but they do not change the basic symmetry of the CuO$_2$ planes (see references within \cite{Tsuei00RMP}). Superconductivity has been concluded to originate in the CuO$_2$ layers. For example, there is strong anisotropy in resistivity, penetration depth, and superconducting coherence length between in-plane directions and the $c$-axis. Moreover, in-plane charge confinement  and interplane dc and ac intrinsic Josephson effects have established the cuprate superconductors to be stacks of two-dimensional superconducting CuO$_2$-based layers (see references within \cite{Tsuei00RMP}).
Moreover, the relatively high ratio of $c$-axis to $a$-axis lattice constant makes for a flattened Brillouin zone with little dispersion along the $z$-axis, further motivating a two-dimensional treatment.

Based on these extensive experimental data, the cuprate superconductors can be viewed as two-dimensional materials. To phenomenologically describe a two-dimensional spin-singlet superconductor, we consider the real-space mean-field BCS Hamiltonian \cite{deGennes66, Balatsky_RMP},
 \begin{align}
 H_{\rm BCS} & =\sum_\alpha\int\diff^2 r\psi_\alpha^\dagger(r)\hat H_0(r)\psi_\alpha(r) \\ \nonumber
 & +\sum_{\alpha,\beta}\int\diff^2 r\diff^2 r'\left(\Delta_{\alpha,\beta}(r,r')\psi^\dagger_\alpha(r)\psi^\dagger_\beta(r')+\hc\right),
\end{align}
in two spatial dimensions. Here, $\psi_\alpha^\dagger(r)$ and $\psi_\alpha(r)$ are the electronic field operators and $\alpha$ is the spin index. Assuming singlet paring $\Delta_{\alpha,\beta}(r,r')=(i\sigma_y)_{\alpha,\beta}\Delta(r,r')$ and introducing the Nambu spinor $\Psi^\dagger(r)=(\psi^\dagger_\uparrow(r),\psi_\downarrow(r))$, we arrive at 
 \begin{align}
H_{\rm BCS}=\int\diff^2 r\Psi^\dagger(r) \tau_z H_0(r)\Psi(r)+\int\diff^2 r\diff^2 r'\Psi^\dagger(r) \tau_x \Delta(r,r')\Psi(r'),
\end{align} 
where $\tau_i$, $i=x,y,z$ are Pauli matrices acting in Nambu space. 
$H_{\rm BCS}$ describes the dynamics of the Bogoliubov quasiparticles in the superconductor. Fourier transforming to the momentum space representation with $\bk = (k_x,k_y)$ and rewriting $H_{\rm BCS}$ in first quantized form, we obtain the Bogoliubov-de Gennes (BdG) Hamiltonian:
\begin{equation}
H_{\rm BdG} = \epsilon_\bk \tau_z + \Delta_{\bk} \tau_x,
\label{EQ:BDG1}
\end{equation}
where $\epsilon_\bk$ is the normal-state band structure and we have assumed time-reversal symmetry such that $\Delta$ can be chosen to be real and $\epsilon_\bk = \epsilon_{-\bk}$.
By diagonalizing $H_{\rm BdG}$ we find the energy spectrum of the Bogoliubov quasiparticles:
\begin{equation}
\label{eq:BQPSC}
E_\bk = \sqrt{\epsilon_\bk^2 + |\Delta_\bk|^2}.
\end{equation}

The above derivation of the energy spectrum of the Bogoliubov quasiparticles is valid for any spin-singlet superconducting order parameter, including $d$-wave superconductors. For the case of the tetragonal cuprate superconductors, the superconducting order has been shown to belong to the $B_{1g}$ irreducible representation, which has $d_{x^2-y^2}$-wave symmetry in the CuO$_2$ plane. We can therefore write the gap function as $\Delta_{\bk} = \Delta_0[\cos(k_xa)-\cos(k_ya)]$. The superconducting gap thus vanishes along the nodal directions $|k_x|=|k_y|$ throughout the Brillouin zone. In the high-temperature cuprate superconductors the normal state dispersion $\epsilon_\bk$ gives closed Fermi surfaces around the Brillouin zone corners at $(\pm \pi,\pm \pi)$, due to hole doping. There are thus four nodal points in the spectrum, where the nodes of the order parameter $\Delta_{\bk}=0$ cross the Fermi surface $\epsilon(\bk)=0$, such that $E_\bk = 0$.
For the orthorhombic cuprate superconductors, the $d_{x^2-y^2}$ order belong together with the $s$-wave order to the $A_{1g}$ irreducible representation. This allows for a real admixture of $d$- and $s$-waves, while still only a single superconducting transition is observed \cite{Tsuei00RMP}. 
In the presence of weak orthorhombic distortion, for example present in YBCO, the nodal points move slightly away from the zone diagonals due to the presence of the subdominant $s$-wave order. This breaks the fourfold rotation symmetry between the Dirac points, but the existence of the four nodal points is never compromised as long as the superconducting gap function has nodes somewhere between $0^\circ$ and $90^\circ$. This makes the appearance of Dirac physics in the quasiparticle spectrum of the $d$-wave cuprate superconductors very robust.

Near the nodal points we can extract the low-energy properties of the Bogoliubov quasiparticles by expanding Eq.~(\ref{eq:BQPSC}) in deviations perpendicular ($k_\perp$) and parallel ($k_\parallel$) to the Fermi surface, respectively. To first approximation we write $\epsilon_\bk \approx \hbar v_F k_\perp$ and $\Delta_\bk \approx \hbar v_\Delta k_\parallel$. Here the Fermi velocity is defined as $v_F = \frac{\partial \epsilon_\bk}{\partial \bk}$, whereas the gap velocity is $v_\Delta = \frac{\partial \Delta_\bk}{\partial \bk}$.
This gives an two-dimensional anisotropic massless Dirac Hamiltonian of the form of Eq.~(\ref{eqn:Dirac}), but with different  dispersion velocities in the two (orthogonal) spatial directions. The anisotropy ratio $v_F/v_\Delta$ changes between different cuprate materials  and also with doping levels for single cuprate compounds. At optimal doping the anisotropy ratio has been measured to be $v_F/v_\Delta \approx 10, 12$, and 19 for the cuprates LSCO, YBCO, and Bi-2212, respectively \cite{Sutherland03}. Due to the particle-hole symmetry of the Bogoliubov spectrum, the Dirac point is always located at the chemical potential, and thus there is no need for external fine tuning to reach the Dirac point in a $d$-wave superconductor. This is different from most other incarnations of Dirac materials.

The low-energy quasiparticles in a $d$-wave superconductor are thus Dirac fermions, just as in graphene and topological insulators. For properties mainly dependent on the quasiparticle energy spectrum there should therefore not be any important differences between all these two-dimensional Dirac materials. However, when other aspects of the low-energy quasiparticles are involved the specific properties of the quasiparticles and the materials background can become important. For example, the different nature of the pseudospin results in different properties for graphene and topological insulators. In the case of $d$-wave superconductors the Dirac fermions are also Bogoliubov quasiparticles satisfying
\begin{align}
\label{eq:BQP}
\gamma_{\bk \uparrow} & = u_\bk c_{\bk \uparrow} - v_\bk c_{-\bk \downarrow}^\dagger, \nonumber \\\gamma_{\bk \downarrow} & = u_\bk c_{-\bk \downarrow} - v_\bk c_{\bk \uparrow}^\dagger,
\end{align}
where the coefficients are related according to $|v_\bk|^2  = 1-|u_\bk|^2 = \frac{1}{2}\left( 1-\frac{\epsilon_\bk}{E_\bk}\right)$. The Bogoliubov quasiparticles still satisfy fermionic anticommutation relations but they are mixtures of electrons and holes as well as up and down spin. For example, creating a quasiparticle using $\gamma_{\bk \uparrow}^\dagger$ means creating an electron with momentum $\bk$ and spin-up together with destroying an electron with momentum $-\bk$ and spin-down. The net effect is to increase the system momentum by $\bk$ and to increase $S_z$ by $\hbar/2$.
This lack of well-defined charge for Bogoliubov quasiparticles has consequences for the properties of the quasiparticles in external electromagnetic fields. Intimately connected to the nature of the Bogoliubov quasiparticles is the existence of the superconducting condensate, which changes the materials background in a superconductor. This has, again, profound effects on the electromagnetic response and also, for example, means that the long-range Coulomb repulsion is fully screened.

Interestingly, the zero-energy Bogoliubov quasiparticles at the Dirac point are their own antiparticles and are thus, per definition, Majorana fermions, see e.g.~\cite{Wilczek09}. However, due to the spin-degeneracy in the high-temperature cuprate superconductors, there is always an even number of Majorana zero-energy states, which comprise nothing else than a regular fermionic excitation. The Majorana nature is only revealed if the spin-degeneracy is broken, which for example has been proposed to happen in a topological insulator$-$$d$-wave superconductor hybrid structure, where the momentum-spin locking in the topological insulator surface state breaks the spin-degeneracy \cite{Linder10PRL}.

%% file: Subsection_Weyl.tex
\subsection{Weyl semimetals}
\label{subsec:Weyl}
Graphene, topological insulators, and the cuprate $d$-wave superconductors are all two-dimensional (or even one-dimensional in the case of two-dimensional topological insulators) Dirac materials. It has, however, been known for a very long time that band crossings are stable also in three dimensions \cite{Neumann29, Herring37}. The bands expanded around such crossing points are to first order necessarily linear in momentum, and there is thus a three-dimensional Dirac-like nodal point at the crossing.
If it is also possible to tune the chemical potential such that it lines up at or near the crossing point, and no other bands are present at this energy, the result is a three-dimensional semimetal with a linear Dirac spectrum, which constitutes a Dirac material. If there are just two single bands crossing, i.e.~no additional band degeneracy at the band crossing point, the low-energy dispersion of such a crossing point resembles the solution to the Weyl equation of particle physics. The Weyl equation is nothing else than the massless limit of the Dirac equation, in which the four-component Dirac solution separates into two independent two-component solutions. 
Materials with no degeneracy in the crossing bands, creating what is called a Weyl point, have therefore started to be referred to as Weyl semimetals \cite{WanTurner11}. Beyond solids, the bulk superfluid $^3$He A phase is also a three-dimensional Weyl system \cite{Volovikbook2}.

For simplicity let us consider the following form of the Hamiltonian for the low-energy dispersion in a Weyl semimetal, consisting of two band crossing points:
\begin{equation}
\label{eqn:Weyl}
H_{\pm} = \pm \hbar v_F (k_x\sigma_x + k_y \sigma_y + k_z \sigma_z).
 \end{equation}
Here $\sigma_i$ are the three Pauli matrices acting in the space of the two bands causing the band crossings and $k_i$ measures the momentum from the two band crossing points ${\bf k}_\pm$. The velocity ${\bf v} = \pm \hbar v_F {\boldsymbol \sigma}$ is either parallel or opposite to the (pseudo)spin and set by the chirality. The above Hamiltonian has an energy spectrum $E = \hbar v_F |{\bf k}|$ around both band crossing points, as expected of a Dirac material. This is a simplification of a general anisotropic Weyl Hamiltonian where there can be three different velocities in three independent, but not necessarily orthogonal, directions.

Even with the simplification of an isotropic velocity Eq.~(\ref{eqn:Weyl}) reveals several special features of Weyl semimetals. 
First of all, the Weyl points are topological objects in momentum space. The Weyl point looks like a hedgehog or a (pseudo)magnetic monopole in momentum space, with the (pseudo)spin vectors oriented towards or away from the Weyl point depending on the chirality. Mathematically this can be seen by constructing the effective vector potential ${\bf A}(\bk)$ and the corresponding magnetic field strength ${\bf B}(\bk)$ for the Bloch states $|u_{n,\bk} \rangle$:
\begin{align}
\label{eqn:AandB}
{\bf A}(\bk) & = -i \sum_{n, occ}\langle u_{n,\bk} | {\bf \nabla}_\bk| u_{n,\bk} \rangle \\
{\bf B}(\bk) & = {\bf \nabla}_\bk \times {\bf A}(\bk),
 \end{align}
where the summation is over occupied bands $n$. ${\bf B}(\bk)$ is also known as the Berry curvature or flux. 
Integrating this flux through a small surface containing the Weyl point yields $\pm 2\pi$, as e.g.~ shown in Ref.~\cite{Turner13}. According to Gauss' law this is then also the flux through any surface containing the Weyl point and the Weyl point can thus be regarded as a magnetic monopole. 
If we instead apply Gauss' law around the whole Brillouin zone we need to obtain a net zero flux; there cannot be an overall magnetic source or sink. This means that Weyl points always come in pairs with opposite chirality. This is known as the fermion doubling theorem \cite{Nielsen81, Nielsen81b} and explains why Eq.~(\ref{eqn:Weyl}) has two crossing points $\bk_\pm$ with different chirality. 

Gauss' law also directly gives the stability of a single Weyl point because it cannot just suddenly disappear. The stability of the Weyl points can also be seen directly from the Hamiltonian in Eq.~(\ref{eqn:Weyl}) since it uses all three Pauli matrices. Thus, there is no $2 \times 2$ matrix left that anticommutes with the Hamiltonian and can open a gap in the spectrum. The only way to destroy a Weyl point is therefore to annihilate it with another Weyl point of opposite chirality. This can be done either by moving the Weyl points in momentum space and finally merging them, leading to a fully gapped insulator, or, alternatively, by breaking translational symmetry by allowing scattering between the two Weyl points. 
The topological protection of the Weyl points is thus closely linked to being able to define a crystal momentum and having separate crystal momentum for each Weyl point. If crystal translation symmetry is broken, the distinction between the two Weyl points might be lost and so might the topological features associated with them. However, as exemplified with graphene, disorder is seldom large enough to destroy well-separated nodal points, and Weyl points therefore do not seem unachievable in crystalline materials.
The above arguments for the stability of the Weyl point do not include the possibility of violating charge conservation with a superconducting order, which can gap the Weyl point without annihilation, just as in any Dirac material. 

The topological stability of the Weyl points crucially requires that the bands involved are non-degenerate. Otherwise there can be terms that cause band hybridization within the degenerate subspaces and in that way produce an energy gap. Non-degenerate bands require either time-reversal symmetry or inversion (parity) breaking. Time-reversal symmetry breaking, which in practice is often achieved by magnetic order, can realize the minimal case of a single pair of Weyl points \cite{WanTurner11}. Inversion symmetry breaking, on the other hand, has been shown to generate a minimum of four Weyl points in the Brillouin zone \cite{Murakami07}.

\begin{figure}
\begin{center}
\includegraphics[width=0.4\columnwidth]{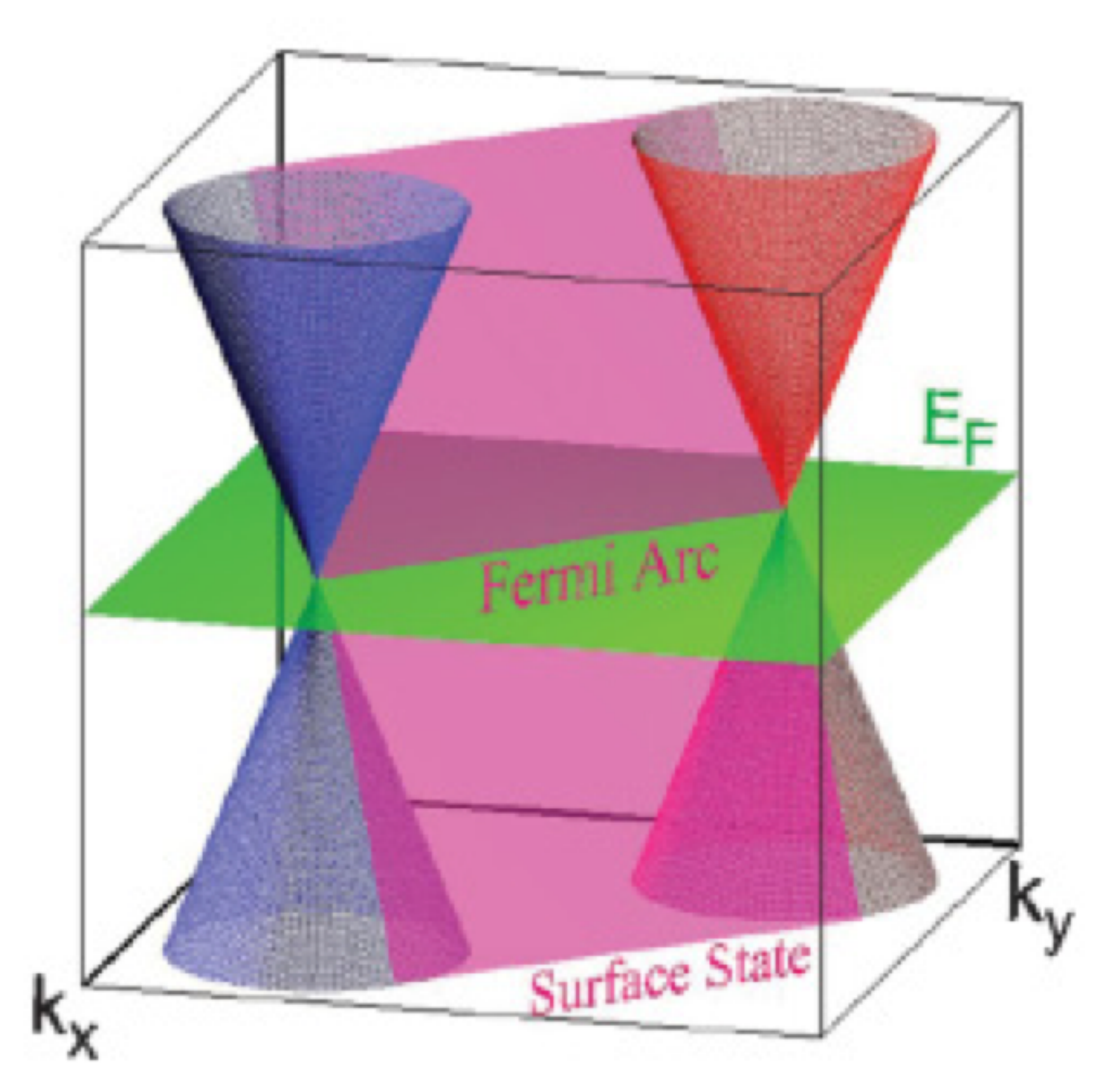}
\end{center}
\caption{The surface states of a Weyl semimetal form a Fermi arc connecting the two Weyl points. The bulk dispersion of the two Weyl cones (blue and red) along with the surface states (pink plane) crossing the horizontal Fermi level and thus forming a Fermi arc. (Reprinted figure by permission from X.~Wan {\it et al.}, Phys.~Rev.~B 83, 205101 (2011) \cite{WanTurner11}. \href{http://link.aps.org/abstract/PRB/v83/p205101}{Copyright \copyright~(2011) by the American Physical Society.})
}
\label{fig:Weylsurf}
\end{figure}

One of the most striking features of Weyl semimetals is that they have surface states forming {\it Fermi arcs} \cite{WanTurner11}. We usually expect the Fermi surface to form closed loops, but in Weyl semimetals this is not true. 
The Fermi arc on the top surface is instead complemented by the Fermi arc on the bottom surface, such that they together form a closed surface, as expected for a two-dimensional system. In a thin film two halves of the Fermi surface will spatially separate to opposite sides of the film as the film thickness is increased. 
These separated Fermi arcs can be understood by the following reasoning. 
Let us assume a three-dimensional thick film of Weyl semimetal with surfaces in the $xy$ plane. For a clean surface we can use translational invariance and label the single electron states by the crystal momentum in this plane. Further, let us assume that we have a single pair of Weyl points as described by Eq.~(\ref{eqn:Weyl}). At the Fermi level we thus find both the surface states and the states associated with the bulk Weyl points. For momenta away from the Weyl points the surface states are well-defined because there is no other bulk excitations available. On the other hand, at the Weyl points the surface states will terminate and they will thus describe an arc between the two Weyl points, as illustrated by the pink plane in Figure \ref{fig:Weylsurf} cutting the green horizontal Fermi level.
The Fermi arc surface states also follow by noting that the Weyl points are monopoles of the Berry flux. 
If we place two two-dimensional momentum space surfaces in, say the $yz$-plane, one between the two Weyl points and one on the outside, there will be a non-zero net flux through the two of them. Thus, the Chern numbers associated with each plane differ by one, and thus at least one plane has a non-zero Chern number. Each  non-zero Chern number plane describes a two-dimensional quantum Hall state or Chern insulator, which have, per definition a chiral surface state which cross the Fermi level. Putting the edge states of all possible two-dimensional momentum space planes together, we arrive at a Fermi arc connecting the two Weyl points. This unique Fermi arc feature of the Weyl semimetals should provide strong evidence for a Weyl semimetal state for surface sensitive probes such as ARPES or STS. 

In order to demonstrate that Weyl points can appear even in simple band structures, it is useful to study a toy model. For example, consider the following half-filled two-band model on a cubic lattice \cite{Yang11, Delplace12,Turner13}: 
\begin{align}
\label{eqn:Weyltoy}
H = \left[2t_x (\cos k_x  - \cos k_0) + m(2-\cos k_y-\cos k_z)
\right]\sigma_x + 2t_y \sin k_y\sigma_y + 2t_z\sin k_z\sigma_z,
 \end{align}
where $\sigma$ is the spin of the electron. The model breaks time-reversal symmetry and it is straightforward to show that it has two band touching points at zero energy in the bulk Brillouin zone at $\bk = \pm k_0 \hat{x}$. These are the only points in the Brillouin zone where the bulk gap closes for large enough $m$. 
Expanding around these two points by setting $\bp_\pm = (\pm k_x \mp k_0,k_y,k_z)$, assuming the expansion parameter $|\bp|$ to be much smaller than $k_0$, we arrive at \cite{Turner13}
\begin{align}
\label{eqn:Weyltoy2}
H_\pm = v_x [p_\pm]_x \sigma_x + v_y [p_\pm]_y \sigma_y + v_z[p_\pm]_z\sigma_z,
 \end{align}
with the velocities $v_x = 2t_x \sin(k_0)$ and $v_{y,z} = -2t_{y,z}$. This is an anisotropic version of the isotropic Weyl Hamiltonian in Eq.~(\ref{eqn:Weyl}), and we thus conclude that the Hamiltonian in 
Eq.~(\ref{eqn:Weyltoy}) describes a Weyl semimetal with two Weyl points located at $\mp k_0 \hat{x}$. 
Since all Pauli matrices are used in Eq.~(\ref{eqn:Weyltoy}), perturbations will only move around the Weyl points, but can never gap them, except if the perturbations are strong enough to eventually merge the two Weyl points, in which case they annihilate each other and the result is an insulating state.

Despite the seemingly simple band structure in Eq.~(\ref{eqn:Weyltoy}) required to produce a Weyl semimetal, no (crystal) material has yet been definitively classified as a Weyl semimetal. However, in the last years several material families have theoretically been predicted to be Weyl semimetals. The first class of materials to be predicted to be Weyl semimetals were the pyroclore iridates, A$_2$Ir$_2$O$_7$, where A is a rare earth element, such as A = Y, Eu, Nd, or Sm \cite{WanTurner11}. These materials all have strong spin-orbit coupling and strong electron correlations. 
For all but weak electron correlations the magnetic moments on the Ir sites order and, in at least one possible phase, they form a noncollinear pattern with a so-called all-in/all-out antiferromagnetic order, where all the spins on a given tetrahedron either point in or out from the center \cite{WanTurner11, Witczak12,ChenHermele12}. For very strong electron correlations an antiferromagnetic insulator phase is reached but in the intermediate correlation regime the electronic ground state might be a Weyl semimetal with linearly dispersing nodes at the chemical potential. All these magnetic orders preserve inversion symmetry but naturally breaks time-reversal symmetry.
Experimentally, $\mu$SR experiments have found long-range magnetic order \cite{Zhao11} and transport data seems consistent with linearly dispersing bands, but there is yet no definitive evidence for the pyroclore iridates being Weyl semimetals. \cite{Yanagishima01,Ueda12,Tafti12}. 

Also the spinel compounds AB$_2$O$_4$, where the B sites form a corner-sharing tetrahedral network with $5d$ elements, have been proposed, using density functional theory calculations, to have a Weyl semimetal phase depending on the strength of the Coulomb correlations among the $5d$ orbitals. One example is the Os-based spinels AOs$_2$O$_4$, with A being Ca or Sr \cite{Wan12}. Another example with a Weyl semimetal phase appearing theoretically when tuning the Coulomb interactions is the ferromagnetic spinel is HgCr$_2$Se$_4$ \cite{Xu11}. In this material the Weyl points have $\pm 2$ magnetic charge.

There also exist proposals to engineer Weyl semimetals in heterostructures based on topological insulators. In stacks of thin magnetically doped topological insulator layers separated by an insulator spacer, a Weyl semimetal with two Weyl points has  theoretically been predicted to exist \cite{Burkov11}. The Weyl points are here separated in momentum space along the growth direction of the heterostructure. This Weyl semimetal breaks time-reversal symmetry due to the magnetic doping. Another related heterostructure hosting Weyl points has no magnetic doping but instead breaks inversion symmetry using an electric field \cite{Halasz12}. In a slightly different proposal, a topological insulator can be turned into a Weyl semimetal by magnetic doping close to the topological phase transition between the topological non-trivial and trivial insulating phases \cite{Cho11}. The Weyl semimetal appears when the magnetization mass is strong enough to close the topological bulk band gap. 
Interestingly, three-dimensional photonic crystals have lately also been shown to be able to host Weyl lines, which by breaking inversion or time-reversal symmetry generate separated Weyl points \cite{Lu13}. The benefit of photonic crystals is that there is no spin-degeneracy to initially break in order to produce non-degenerate bands, since there is no Kramers degeneracy for photons.

\subsubsection{Three-dimensional Dirac semimetals}
The non-degeneracy of the bands at a Weyl point generates topological protection of the three-dimensional linear Dirac spectrum in Weyl semimetals, as discussed above. Weyl semimetals only exists if either time-reversal or inversion symmetry is broken in the system. On the other hand, if both of these symmetries are present, the Weyl point must be degenerate according to the following argument \cite{Young12}: Assume a Weyl point appears at some Brillouin zone momentum $\bk$. We can quantify the Weyl point by the Chern number of the valence band on a sphere surrounding the point, which takes values $\pm1$. Time-reversal symmetry dictates that another Weyl point with the same chirality, or Chern number, appears at $-\bk$. However, the overall Chern number should cancel in the system as a whole. There must thus be another Weyl point with opposite chirality at a point $\bk'$, and due to time-reversal symmetry another one at $-\bk'$. Now, if inversion symmetry is also present it requires that the Weyl points at $\bk$ and $-\bk$ has opposite chiralities. Thus, in a system with both time-reversal and inversion symmetries $\bk = \bk'$ is necessary and the system has at least one pair of nodal points where the valence and conduction bands touch, which each hosts four linearly dispersing bands (two copies of a Weyl point). This system is thus described by the $4\times 4$ Dirac Hamiltonian \cite{Wang13}:
\begin{align}
\label{eqn:3DDirac}
H = \hbar v_F\left(
\begin{array}{cc}
 {\bm \sigma} \cdot \bk & 0 \\
 0 & -{\bm \sigma} \cdot \bk \\
\end{array}
\right),
 \end{align}
which can be seen as two copies of the $2 \times 2$ Weyl Hamiltonian with opposite chiralities. A material with this low-energy Hamiltonian has been referred to as a three-dimensional Dirac semimetal \cite{Young12}.
The three-dimensional Dirac semimetals naturally have many of the same characteristics as the Weyl semimetals, since they have  the same three-dimensional Dirac-like energy dispersion. On surfaces Dirac semimetals have Fermi arcs touching each other at the surface projections of the Dirac points, where the Fermi surface becomes ill-defined \cite{Wang12,Wang13}. These touching Fermi arcs can be seen as built up by the disconnected Fermi arcs found in the Weyl semimetals \cite{WanTurner11}.

\begin{figure}
\begin{center}
\includegraphics[width=0.8\columnwidth]{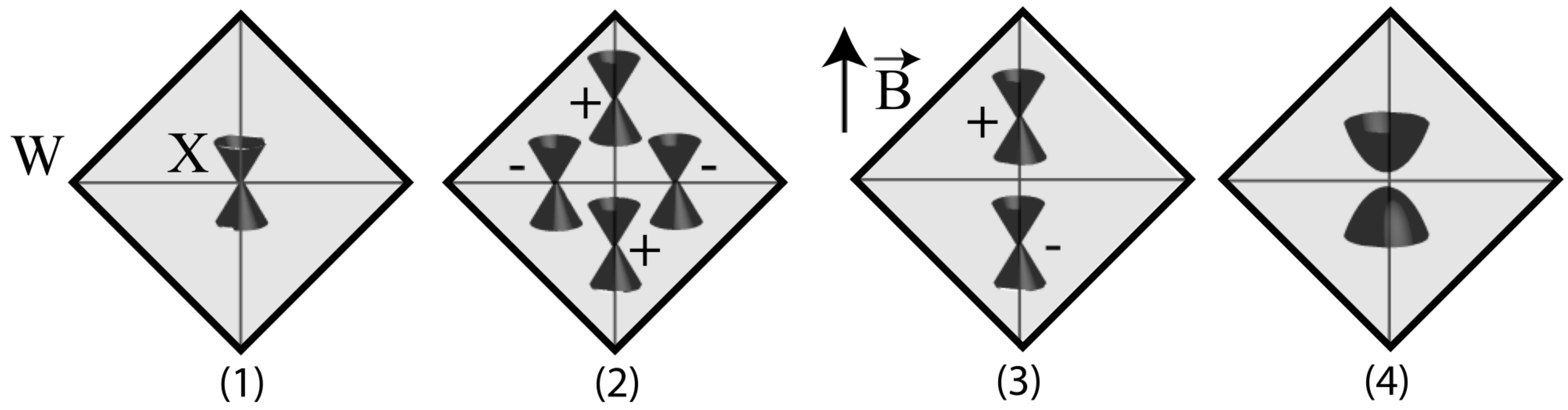}
\end{center}
\caption{The Dirac, Weyl, and insulating phases of the so-called Fu-Kane-Mele $s$-state model on the diamond lattice \cite{Kane_PRL07}. (1) The states at the Dirac point X are fourfold degenerate. (2) Four momentum separated Weyl points appear with the application of a small inversion symmetry breaking perturbation. The chirality of the Weyl points (Chern number) is indicated with $\pm$. (3) Two Weyl points appear for a time-reversal symmetry breaking Zeeman term ${\bf B}$ oriented along $X$ to $W$. Any other magnetic field orientations break rotation symmetry, which produces a fully gapped state (4).
(Reprinted figure by permission from S.~M.~Young {\it et al.}, Phys.~Rev.~Lett.~108, 140405 (2012) \cite{Young12}. \href{http://link.aps.org/abstract/PRL/v108/p140405}{Copyright \copyright~(2012) by the American Physical Society.})
}
\label{fig:DiracWeyl}
\end{figure}

While a single Weyl point is topologically protected, since there are no other Pauli matrices available to gap the spectrum, the Dirac point generated in Eq.~(\ref{eqn:3DDirac}) is not generally robust against perturbations since there are several additional $4 \times 4$ Dirac matrices that can be used to open a gap at the Dirac point. Alternatively, the relative instability of a three-dimensional Dirac point compared to a Weyl point can be seen by simply noting that a Dirac point consists of two Weyl points of opposite chirality. These two Weyl points in general annihilate each other unless additional symmetries protect the Dirac point degeneracy.  
The much more stable non-degenerate Weyl points can be directly generated in a three-dimensional Dirac semimetal by breaking either time-reversal or inversion symmetry. Thus the three-dimensional Dirac semimetal state can be seen as a precursor to the Weyl semimetal.
An illustration of this close relationship between three-dimensional Weyl and Dirac points is given in Fig.~\ref{fig:DiracWeyl}, where the fourfold degenerate Dirac point in an $s$-state model on the diamond lattice \cite{Kane_PRL07} is split into four separate Weyl points by a small inversion breaking perturbation and into two Weyl points with a time-reversal symmetry breaking magnetic field. A fully gapped state can also be reached by breaking the fourfold rotation symmetry or with a magnetic field not aligned along the primary axes. 

In terms of real material realizations of three-dimensional Dirac semimetals, bulk bismuth has been known for quite some time to have massive Dirac bands \cite{Cohen60,Wolff64,Li08Bismuth}. Massless Dirac semimetals with a Dirac point have, however, only just recently been proposed. 
It has, for example, been shown that the critical point between a topological insulator and a normal insulator can constitute a Dirac semimetal \cite{Murakami07,Xu11Science}. However, to reach this state extreme fine-tuning of the alloying composition or spin-orbit strength is needed. This makes such a Dirac semimetal realization very hard to control, especially since the spin-orbit strength is not usually an experimentally continuously tunable parameter.

Even more recently, symmetry analyses of crystal structures hosting the necessary symmetry protection for three-dimensional Dirac points have appeared. A general symmetry analysis of crystal structures able to host Weyl points for spinless electrons was preformed in Ref.~\cite{Manes12}. But more specifically, several different material families have theoretically been identified to host three-dimensional Dirac points, of which a few have also already been experimentally confirmed. 
For example, three-dimensional nonsymmorphic double space groups with four-dimensional irreducible representations at some $\bk$ points have been shown to be candidates to host Dirac points at these $\bk$ points \cite{Young12}. Physical and chemical considerations of the particular material determine if the Dirac points also appear at the Fermi level without any additional non-Dirac-like Fermi surface pockets. First-principles calculations have verified that $\beta$-cristobalite BiO$_2$ fulfills these crystallographic criteria and it has been shown to only have the Dirac points at the Fermi level \cite{Young12}.

A three-dimensional Dirac semimetal state has also been shown to exist by first-principles calculations in the simple stoichiometry alkali pnictides A$_3$Bi with A = Na, K, Rb, which all have hexagonal $P6_3/mmc$ ($D_{6h}^4$) symmetry  \cite{Wang12}. The Dirac points are here protected by the fact that the two bands crossing at the Dirac points belong to different irreducible representations under three-fold rotational symmetry. A small strain breaking this symmetry will open a gap at the Dirac points \cite{Wang12}. Very recent ARPES measurements on Na$_3$Bi have confirmed that it is a three-dimensional Dirac semimetal with only a small in-plane anisotropy, $v_x \approx v_y = 3.7 \times 10^5$~m/s, but a large out-of-plane anisotropy, $v_z = 2.9 \times 10^4$~m/s in the Dirac cone \cite{Liu13Na3Bi}. By modifying the surface by {\it in-situ} evaporation of K atoms it was also found that the Dirac points remain intact, indicating symmetry protection of the Dirac points.

Also a member of the II$_3$-V$_2$ narrow gap semiconductors, Cd$_3$As$_2$, has very recently been shown to be a Dirac semimetal \cite{Wang13, Borisenko13, Neupane13}. Cd$_3$As$_2$ has a complicated crystal structure related to the anti-flourite structure with Cd in tetragonal coordination and with Cd vacancies. Upon cooling, the Cd vacancies order, forming first a centrosymmetric tetragonal unit cell with $P4_2/nms$ ($D_{4v}^{15}$) symmetry and then a non-centrosymmetric body-centered unit cell with $I4_1cd$ ($D_{4v}^{12}$) symmetry. Both structures have three-dimensional Dirac points protected by fourfold rotation symmetry \cite{Wang13}. Recent ARPES results have shown the existence of Dirac points in this material, with the Fermi level located in the conduction band slightly above the Dirac point, suggesting $n$-doping effects \cite{Borisenko13,Neupane13}. The measured Fermi velocity of about $1.6 \times 10^6$~m/s in Cd$_3$As$_2$ is notably higher than, for example, the topological insulator Bi$_2$Se$_3$ and graphene, and explains the very high electron mobility in Cd$_3$As$_2$, which has been known for a long time \cite{Rosenberg59}.
Three-dimensional Dirac materials are currently a very rapidly developing field and there will likely be many more three-dimensional Dirac and Weyl semimetals discovered in the near future.

%% file: Section_Stability.tex
\section{Stability of Dirac materials}
\label{sec:stability}
After having reviewed how an effective low-energy Dirac Hamiltonian appears in a wide range of different condensed matter systems in the last section we now turn to a general discussion on the stability of the energy spectrum in Dirac materials. We have already briefly touched on this subject for individual Dirac materials in the last section, but here we will focus more generally on the symmetries which protect the Dirac point degeneracy and thus give a gapless state. 
This protection is a consequence of material specific symmetries and the spectrum can only be gapped if these protective symmetries are broken. 
The symmetries protecting the Dirac points also determine which physical properties are important for generating the Dirac Hamiltonian, even for Dirac materials with a finite energy gap.
Knowing which symmetries protect the Dirac points thus naturally leads to an understanding of the origin of Dirac materials. Beyond a general discussion on the symmetry protection we will here also review multiple theoretical and experimental results on mass generation in Dirac materials.
Finally, we will also discuss the quite intrusive possibility of destroying the full Dirac spectrum by annihilating two cones with opposite chirality with each other.

\subsection{Symmetry protection}
A prerequisite for the existence of gapless Dirac fermion excitations is the degeneracy point encountered in the Hamiltonian in Eq.~(\ref{eqn:Dirac}) for $\bp\to 0$. At the Fermi level degenerate states either form a Fermi surface, as in the case of conventional metals, or develop an insulating, or many-body, gap. If instead single degeneracy points emerge, as is the case of Dirac materials, this can be traced back to material specific symmetries.

In many materials, time-reversal symmetry and crystal symmetries such as degenerate sublattices, lead to Dirac cones emerging in the energy spectrum.
In graphene, the Dirac points are a direct consequence of the two equivalent sublattices of carbon atoms making the graphene honeycomb lattice. These two sublattices transform into each other under spatial inversion. In combination with time-reversal symmetry this leads to the only possible gap opening terms in the Hamiltonian in Eq.~(\ref{eqn:Dirac}), which are $\sim \sigma_z$, being forbidden \cite{Manes_07} and thus the spectrum necessarily contains Dirac points. If, on the other hand, a sublattice asymmetry is present in graphene the Dirac spectrum is gapped. However, for small asymmetries a linear Dirac spectrum, extrapolating down to the original Dirac points, is still present above the gap.

In topological insulators it is time-reversal symmetry alone which guarantees the existence of Kramers doublets of states at the time-reversal invariant points of the Brillouin zone \cite{Zhang_PhysToday2010,Kane_RMP2010}. These Kramers doublets cannot mix due to time-reversal symmetry and necessarily cross each other in reciprocal space, producing a Dirac point and its accompanied linear Dirac spectrum. While the Dirac point is located at the Fermi surface in suspended graphene, this is in general not the case for the heretofore known topological insulators, where the Fermi level is often found significantly away from the Dirac point due to intrinsic doping. Still, only the Dirac surface states often cross the Fermi level, although the bulk band gap can sometimes be rather small, such as in Bi$_{1-x}$Sb$_x$, the first experimentally discovered three-dimensional topological insulator \cite{Hsieh08}.

In the three-dimensional Dirac semimetals, material specific crystal symmetries are crucial for avoiding producing a gap at the Dirac band degeneracy point, as it consists of two Weyl points with opposite chirality, which easily annihilate each other if not protected by symmetry. For example, in the recently discovered three-dimensional Dirac semimetal Na$_3$Bi, the two bands crossing each other at the Dirac point belong to different irreducible representations under threefold rotational symmetry \cite{Wang12}. In general, materials with two doubly degenerate bands with distinct two-dimensional representations and an avoided crossing are potential Dirac semimetals candidates \cite{Wang13}.

Symmetry also enforces the Dirac degeneracy points in unconventional superconductors and superfluids. Not only is the original global one-dimensional gauge symmetry U(1) broken, but additional symmetries, present in the normal-state Hamiltonian, are also broken in unconventional superconductors/superfluids. These additional symmetries are usually crystal/spatial symmetries and spin-rotation symmetry, but can also include time-reversal symmetry \cite{Sigrist91, Tsuei00RMP}.
However, a combination of U(1) and the broken crystal/spatial symmetries often prevails. 
For example, in the A-phase of superfluid $^3$He, U(1) gauge symmetry broken, but the residual symmetry of the superfluid phase still contains a combined gauge-orbit symmetry: $U(1)^{\text{combined}}=U(1)SO_2^L$. Here, $SO_2^L$ stands for orbital rotations about an axis $\bl$. As a consequence, the superfluid gap vanishes along a nodal line parallel to the axis $\bl$ \cite{Volovikbook}.
Similarly, in unconventional superconductors the symmetry is broken in such a way that the residual symmetry of the superconducting phase still contains elements of the gauge group combined with discrete point group symmetries \cite{Volovikbook}. In the example of $d$-wave superconductors, the residual symmetry is $C_{4v}^{\text{combined}}=U(1)C_{4v}$, since a $\pi/2$ spatial in-plane rotation multiplied by a $\pi$ phase is still invariant in a $d$-wave superconductor. Thus the $d$-wave order parameter changes sign under in-plane $\pi/2$-rotations but remains unchanged under reflections about the in-plane $x$- and $y$-axes. As a consequence, the order parameter has nodes along the lines $|k_x|=|k_y|$, as these directions are mapped onto themselves upon rotation by $\pi/2$ (around the $z$-axis) and consecutive mirror reflection in the $x$- or $y$-axis. If the normal-state Fermi surface cross these $|k_x|=|k_y|$ lines, gapless quasiparticles are found at the crossing points. For a square or rectangular Brillouin zone with a Fermi surface centered around  $\Gamma$ (electron doping) or $(\pi,\pi)$ (hole doping), this necessarily means four Dirac points.

The exception to material specific symmetries protecting the Dirac points are the three-dimensional Weyl semimetals. Here it is instead topology protecting the Weyl points. The Weyl points are (pseudo)magnetic monopoles and can therefore not just disappear by perturbations inducing a gap in the spectrum. This conclusion can also be drawn directly by studying the Hamiltonian Eq.~(\ref{eqn:Weyl}) and noticing that there is no more Pauli matrix left to induce perturbations for a three-dimensional Weyl semimetal. The only symmetry needed to ensure this topological protection is crystal translation symmetry, since the Weyl points appear in pairs at specific crystal momenta. But, as exemplified by graphene, disorder is in practice seldom strong enough to completely destroy two well-separated nodal points.

\begin{figure}[tb]
\includegraphics[width=\columnwidth]{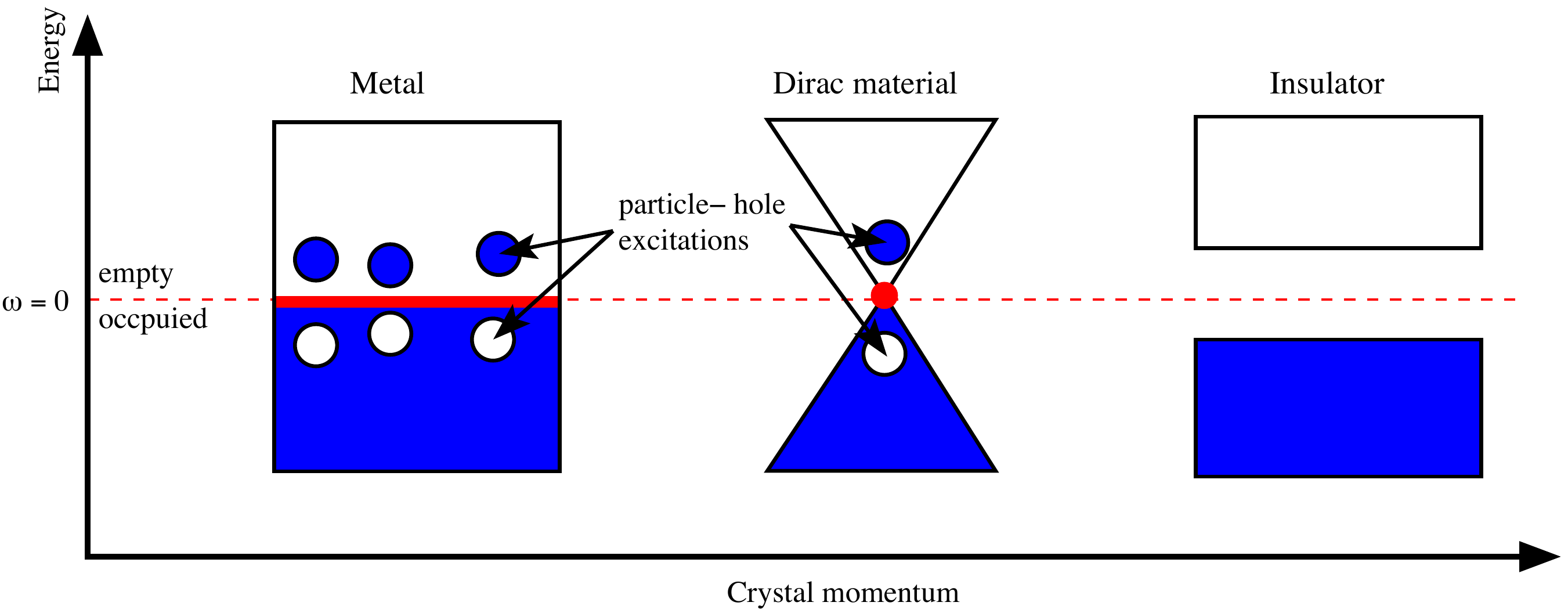}
\caption{\label{fig:fermi_surfaces} The Fermi sea in metals, massless Dirac materials, and insulators. In $d$-dimensional metals, the Fermi surface has dimension $d_{\rm FS}=d-1$. Dirac materials with $d\geq 2$ have dimensionality of the Fermi surface of at least one less, $d_{\rm Dirac}\leq d-2$ when the Fermi level coincides with the Dirac point. For $d = 2$, as is the case of graphene, this means that, if the Fermi surface is located at the Dirac point, it has shrunk to a single point. Note, however, that such Dirac materials still exhibit quasiparticles with arbitrarily low energies both above and below the Fermi energy.}
\end{figure}

In summary, we can thus conclude that symmetry enforced degeneracies present a very general mechanism for creating stable Dirac points.
These symmetry-induced degeneracies also answer a related puzzle: One can define the Fermi surface as the set of $\bk$-points in reciprocal momentum space where on-shell excitations with zero energy, $\omega=0$, exist. For conventional metals, the dimensionality of the Fermi surface is $d_{\rm FS} = d-1$ where $d$ is the dimensionality of the reciprocal space. Insulators do not allow for any on-shell excitations. Gapless semiconductors are at the boundary between these two cases, where there are zero-energy excitations, but the dimensionality of the zero-energy states is at least one less than for conventional metals \cite{Tsildilkovski_book}. Also for Dirac materials with the Dirac point(s) at the Fermi level, the set of $\bk$-points at zero energy has dimension $d_{\rm Dirac}< d-1$.  Physically this means that the phase space available for particle-hole excitations is reduced in both Dirac materials and gapless semiconductors, as depicted in Figure \ref{fig:fermi_surfaces}.
This reduction of Fermi surface phase space cannot be a general case. It can occur by accident as is the case in gapless semiconductors like HgTe \cite{Tsildilkovski_book}. However, in Dirac materials it occurs due to symmetries enforcing the existence of the degeneracy points. Gapless semiconductors are thus not necessarily Dirac materials: the former usually have a parabolic low-energy dispersion with different effective masses for the valence and conduction band \cite{Tsildilkovski_book}.

\subsection{Mass generation}
Directly related to the symmetry protection of the Dirac points is the generation of mass terms in the Dirac Hamiltonian. As long as the symmetry protecting a Dirac point remains unbroken, such mass terms are forbidden in the bulk of the system and the Dirac spectrum remains gapless. Conversely, a mass term and, consequently an energy gap, can be generated by breaking the symmetries protecting the Dirac spectrum. 

\subsubsection{Graphene}
For graphene, terms breaking the (chiral) sublattice symmetry, i.e.~explicitly containing $\sigma_z$ in the sublattice psudospin space, will directly give rise to an energy gap. Theoretically this can be implemented by the Semenoff gap \cite{Semenoff-1984}, where a staggered scalar potential differentiates between the two sublattice sites, or the Haldane gap \cite{Haldane88}, where additionally time-reversal symmetry is broken due to the presence of a staggered magnetic field. From a more experimental viewpoint, sublattice-dependent substrate effects can easily cause sublattice symmetry breaking. Early experimental results showed an energy gap of 0.26~eV for epitaxial-grown graphene on a silicon-carbide (SiC) substrate, attributed to sublattice symmetry breaking \cite{Zhou07}. Another promising substrate candidate is hexagonal boron-nitride (h-BN), whose lattice constant differ by less than $2\%$ from that of graphene, but due to the difference between B and N has a band gap of $6.0$~eV \cite{Watanabe04}. Early first-principles calculations showed a gap opening of $53$~meV at the Dirac points in graphene for the optimal configuration of one carbon atom on top of a boron atom \cite{Giovannetti07}. However, later results \cite{Sachs11} have shown the interlayer bonding being due to the much  weaker but long-range van der Waals force, likely causing both stacking disorder and moir\'{e} structures. The band gap was then found to decrease by at least an order of magnitude. This latter result is in close agreement with scanning tunneling microscopy results, which found quasi-randomly oriented stacking and no band gap \cite{Dean10}.

Spin-orbit coupling will also open a gap in the Dirac spectrum in graphene \cite{Dresselhaus65}. 
Here the original four-fold degeneracy (band and spin) at the Dirac points is partially removed and only the twofold Kramers degeneracy is left, which turns out to result in an energy gap. More specifically, the intrinsic spin-orbit coupling is proportional to $\sigma_z\tau_z s_z$ \cite{Kane_PRL05_2}, where $\tau_z, s_z$ act on valley degree of freedom and in spin space, respectively. Thus spin-orbit coupling can directly be seen to generate an energy gap. This is the only spin-orbit coupling present when mirror symmetry about the graphene plane is preserved. However, the leading contribution is only quadratically dependent on the already small atomic carbon spin-orbit coupling strength and is estimated to only 0.01~K \cite{Min06}. It is this intrinsic spin-orbit coupling that turns the honeycomb lattice with nearest neighbor hopping into a two-dimensional topological insulator, or a quantum spin Hall insulator \cite{Kane_PRL05_2}, but its small intrinsic value for carbon makes graphene not a viable candidate for topological order. If the mirror symmetry is broken by an electric field, a Rasbha spin-orbit coupling term is also generated. The Rashba spin-orbit coupling generates a term of the form $(\sigma_x \tau_z s_y - \sigma_y s_x)$ in the Hamiltonian, and will thus not cause an energy gap \cite{Kane_PRL05_2}. The Rashba spin-orbit coupling is linearly proportional to the atomic carbon spin-orbit coupling, and is in general much larger than the intrinsic spin-orbit coupling in the presence of an electric field \cite{Min06}.

There is also a possibility of spontaneous gap generation in graphene due to long-range Coulomb interactions producing an excitonic state, which has received a fair amount of attention \cite{Kotov12}. The study of this phenomenon dates back to chiral symmetry breaking in QED in 2+1 dimensions \cite{Pisarski84, Appelquist86}. Many body effects are treated in more detail in section \ref{sec:MP}, but, in short, a transition to a gapped state can be found above a critical fine-structure constant $\alpha_c$ \cite{Gorbar02, Khevshchenko01, Khveshchenko09, Liu_Li09, Gamayun10}. Different levels of approximations, including Monte Carlo results \cite{Drut09}, put $\alpha_c$ in the vicinity of the physical value for graphene. The physical nature of the excitonic state has been proposed to be of several varieties, ranging from N\'{e}el and staggered-density states \cite{Herbut06, Khevshchenko01} to Kekule dimerization \cite{Hou_Chamon07} and different time-reversal symmetry breaking phases \cite{Herbut09}, all generating a gap in the energy spectrum.
In the presence of strong short-range repulsion, a time-reversal symmetry broken antiferromagnetic state, or possibly even a gapped spin-liquid state, seems to be favored \cite{Sorella92, Raghu08, Honerkamp08, Meng10,Sorella12}.

Beyond breaking the sublattice symmetry protecting the Dirac points, finite-size confinement has experimentally been shown to generate a finite energy gap in graphene in both ribbon \cite{Han07} and dot \cite{Ponomarenko08} configurations. 
Finite-size confinement breaks the translational symmetry, which is implicitly assumed when obtaining the massless Dirac spectrum in graphene, and can thus produce an energy gap.
Due to the linear energy spectrum in Dirac materials the typical level spacing in a quantum dot of diameter $D$ is $\delta E \approx hv_F/(2D)$, instead of $\delta E \approx h^2/(8m^\ast D^2)$ found for massive carriers in metals with an effective mass $m^\ast$ \cite{Ponomarenko08}.
Thus confinement effects can become important, and generate finite energy gaps due to momentum quantization, even for modest confinements in Dirac materials. This is a direct consequence of the already reduced quasiparticle phase space in Dirac materials. 
In fact, any coupling or scattering between the two Dirac cones at $K$ and $K'$ in graphene also breaks the translational symmetry and can thus generate an energy gap. One example is Kekule dimerization, which has been shown to generate massive Dirac fermions in graphene \cite{Hou_Chamon07}.
In aggregate, including all spin, valley, and pairing channels, there has been shown to be a total of 36 different gap-opening instabilities in graphene \cite{Ryu09}.

\subsubsection{Topological insulators}
\begin{figure}[tb]
\includegraphics[width=\columnwidth]{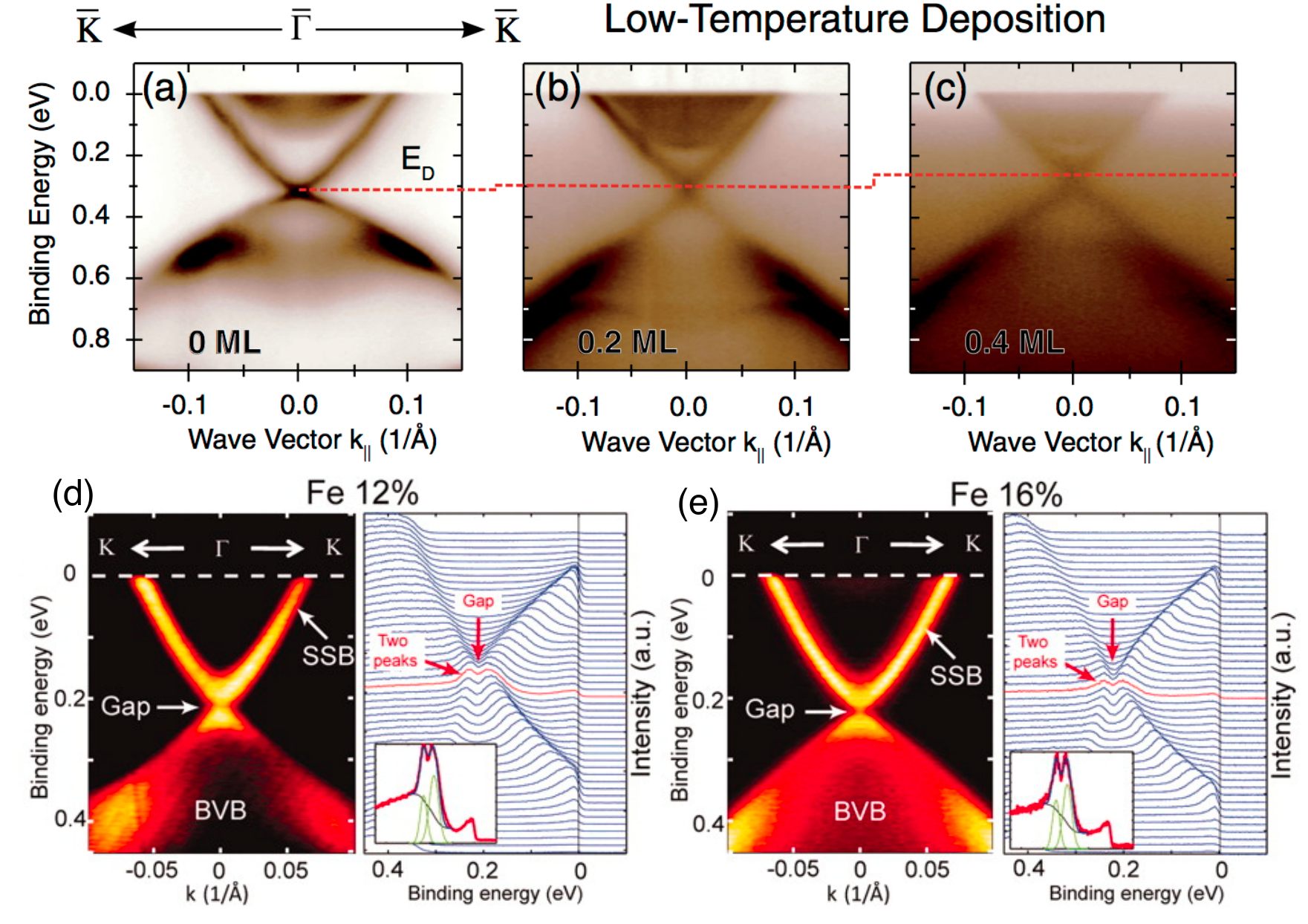}
\caption{\label{fig:TIgaps} ARPES surface spectra of Fe doped topological insulator Bi$_2$Se$_3$. (a)-(c) show the evolving electron spectrum with on-surface deposition of Fe at 8 K ranging form 0 to 0.4 monoatomic layers (ML) of Fe. There is no indication of an energy gap at the Dirac point although there is a shift of the Dirac point (red dashed line) to lower binding energy  with increased Fe coverage. (Reprinted figure with permission from M.~R.~Scholz {\it et al.}, Phys.~Rev.~Lett.~108, 256810 (2012) \cite{Scholz12}. \href{http://link.aps.org/abstract/PRL/v108/p256810}{Copyright $\copyright$~(2012) by the American Physical Society.})
(d)-(e) show the surface energy spectrum for bulk Fe doping of $12\%$ and $16\%$, respectively. The reduced spectral intensity (left panels) and the twin-peak structure in the energy distribution curves (right panels) at the Dirac point signal the opening of an energy gap. Such gap was not found either in pristine or nonmagnetically doped samples. (Adapted from Y.~L.~Chen {\it et al.}, Science 329, 659 (2010) \cite{Chen10}. Reprinted with permission from AAAS.)
}
\end{figure}

In topological insulators a gap can be opened in the Dirac spectrum from time-reversal symmetry breaking perturbations. Magnetic fields \cite{Kane_PRB07}, as well as proximity to a magnetic material or magnetic impurities \cite{Qi08, SCZhangimp} have been theoretically proposed to open a gap. However, experimental studies of magnetic impurities on the surface and in the bulk of topological insulators, using mainly ARPES and STM, are somewhat contradictory at the present time with regards to the existence of an energy gap \cite{Scholz12,Valla12, Xu12, Wray11, Chen10, Schlenk13PRL, Honolka12PRL}. This is exemplified by Figure \ref{fig:TIgaps}, where no energy gap is found to appear at the Dirac point for even large on-surface deposition of Fe. However, for bulk Fe doping, a magnetic state is reached in the bulk and there is a signal of a gap at the Dirac point in the surface state.
One complication at the moment might be that magnetic impurities on the surface of a topological insulator seems to have similar features to nonmagnetic impurities \cite{Bianchi11,Valla12}. For nonmagnetic impurities there is an apparent gap opening due to valence band quantization caused by strong band bending at the surface, which results in confinement effects \cite{Bianchi10, King11}.
Time-reversal symmetry breaking can also be achieved by shining circularly polarized light on the surface of a topological insulator. Time-resolved ARPES studies have recently confirmed that a gap opens at the Dirac point in this case \cite{WangGedik13}.
In the closely related crystalline topological insulators, where crystal symmetries, such as mirror symmetries, give rise to the non-trivial topological state, a tunable gap in the surface Dirac spectrum has been shown to be achievable by elastic strain engineering breaking the protecting crystal symmetries \cite{HsiehTim11}.

\subsubsection{$d$-wave superconductors}
In the $d$-wave superconductors it is also possible to induce a gap in the low-energy spectrum. Here it is the combined $U(1)C_{4v}$ symmetry that needs to be broken by an additional subdominant superconducting order parameter. On very general grounds, based on a Ginzburg-Landau free energy argument, two different superconducting orders favor a $\pi/2$ out-of-phase alignment. The two most widely discussed complex superconducting orders in the high-temperature $d$-wave cuprate superconductors are $d_{x^2-y^2} + is$ and $d_{x^2-y^2}+id_{xy}$, where the $s$ and $d_{xy}$ are the subdominant parts, respectively \cite{Fogelstrom97, Volovik97, Laughlin98, Balatsky98}. Both of these order parameters produce a fully gapped state. Experimental data have, however, largely been contradictory, invoking large imaginary subdominant orders to explain tunneling experiments in YBa$_2$Cu$_3$O$_{7-\delta}$ (YBCO) \cite{Covington97, Elhalel07} and La$_{2-x}$Sr$_x$CuO$_4$ \cite{Gonnelli01} or thermal conductivity in Bi$_2$Sr$_2$CaCu$_2$O$_8$ \cite{Krishana97}, but, on the other hand, only very small imaginary components seem compatible with the absence of any measured spontaneous magnetization \cite{Tsuei00RMP, Carmi00, Neils02, Kirtley06, Saadaoui11}. Recent experiments on nanoscale YBCO islands have shown a small, but measurable, energy gap of size 20-40~$\mu$eV \cite{Gustafsson13}, which was subsequently theoretically identified to be due to a subdominant $s$-wave order supported on the island surfaces \cite{Black-Schaffer13}.
Gapped nodal quasiparticles have also been measured in the bulk in the deeply underdoped regime and seem to be connected to the pseudogap state above $T_c$  \cite{Vishik12, Razzoli13}, but the origin of this gap is at present not understood.

\subsubsection{Three-dimensional Dirac and Weyl semimetals}
Weyl semimetals described by the Weyl Hamiltonian have a spectrum and nodal points protected purely by topology, since all $2 \times 2$ Pauli matrices are already in use in the Hamiltonian building up the spectrum and can thus not be part of a perturbative mass term. There is thus no perturbative approach to gap out a single Weyl node well-separated from other Weyl points in momentum space.
In the three-dimensional Dirac semimetals, however, the nodal points consists of two, in momentum space, overlaying Weyl points with opposite chirality. They will thus annihilate and open up a gap as soon as the (crystal) symmetry protecting the degeneracy at the Dirac points is broken. In the alkali pnictides A$_3$Bi with A = Na, K, Rb breaking the threefold rotation symmetry protecting the band degeneracy at the Dirac point makes the material insulating. First-principles calculations have shown that even a $1\%$ compression along the $y$-axis directly open a 6~meV energy gap \cite{Wang12}. In another recently discovered Dirac semimetal, Cd$_3$As$_2$, breaking the protective fourfold rotation symmetry couples the two degenerate Weyl nodes and results in massive Dirac fermions \cite{Wang13}. However, in contrast to graphene, spin-orbit coupling will in general not generate a gap in the three-dimensional Dirac semimetals \cite{Young12,Wang12,Wang13}.
The three-dimensional Dirac semimetals can also easily be turned into Weyl semimetals by either breaking inversion or time-reversal symmetry \cite{Halasz12, Burkov11,Young12}. This mechanism in fact provides an interesting design approach for generating Weyl semimetals.
For example, lowering the crystal symmetry from $C_{4v}$ down to $C_4$ in Cd$_3$As$_2$ breaks inversion symmetry and splits the Dirac point into two separate Weyl points \cite{Wang13}. Time-reversal symmetry can likewise be broken by a magnetic field, which in the case of Cd$_3$As$_2$ can be quite small and still have a large effect, due to a large $g$-factor \cite{Wang13}.

\subsubsection{Superconducting gaps}
The above discussed mechanisms for mass generation in graphene, topological insulators, and three-dimensional Dirac materials are built upon analysis in the particle-hole channel. In addition, breaking the charge conservation $U(1)$ symmetry with a superconducting electron pair interaction term can also produce an energy gap in Dirac materials at the Fermi level, as it does in normal metals. This is always true for conventional spin-singlet $s$-wave superconductivity, which, per definition, does not have any nodes in the energy spectrum, but other superconducting order parameters can also produce a fully gapped spectrum. For example, several spin-triplet $p$-wave order parameters are fully gapped on the surface of a three-dimensional topological insulator due to an induced spin-singlet $s$-wave pair amplitude \cite{Black-Schaffer13Bi2Se3SC}. 
Complex combinations of nodal superconducting order parameters are also often fully gapped at the normal state Fermi surface. The latter is especially relevant for the sixfold symmetric honeycomb and triangular lattices, where the two $d$-wave solutions are degenerate and thus give a $d_{x^2-y^2} + i d_{xy}$ fully gapped state, if the normal-state Fermi surface is located away from the $\Gamma,K,K'$ points \cite{Black-Schaffer07, Gonzalez08, Nandkishore12}. The same group theory driven degeneracy is present for the fully gapped spin-triplet $p_x+ip_y$-wave state on the square lattice.
It is here, however, important to remember that a superconducting gap always appears at the Fermi level. If the Fermi level coincides with the Dirac point, there will thus be an energy gap at the Dirac point, but otherwise the superconducting energy gap will be located in the linear part of the spectrum, away from the Dirac point. 

In terms of experimental progress on superconductivity in Dirac materials, the proximity effect, where a normal material is placed in close contact with an external superconductor, has proven to be a successful way to introduce superconductivity in Dirac materials. In graphene, proximity-induced superconductivity was first demonstrated in a Josephson junction with Ti/Al superconducting contacts \cite{Heersche07}. 
In topological insulators superconductivity has also been proximity-induced from both conventional \cite{Sacepe11, Veldhorst12, Wang12SC,Yang12SC} and high-temperature $d$-wave superconductors \cite{Zareapour12,Wang13TISC}. Most experiments have so far been focusing on transport properties such as the Josephson current, which often also can extract the induced superconducting gap. The superconducting gap has also explicitly been measured in thin Bi$_2$Se$_3$ films with conventional superconducting contacts using STM/STS \cite{Wang12SC}, and most recently, using ARPES the topological surface state in Bi$_2$Se$_3$ was characterized in the presence of the $d$-wave high-temperature superconductor Bi-2212 \cite{Wang13TISC}. Very intriguingly, the gap in the latter case was seen to be largely isotropic across the Fermi surface, which has been argued to be due to misalignment and disorder \cite{Wang13TISC, Li14}.

\begin{figure}[tb]
\includegraphics[width=\columnwidth]{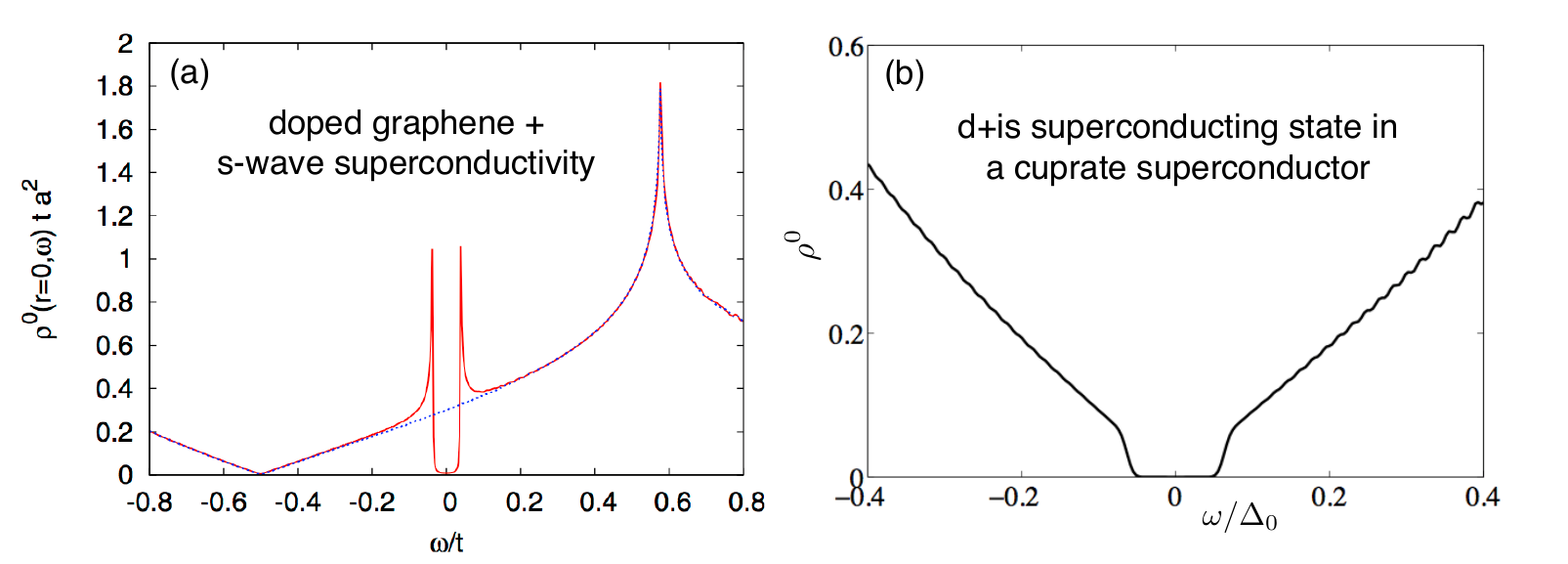}
\caption{\label{fig:cpeaks} 
(a) Density of states ($\rho^0$) for $s$-wave superconducting doped graphene (red line) compared to the normal state (blue dotted line).  The Dirac point is located at $-0.5t$, whereas the superconducting $s$-wave gap appears at the Fermi level and is accompanied by clear superconducting coherence peaks.
(Reprinted by kind permission from F.~M.~D. Pellegrino {\it et al.}, Eur. Phys. J. B 76, 469 (2010) \cite{Pellegrino10}. Copyright $\copyright$~(2010) by Springer Science and Business Media.)
(b) Density of states ($\rho^0$) for a two-dimensional model on a square lattice of a high-temperature $d$-wave cuprate superconductor with a subdominant $is$-wave order, producing a full energy gap at the Dirac point. $\Delta_0$ is the energy of the $d$-wave coherence peaks. Note the absence of coherence peaks when the superconducting gap is at the Dirac point.}
\end{figure}

Directly related to the superconducting gap is the existence of superconducting coherence peaks at the gap edges. In a conventional superconductor these coherence peaks are built up by the density of states expelled from the superconducting gap. Superconducting Dirac materials at finite doping behave similarly, i.e.~they have an energy gap at the Fermi level (located away from the Dirac point) with accompanied coherence peaks \cite{Black-Schaffer08, Pellegrino10, Black-Schaffer13Bi2Se3SC}, as illustrated in Figure \ref{fig:cpeaks}(a) for an $s$-wave superconducting gap in doped graphene. On the other hand, if the Fermi level is aligned with the Dirac point there will, for an $s$-wave order parameter $\Delta$, still be an energy gap for energies $|\omega|<\Delta$, but a density of states $N_0(\omega) = 4|\omega|/W^2$, where $W$ is the bandwidth, beyond the energy gap \cite{Wehling08SC}. There is thus {\it no} coherence peaks in the spectrum for a superconducting gap at the Dirac point. The expelled states in the gap region, although reduced due to the low density of states, are instead recovered at higher energies of the order of the bandwidth. In both graphene and topological insulators a superconducting gap at the Dirac point often requires fine tuning. This is especially true for proximity-induced superconductivity into these materials, as the external superconductor will also affect the doping level of the Dirac material. The situation is, however, qualitatively different in $d$-wave superconductors, as there the Fermi level, per definition, is located at the Dirac point. In the presence of a subdominant order parameter, such that the total superconducting order parameter is fully gapped, there will be an energy gap at the Dirac point and no accompanied coherence peaks. The original $d$-wave order, responsible for the Dirac spectrum in the first place, of course have pronounced coherence peaks where the nodal quasiparticle spectrum ends, but the additional fully gapped state produced by the subdominant order does not have any coherence peaks. This is exemplified in Figure \ref{fig:cpeaks}(b) where we plot the density of states for a high-temperature $d$-wave cuprate superconductor with additional subdominant $s$-wave pairing, producing a $d+is$ state. 
%

\subsection{Dirac cone annihilation}
Apart from mass terms induced by symmetry breaking it is also possible, in the presence of multiple Dirac cones, to induce a gap by merging two Dirac cones. In this process the Berry phases associated with each Dirac cone annihilate each other and thus not only the Dirac points but also the full Dirac cones disappear in a full merge. In terms of the energy spectrum this results in a semi-metal to insulator transition.

\begin{figure}[tb]
\begin{center}
\includegraphics[width=0.5\columnwidth]{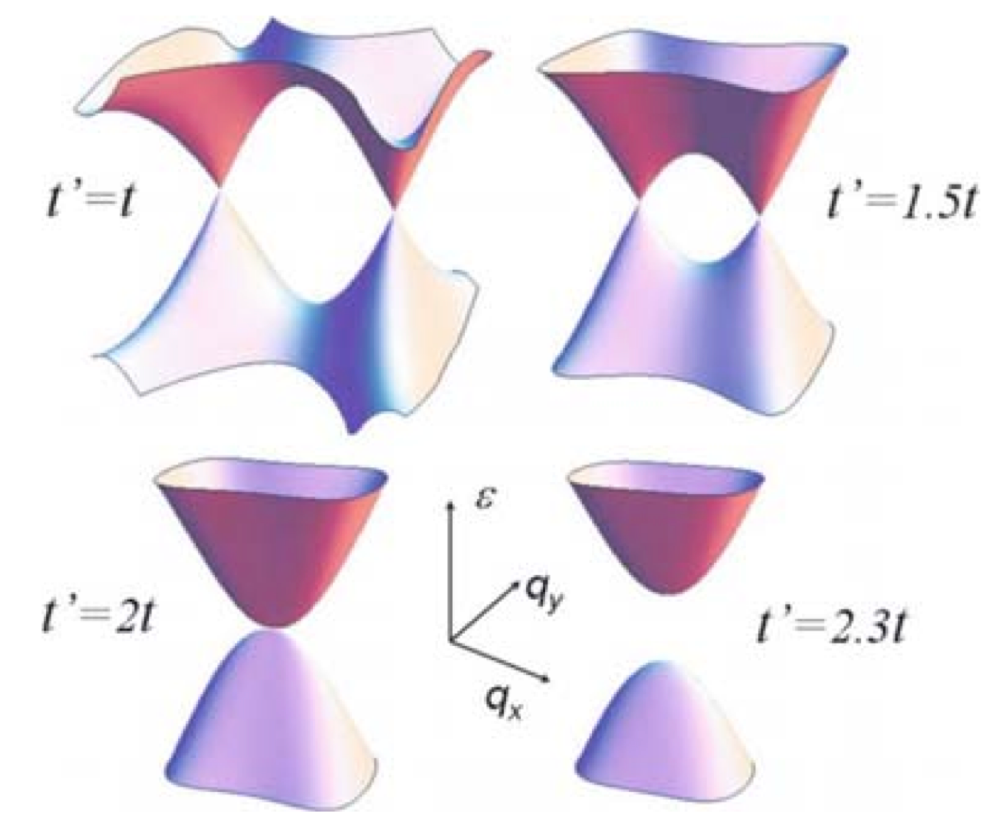}
\caption{\label{fig:DPmerging} 
Evolution of the low-energy spectrum when one of the three nearest neighbor hopping parameters $t'$ increases from the original value $t$ to beyond the Dirac point merging at $t' = 2t$ on the undoped honeycomb lattice. 
(Reprinted by kind permission from G.~Montambaux {\it et al.}, Eur. Phys. J. B 72, 1434 (2009) \cite{Montambaux09b}. Copyright $\copyright$~(2009) by Springer Science and Business Media.)
}
\end{center}
\end{figure}

A prototype for this cone annihilation process can be constructed on the half-filled honeycomb lattice with only nearest neighbor hopping, in a model which imitates strained graphene. For equally strong hopping $t$ on all three nearest neighbor bonds, as is the case in graphene, Dirac cones appear at $K$ and $K'$, with the Fermi level located at the Dirac point. If now one of the three hopping parameters $t'$ gets progressively stronger, the Dirac cones move toward each other, until they finally merge when $t' = 2t$ \cite{Hasegawa06}, see Figure \ref{fig:DPmerging} for an illustration. 
%
At the merging point the low-energy spectrum is linear in one direction but quadratic in the other direction \cite{Dietl08}. This gives rise to a square-root dependence of the DOS on the energy instead of the linear dependence for the individual cones \cite{Hasegawa06}. Moreover, it produces a novel $B^{2/3}$ dependence of the Landau levels on the magnetic field \cite{Dietl08}.
Past the merging point a full gap is present and there is no remnant Berry phase from the original Dirac cones left. 
Beyond this toy model, the merging of two Dirac points in a two-dimensional crystal has also been studied using a universal Hamiltonian \cite{Montambaux09}.

To actually achieve such a Dirac cone annihilation in graphene requires very large deformations of the lattice.
Threshold deformations in the excess of $20\%$ have theoretically been predicted \cite{Pereira09}. The zigzag direction is the most effective for overcoming the threshold, with the armchair direction requiring an even higher strain for producing an energy gap by Dirac cone merging. These very large threshold strains make lattice deformations not a every likely way to gap the graphene spectrum. However, in artificial honeycomb lattices where other methods than strain are available to manipulate the bond lengths, Dirac cone merging is more likely achievable, as for example recently experimentally demonstrated in a tunable ultracold atom honeycomb lattice \cite{Esslinger_2012}.

In strong three-dimensional topological insulators there is per definition an odd number of Dirac cones on each surface, and it is then impossible to annihilate all Dirac cones on the same surface. In contrast, in weak three-dimensional topological insulators the even number of surface Dirac cones could in principle all be mutually annihilated \cite{Kane_PRL07}. The latter is also the case for crystalline topological insulators, where the even number of surface Dirac cones are not protected by time-reversal symmetry but by mirror symmetries \cite{Fu11crystTI,HsiehTim11}.
A related gap opening mechanism is, however, present in very thin strong topological insulator films, where a hybridization gap is present in the surface spectrum due to quantum tunneling between the single Dirac cones on opposite film surfaces \cite{Linder09, Zhang_Xue10}. This gap is due to a cross-over from three-dimensional to two-dimensional behavior.
In the presence of Coulomb interactions a topological exciton condensate gap has also been proposed for thin topological insulator films \cite{Seradjeh09}.

For Weyl semimetals, cone annihilation is the only process, beyond superconductivity, that can gap the Weyl points and even destroy the whole linear Dirac dispersion. Two Weyl points with opposite chirality will annihilate each other if brought together and allowed to merge in momentum space. 
In fact, the same physical process is responsible for gap opening in three-dimensional Dirac semimetals, since Dirac points in three dimensions consists of two Weyl points with opposite chirality.

%% file: Subsection_ARPES.tex
\subsection{Angle-resolved photoemission spectroscopy}
\label{sec:ARPES}
The shared low-energy band structure of Dirac materials naturally makes for very similar results using ARPES. In ARPES a photon is used to eject an electron from a crystal. By analyzing the energy and momentum of the emitted electron the band structure below the Fermi level can be determined. The advantage of ARPES is thus the direct investigation of the momentum-resolved quasiparticle band structure and with it, the Fermi surface topology. This applies equally well in the normal state as in a superconductor.
The ARPES technique has experienced somewhat of a revolution in the last decade, where significantly improved energy and momentum resolution has turned it into an indispensable tool for investigating complex materials \cite{Lu12review}.  
ARPES is mainly a surface sensitive probe with best resolution for in-plane momenta. It can also probe bulk properties, but it is then dependent on the penetration depth of the incoming photon, which vary with photon energy. Further, by investigating and exploiting the dependence of momenta perpendicular to the surface, bulk and surface bands can be distinguished, since surface states will not have any dispersion in momentum perpendicular to the surface. Moreover, APRES can be made spin-sensitive by analyzing the spin of the outgoing electron. Thus, the spin orientation on the Fermi surface can be determined. This has for example proven very useful for measuring the spin-momentum locking in the surface states of topological insulators. 
We will here review some key works leading to the experimental discovery and verification of the linear low-energy Dirac spectrum in graphene, topological insulators, high-temperature cuprate superconductors, and three-dimensional Dirac semimetals. These experimental results explicitly illustrate how these widely different materials all share the common characteristics of Dirac materials. We will also review some ARPES results demonstrating effects beyond the simple low-energy Dirac spectrum, resulting from such effects as symmetry breaking, doping, and many-body interactions.

\subsubsection{Graphene}
\begin{figure}[tb]
\begin{center}
\includegraphics[width=0.7\columnwidth]{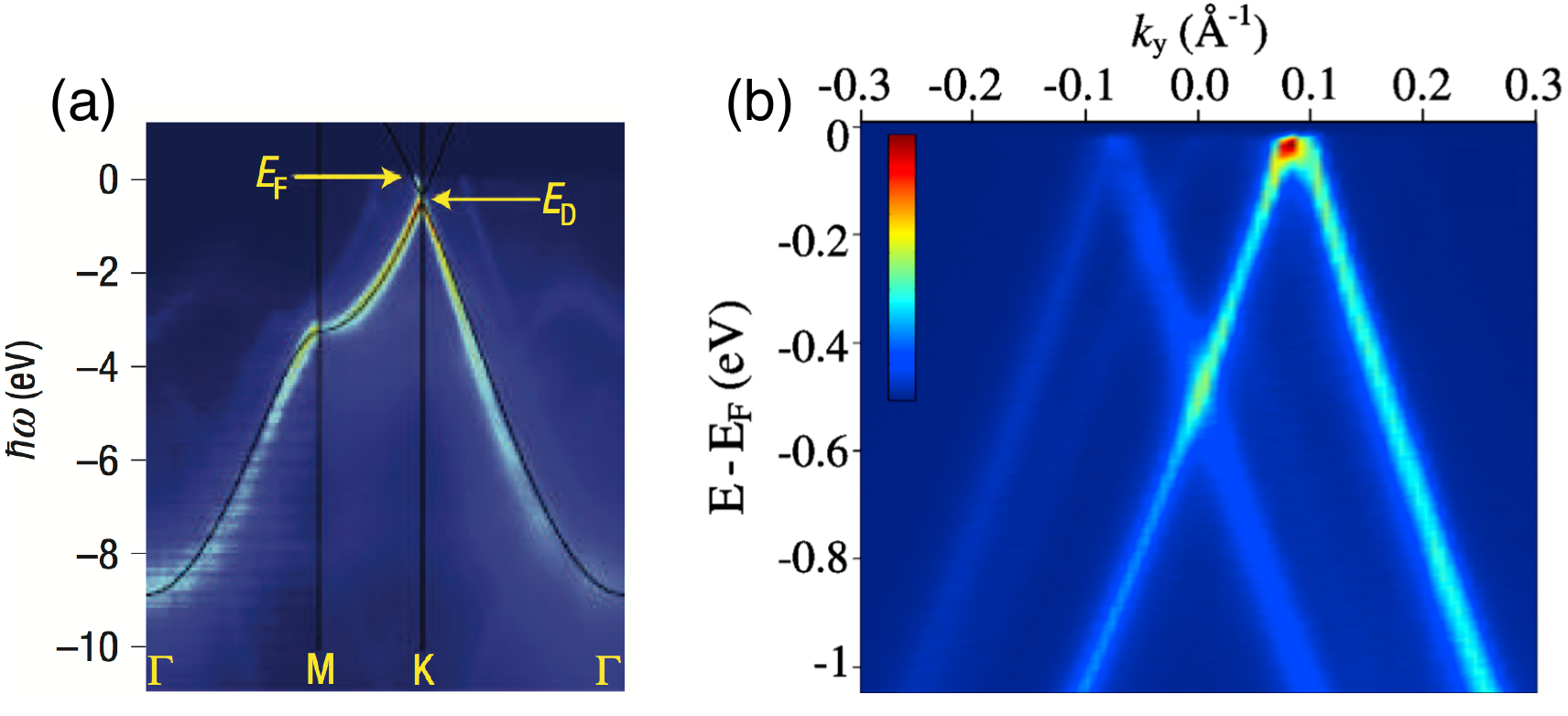}
\end{center}
\caption{\label{fig:ARPES_graphene} (a) ARPES measured dispersion of single-layer graphene grown on the SiC(0001) Si surface. $E_D$ marks the Dirac point and $E_F$ marks the Fermi level, located approximately 0.4~eV above the Dirac point. Black lines are the bands expected from the single orbital model with nearest neighbor hopping $t = 2.82$~eV on the honeycomb lattice. 
(Adapted by permission from Macmillan Publishers Ltd: Nature Physics, A.~Bostwick {\it et al.}, Nature Phys. 3, 36 (2007) \cite{Bostwick07}, copyright \copyright ~(2007).)
(b) ARPES measured dispersion of single-layer graphene grown on the SiC($000\bar{1}$) C surface. Here the Dirac point is  $\sim 30$~meV above the Fermi surface. Ghost Dirac bands are from multiple graphene monolayers which are electronically decoupled due to a small angle rotational stacking disorder. 
(Reprinted figure with permission from M.~Sprinkle {\it et al.}, Phys.~Rev.~Lett.~103, 226803 (2009) \cite{Sprinkle09}. \href{http://link.aps.org/abstract/PRL/v103/p226803}{Copyright \copyright~(2009) by the American Physical Society.})
}
\end{figure}

ARPES has been used extensively on graphene to experimentally verify the linear Dirac dispersion around the $K,K'$ points in the Brillouin zone, as well as to investigate deviations from a perfect Dirac dispersion. Relevant early ARPES experiments studied both graphite, few-layer, bilayer, and monolayer graphene. In graphite, ARPES results have established the existence of Dirac fermions with a linear spectrum at the $H$ corners of the Brillouin zone \cite{Zhou06}. These linearly dispersing bands were shown to co-exist with parabolic bands near the $K$ point due to the interlayer interaction in graphite. Early ARPES measurements on few-layer graphene films grown on 6H-SiC also demonstrated increasing energy contours around the $K,K'$ points with decreasing energies \cite{Rollings06}, consistent with a Dirac spectrum.
For bilayer graphene, the spectrum was confirmed to no longer be linear closest to what would be the Dirac point due to interlayer coupling, while the linear spectrum was shown to recovered at larger energies \cite{Ohta06}.
Finally, measurements on single-layer graphene have shown the existence of Dirac cones with a linear dispersion around the Dirac cone \cite{Bostwick07}, see Figure \ref{fig:ARPES_graphene}(a).

The spectrum in Figure \ref{fig:ARPES_graphene}(a) is for epitaxially grown graphene on the Si face of SiC. While it displays a beautiful Dirac cone, it is significantly doped with the Dirac point $\sim 0.4$~eV below the Fermi surface. This energy shift is attributed to carrier doping from the SiC substrate. If instead growing graphene epitaxially on the C face of SiC, more characteristics associated with an isolated graphene sheet has been achieved. ARPES measurements on graphene on the C face of SiC is shown in Figure \ref{fig:ARPES_graphene}(b), where there is only a $\sim 30$~meV $p$-type doping of the graphene induced from the substrate \cite{Sprinkle09}. Moreover, single monolayers of graphene electronically decouple from each other when grown on the C face of SiC, due to small angle rotational stacking disorder. This results in single-layer properties even for few-layer graphene films. The ghost Dirac cones in Figure \ref{fig:ARPES_graphene}(b) are evidence for the presence of several decoupled layers.
ARPES has also proven to be useful to determine the thickness of multilayer graphene films. Films with one to four layers have be characterized by ARPES in terms of stacking order, interlayer screening, and coupling, whereas thicker films resemble bulk graphite \cite{Ohta07, Siegel10}.

The carrier concentration induced by the substrate effect can be modified by chemical doping. Alkali metal deposition has been shown to induce significant electron doping \cite{Ohta06, Bostwick07}, reaching even up to the van Hove singularity where the Dirac cones at $K,K'$ overlap and form a Fermi surface around the $\Gamma$ point \cite{McChesney10}. Also molecular doping in the form of NO$_2$ adsorption has been investigated using ARPES, showing on an induced hole doping able to tune the Fermi level through the Dirac point energy \cite{Zhou08PRL}.

\begin{figure}[tb]
\begin{center}
\includegraphics[width=0.7\columnwidth]{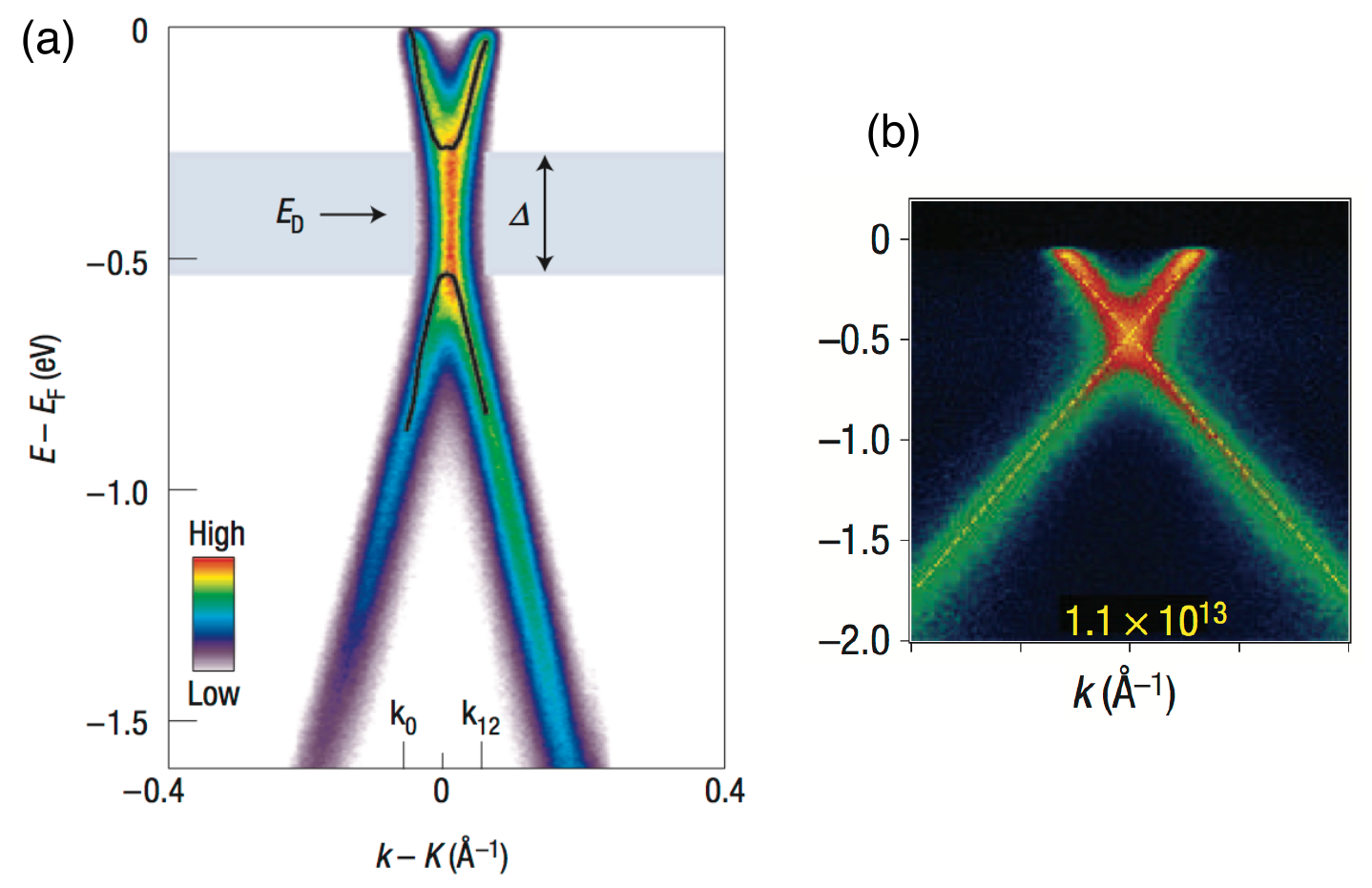}
\end{center}
\caption{\label{fig:ARPES_graphenegap} ARPES dispersion close to the Dirac point for graphene grown epitaxially on the Si face of SiC showing a 0.26~eV gap (a) and no gap, but subtle renormalization from many-body effects (b). Black lines in (a) mark the locations of the peak positions in the energy distribution curves. Dashed lines in (b) are an extrapolation of the linear lower bands.
(a) (Adapted by permission from Macmillan Publishers Ltd: Nature Materials, S.~Y.~Zhou {\it et al.}, Nature Mater.~6, 770 (2007) \cite{Zhou07}, copyright \copyright~(2007).)
(b) (Adapted by permission from Macmillan Publishers Ltd: Nature Physics, A.~Bostwick {\it et al.}, Nature Phys. 3, 36 (2007) \cite{Bostwick07}, copyright \copyright~(2007).)
}
\end{figure}

Substrates do not only have a doping effect on the graphene but can also break the sublattice symmetry between the two atomic sites in the honeycomb lattice of graphene. Such symmetry breaking will, at least theoretically, always result in a gap opening in the energy spectrum at the Dirac point. For graphene grown on the Si face of SiC there are contradictory ARPES reports on gap opening at the Dirac point. In Ref.~\cite{Zhou07} a 0.26~eV gap is reported at the Dirac point. This gap was shown to decrease with graphene layer thickness and it was found to disappear for more than four layers. There is also data showing breaking of the sixfold symmetry down to threefold symmetry around the $K,K'$ points, which would be consistent with sublattice symmetry breaking. On the other hand, in Ref.~\cite{Bostwick07} no gap was found for the same substrate, instead electron-plasmon coupling renormalization of the bands was reported, which induces subtle band changes close to the Dirac point. These two cases are contrasted in Figure~\ref{fig:ARPES_graphenegap}(a)-(b).
Another graphene substrate investigated using ARPES for possible gap formation is Ru(0001). For graphene grown directly on the Ru(0001) surface, with a graphene buffer layer, no gap has been found \cite{Sutter09, Enderlein10}. However, if the interface is also intercalated with Au, a gap of the order of $200$~meV has been reported \cite{Enderlein10}.

Beyond substrate and doping effects, many-body interactions in graphene have also been investigated using ARPES. 
Electron-phonon coupling has been identified through a kink in the linear spectrum at around 200~meV \cite{Bostwick07, Zhou08PRB}, where also electron-electron interactions were evoked to explain the strong electron-phonon coupling \cite{Zhou08PRB}. Coupling to plasmons has also been reported to produce band effects, such that the lower bands do not smoothly pass through the Dirac point but with an additional kink close to the Dirac point \cite{Bostwick07}. Moreover, bound states between charge carriers and plasmons, so-called plasmarons, have also been observed by ARPES close to the Dirac crossing in doped graphene \cite{Bostwick10}. These plasmaron states were found to split the original Dirac point band crossing into three crossings, one between the pure charge bands, one between the pure plasmaron bands, and a third ring-shaped crossing between the charge and plasmaron bands.

ARPES measurements have also been preformed on other graphene-like materials. For example, ARPES has established a linear energy dispersion in silicene grown on Ag(111), although this system has a large substrate-induced energy gap \cite{Vogt12}.

\subsubsection{Topological insulators}
Since the Dirac state is located on the surface in topological insulators and many known three-dimensional topological insulators are layered materials, which easily cleave with clean surfaces, ARPES is an ideal tool for probing topological insulators. APRES has thus naturally played a very instrumental role in the development of the field of three-dimensional topological insulators. In fact, transport measurements have proven hard in these materials, due to difficulty in distinguishing surface from bulk conductivity, and ARPES has emerged as one of the few defining experimental techniques of the field. 
This is in contrast to the two-dimensional topological insulators, or quantum spin Hall insulators, where the surface is only a one-dimensional edge and, as in the case of the HgTe/CdTe quantum wells, the edge states are also often buried inside a heterostructure, hindering ARPES studies. For such two-dimensional quantum wells, transport measurements have instead been instrumental in determining the existence of the (one-dimensional) Dirac edge states.

The first three-dimensional topological insulator experimentally identified was the semiconducting alloy Bi$_{1-x}$Sb$_x$, whose five Dirac surface states were mapped using ARPES \cite{Hsieh08}. Subsequent spin-resolved ARPES measurements proved that the surface states were non-degenerate and strongly spin-polarized, with the spin polarization rotating $360^\circ$ around the central Fermi surface \cite{Hsieh09BiSb}. This confirmed the nontrivial topological order of the surface states. However, since Bi$_{1-x}$Sb$_x$ has a complicated surface band structure, with five branches of surface states and a small bulk band gap, it is not an ideal candidate for Dirac physics. 

\begin{figure}[tb]
\includegraphics[width=\columnwidth]{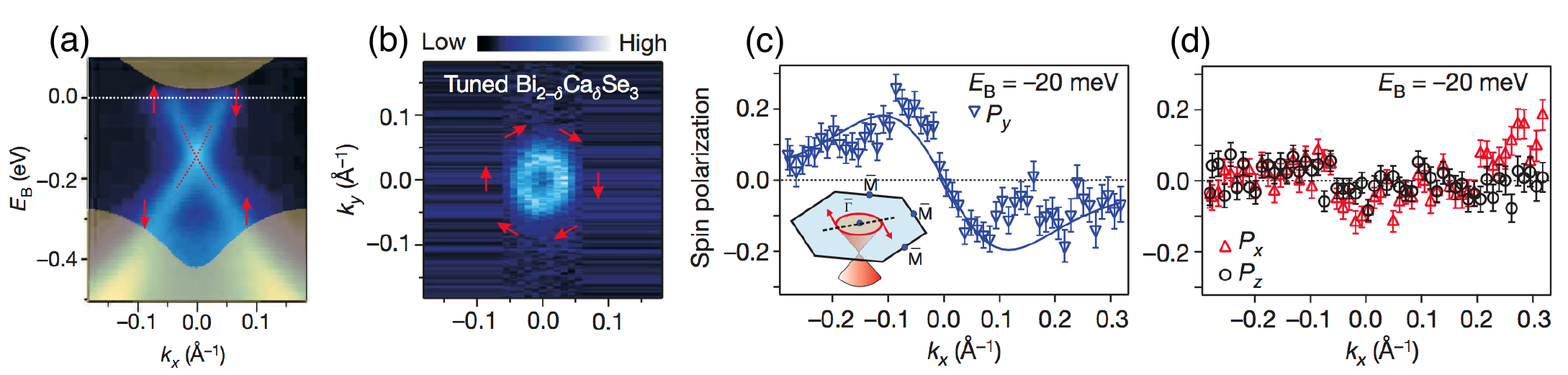}
\caption{\label{fig:ARPES_Bi2Se3} (a) ARPES measured dispersion of Ca-doped Bi$_{2}$Se$_3$ (111) surface along the $k_x$ cut. Ca-doping tunes the Fermi surface to the bulk energy gap.
The shaded regions are the projections of the bulk bands of pure Bi$_2$Se$_3$ onto the (111) surface and dotted red lines are guides to the eye.
(b) APRES intensity map at the Fermi level with red arrows marking the direction of the spin-projection around the Fermi surface. 
(c) Measured $y$-component of the spin-polarization along the $\bar{\Gamma}-\bar{M}$ $k_x$ direction (see inset) at 20~meV binding energy which only cuts through the surface states. Solid line is a numerical fit. Error bars mark one standard deviation.
(d) Measured $x$- (red triangles) and $z$-components (black circles) of the spin polarization for same $k_x$ and energy as (c). 
(Adapted by permission from Macmillan Publishers Ltd: Nature, D.~Hsieh {\it et al.}, Nature 460, 1101 (2009) \cite{Hsieh09spin}, copyright \copyright~(2009).)
}
\end{figure}

APRES results on a second generation of topological insulators, with only one single Dirac surface state and band gaps as large as $\sim 0.3$~eV soon followed. These included especially Bi$_2$Se$_3$ \cite{Hasan_09}, but also Bi$_2$Te$_3$ \cite{Chen09} and Sb$_2$Te$_3$ \cite{Hsieh09PRL}. In all these three-dimensional topological insulators a single Dirac cone with a linear dispersion within the bulk band gap was found, in agreement with theoretical predictions. 
Later spin-resolved measurements also confirmed the spin-momentum locking in the Dirac surface state \cite{Hsieh09spin}. This is illustrated in Figure \ref{fig:ARPES_Bi2Se3}(c,d) for Bi$_2$Se$_3$, where the spin polarization in the surface state along the $k_x$ direction is displayed. Only the $y$-component of the spin polarization is present in this momentum direction and it has opposite spin-polarization for oppositely oriented $\bk$ states, which gives the expected helical spin ordering $({\bm \sigma}\times \bk)\cdot \hat{z}$ in the surface state, as illustrated by the red arrows in Figure~\ref{fig:ARPES_Bi2Se3}(b). However, it should be noted that the amount of spin polarization seen experimentally in the surface state has been varying widely \cite{Hsieh09spin, Souma11PRL, Xu11Science, Pan11PRL}, possibly due to a layer-dependent entangled spin-orbit texture \cite{Zhu13PRL}.

One problem often present even in the second generation of topological insulators, is a finite intrinsic doping, often caused by vacancies in the materials. The position of the Fermi level in the surface states also depends on the detailed electrostatics of the surface and it is often not at the Dirac point. For example, Bi$_2$Se$_3$ has a finite density of $n$-type carriers, believed to be caused by Se vacancies. This places the Fermi level in the upper part of the Dirac cone and also above the bulk conduction band bottom, making Bi$_2$Se$_3$ a bulk metal instead of a bulk insulator \cite{Hasan_09}. Beyond causing conceptual problems for actually defining a topological insulator, not having the Fermi level at or near the Dirac point severely limits the applicability of such a topological insulator as a Dirac material. Careful additional doping has in several cases been shown to overcome these intrinsic doping effects. In Bi$_2$Se$_3$ small levels of Ca can compensate for the intrinsic Se vacancies, placing the Fermi level within the bulk gap \cite{Hsieh09spin}, as seen in Figure~\ref{fig:ARPES_Bi2Se3}(a). Further chemical modification of the surface with exposure to NO$_2$ gas has been found to induced enough hole doping to move the Fermi surface to the Dirac point \cite{Hsieh09spin}. In Bi$_2$Te$_3$ a similar bulk insulating state, by avoiding bulk bands at the Fermi level, has been achieved by $0.67\%$ Sn bulk doping \cite{Chen09}.

\begin{figure}[tb]
\begin{center}
\includegraphics[width=0.7\columnwidth]{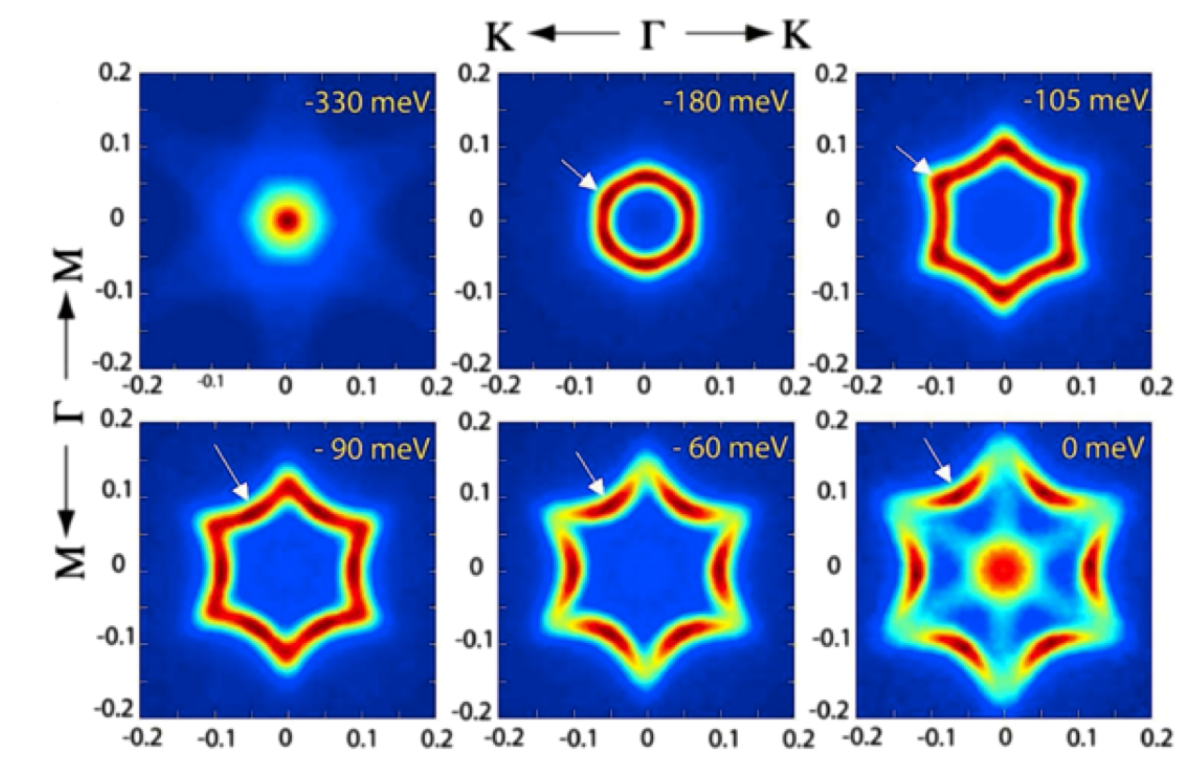}
\end{center}
\caption{\label{fig:ARPES_warping} Constant energy contours measured by ARPES for the surface state in undoped Bi$_2$Te$_3$, starting from the Dirac point located at -335~meV up the Fermi level, clearly indicating a growing hexagonal warping away from the Dirac point. At the Fermi surface a bulk conduction band also appears in the center of the Brillouin zone.
The strength of the DOS grows from no DOS (blue) to large DOS (dark red) and arrows indicate position of maximum DOS, which shows a sixfold symmetry.
(Reprinted figure by permission from Z.~Alpichshev {\it et al.}, Phys.~Rev.~Lett. 104, 016401 (2010) \cite{KapitulnikPRL}. \href{http://link.aps.org/abstract/PRL/v104/p016401}{Copyright \copyright~(2010) by the American Physical Society.})
}
\end{figure}

A simple Dirac cone with linear dispersion captures most of the surface-state behavior in three-dimensional topological insulators, but ARPES and also STS results have found a notable hexagonal warping in especially Bi$_2$Te$_3$ \cite{Chen09, KapitulnikPRL}, as illustrated in Figure~\ref{fig:ARPES_warping}. 
This effect is due to a small bulk band gap and a strong trigonal potential. It can easily be taken into account by adding quadratic and cubic terms to the Dirac Hamiltonian, such that the surface Hamiltonian for Bi$_2$Te$_3$ is instead given by \cite{Fu09warping}:
\begin{align}
\label{eq:warping}
H(\bk) = \bk^2/(2m^\ast) + v_\bk (k_x\sigma_y - k_y \sigma_x) + \frac{\lambda}{2}(k_+^3 + k_-^3)\sigma_z.
\end{align}
Here the first term breaks the particle-hole symmetry, whereas the Dirac velocity $v_{\bk}$ acquires a quadratic dependence on $\bk$ along with the expected constant term. The last term is responsible for the hexagonal warping. Bi$_2$Te$_3$ forms a hexagonal lattice and with $k_\pm = k_x\pm ik_y$, the last term is only invariant under threefold rotation and thus gives rise to hexagonal warping of an otherwise circular Fermi surface \cite{Fu09warping}. Note, however, that close to the Dirac point the spectrum will despite this additional warping still be essentially linear in momentum, as also demonstrated in the experimental ARPES data in Figure~\ref{fig:ARPES_warping}.

There also exist several ARPES studies on possible mass gap openings in the surface Dirac spectrum in topological insulators.
For example, the gap opening in topological insulator thin films, due to a crossover to two-dimensional behavior, where the two surfaces start hybridizing with each other, has been measured by ARPES \cite{Zhang_Xue10, Sakamoto10, Li10thick}.
\begin{figure}[tb]
\begin{center}
\includegraphics[width=1\columnwidth]{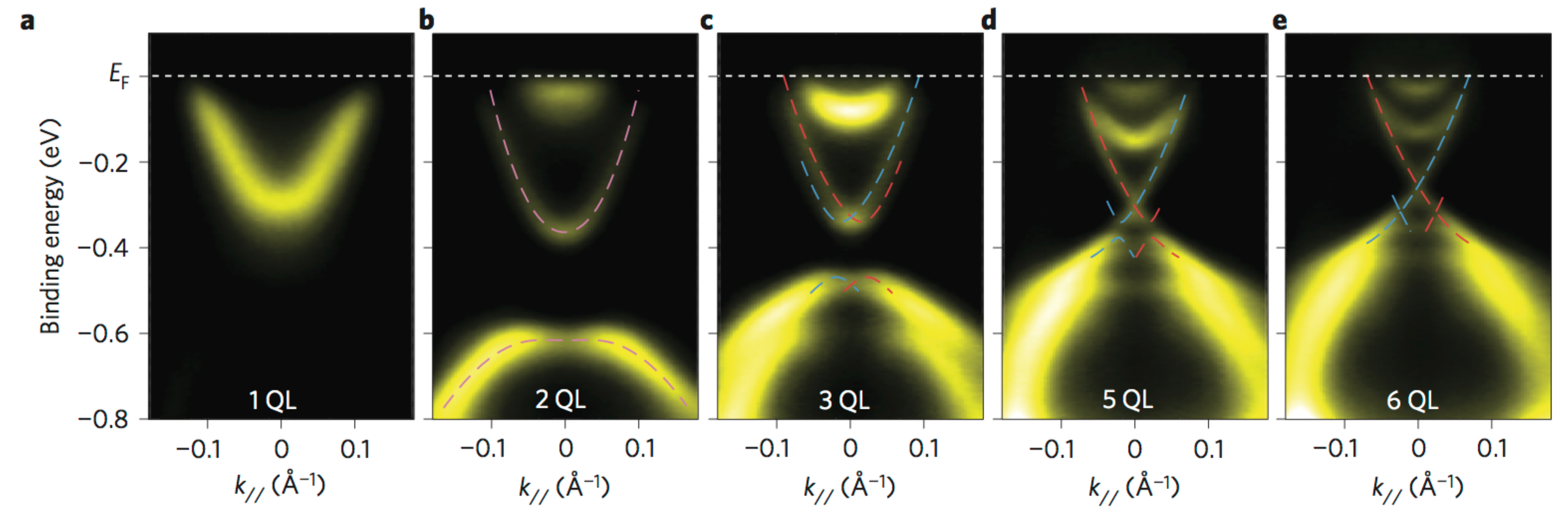}
\end{center}
\caption{\label{fig:ARPES_TIthinfilms} 
ARPES spectra of Bi$_2$Se$_3$ thin films with thicknesses ranging from one to six quintuple layer (QLs) measured in room temperature along the $\bar{\Gamma}-\bar{K}$ direction. The small splitting of the bands is a Rashba spin-orbit coupling effect. The splitting increases for larger wave vectors but disappears at the Dirac point at $\bar{\Gamma}$. Blue and red dashed lines are fitted dispersion lines for the surface states.
(Adapted by permission from Macmillan Publishers Ltd: Nature Physics, Y.~Zhang {\it et al.}, Nature Phys. 6, 584 (2010) \cite{Zhang_Xue10}, copyright \copyright~(2010).)
}
\end{figure}
In Figure~\ref{fig:ARPES_TIthinfilms} the evolution of the surface gap with film thickness is shown for Bi$_2$Se$_3$ thin films \cite{Zhang_Xue10}. At six quintuple layers (QLs) the gap is closed and the film has approached bulk conditions. For thinner films there is a pronounced gap at the Dirac point, due to inter-surface hybridization. Note that this gap appears at the Dirac point, not at the Fermi level.
Gap opening from breaking time-reversal symmetry by introducing magnetic impurities has also been investigated using ARPES, but with so far seemingly contradictory results \cite{Scholz12,Valla12, Xu12, Wray11, Chen10, Schlenk13PRL}, as exemplified by the experimental data presented earlier in Figure~\ref{fig:TIgaps}. A finite energy gap has been measured at the Dirac point \cite{Chen10} for magnetic Fe doping in the bulk, but for Fe surface doping the situation is less clear.
%

Similar to graphene, very recent ARPES measurements have also seen strong mode couplings at low energies \cite{Kondo13PRL}, consistent with phonon modes or possible spin-plasmons \cite{Raghu10PRL}. Despite the very large mass enhancement factor, no evidence of band reconstruction was found. This result is to be expected when evoking the topological protection of the helical Dirac cone.

\subsubsection{High-temperature cuprate superconductors}
\begin{figure}
\begin{center}
\includegraphics[width=0.8\columnwidth]{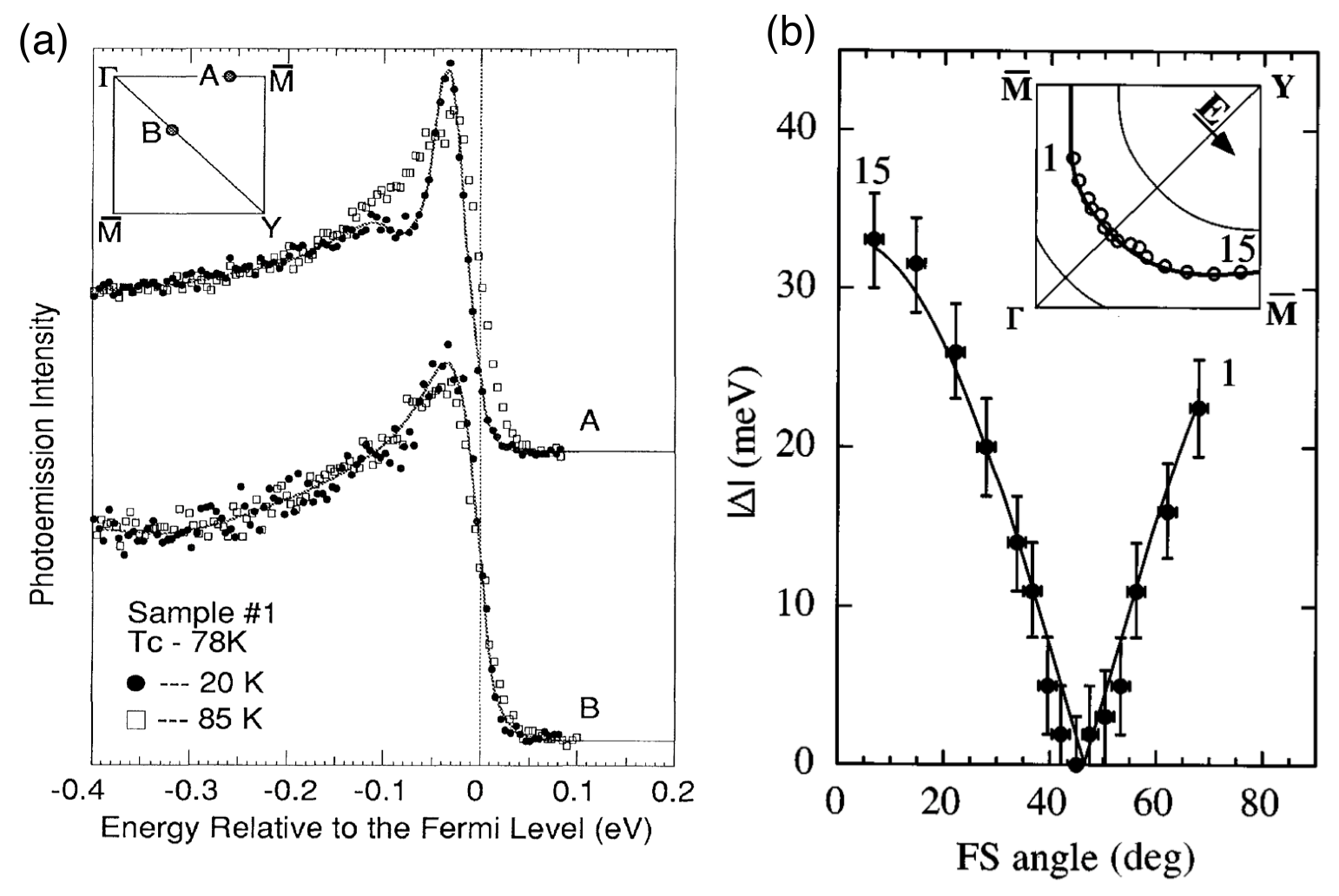}
\caption{\label{fig:ARPEScuprate} ARPES spectra 0n Bi$_2$Sr$_2$CaCu$_2$O$_{8+x}$ (Bi-2212). (a) Spectra taken in the nodal (B) and anti-nodal (A) regions below (black circles) and above (white squares) $T_c = 78$~K. There is no significant change in the spectra at the B point, whereas at the A point there are both line-shape changes as well as a leading edge shift, indicating an energy gap at the Fermi surface (vertical dotted line) in the superconducting phase. Extracted gap is about 12~meV. 
(Adapted with permission from Z.-X.~Shen {\it et al.}, Phys.~Rev.~Lett. 70, 1553 (1993) \cite{Shen93}. \href{http://link.aps.org/abstract/PRL/v70/p1553}{Copyright \copyright ~(1993) by the American Physical Society.})
(b) Extracted superconducting gap at 13~K ($T_c = 87$~K) plotted as function of angle along the normal state Fermi surface, as displayed in the inset. Solid circles mark data, whereas the line is a fit to a $d$-wave gap of the form $\cos(k_x a) - cos(k_y a)$. 
(Reprinted figure by permission from H.~Ding {\it et al.}, Phys.~Rev.~B 54, R9678 (1996) \cite{Ding96PRB}. \href{http://link.aps.org/abstract/PRB/v54/pR9678}{Copyright \copyright~(1996) by the American Physical Society.})
}
\end{center}
\end{figure}

Despite energy resolution and surface sensitivity hampering early ARPES results on the cuprate high-temperature $d$-wave superconductors \cite{Tsuei00RMP, Damascelli03RMP}, the inhomogeniety of the gap in the Brillouin zone along different directions in reciprocal space could be established relatively early \cite{Shen93, Ma95, Ding96PRB, Damascelli03RMP}. Figure~\ref{fig:ARPEScuprate}(a)  shows early ARPES data on overdoped Bi-2212 at momenta A and B, corresponding to the anti-nodal and nodal regions, respectively \cite{Shen93}. In the nodal B region there is no notable difference between spectra taken below and above $T_c$. However, near $(\pi,0)$ at the A point there is a clear difference between the normal and superconducting spectra, both in terms of line-shape evolution but also in a shift of the leading edge, reflecting the presence of a superconducting gap at the Fermi level. These and similar results already early on strongly suggested that the superconducting gap is anisotropic in the cuprates \cite{Damascelli03RMP}.

Further extraction of the gap momentum structure would require accurate measurements of not only the quasiparticle dispersion, but also the band structure in the normal state, since within the BCS framework the quasiparticle energy is given by $E_\bk = \sqrt{\epsilon_\bk^2 +|\Delta_\bk|^2}$. However, due to the non-Fermi liquid behavior of the cuprates in the normal state, $\epsilon_\bk$ is not easily determined to sufficient accuracy. Still, by fitting ARPES data to a phenomenologically broadened BCS spectral function, a quantitative agreement with a $d$-wave gap was reached relatively soon \cite{Ding95PRL, Ding96PRB}, as displayed in Figure~\ref{fig:ARPEScuprate}(b).
Also in other cuprates, including also electron-doped Nd$_{2-x}$Ce$_x$CuO$_4$, ARPES measurements have shown anisotropic superconducting gaps along the Fermi surface, see e.g.~Ref.~\cite{Damascelli03RMP} for a review. However, the lack of phase sensitivity in ARPES prevented a determination of the order parameter symmetry from ARPES alone \cite{Tsuei00RMP, Damascelli03RMP}.

Also after it was established that the cuprate superconductors have a superconducting order parameter with $d$-wave symmetry,  ARPES has continued to play a pivotal role in the study of the cuprates and the quest for the mechanism behind high-temperature superconductivity. Examples of highly topical questions that have been and are studied using ARPES include the pseudogap and its relation to superconductivity, the Mott insulator and the lightly doped region of the phase diagram, and several different dispersion anomalies with possible connections to various bosonic modes, see for example a recent review in Ref.~\cite{Lu12review}. Most of this is beyond the scope of this review article, as it is not directly relevant to the Dirac physics in the cuprate superconductors. However, one recently discovered dispersion anomaly is a very low-energy kink around 10~meV in the nodal dispersion of Bi-2212 \cite{Rameau09, Vishik10, Plumb10, Anzai10}. This is well within the energy regime of the superconducting gap, which is $\sim 40$~meV, and this kink is thus in the linear Dirac spectrum energy regime. This kink appears in underdoped Bi-2212, becomes stronger with underdoping, and leads to a decrease in the nodal Fermi velocity (orthogonal to the Fermi surface) with underdoping and decreasing temperature. The kink leads to a renormalization of the near-nodal low-energy dispersion and has recently been suggested to be due to a coupling to an acoustic phonon \cite{Johnston12}.

\subsubsection{Three-dimensional Dirac semimetals}
ARPES measurements have very recently also been preformed on three-dimensional Dirac semimetals. However, since ARPES is foremost a surface-sensitive technique, acquiring $\bk$-resolved data perpendicular to the surface depends on the penetration depth of the incoming photons, which limits the $\bk$-space resolution in this third spatial direction. Nevertheless, the linear spectrum around the Dirac points in Na$_3$Bi \cite{Liu13Na3Bi} and Cd$_3$As$_2$ \cite{Borisenko13, Neupane13} was very recently established using ARPES.

\begin{figure}[tb]
\includegraphics[width=\columnwidth]{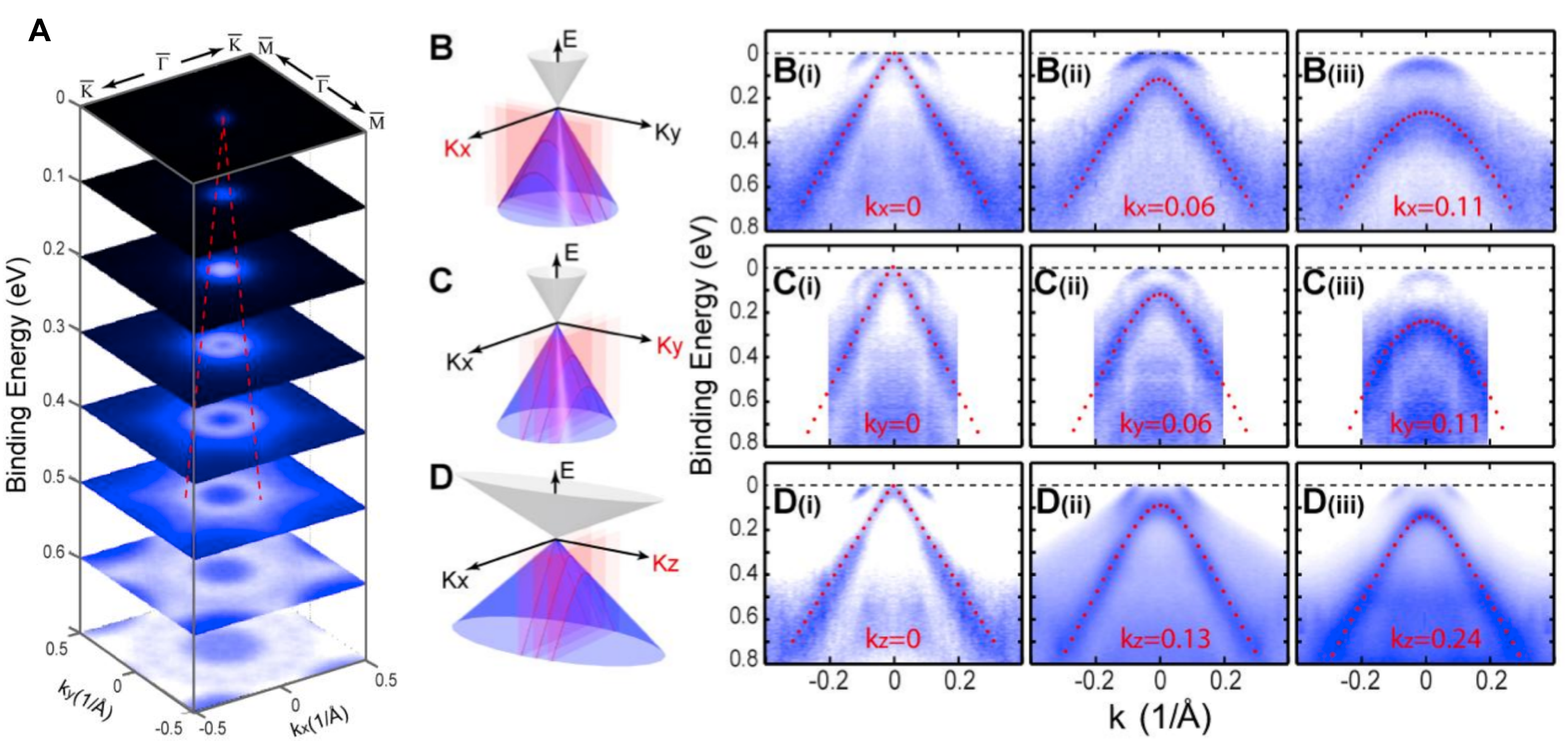}
\caption{\label{fig:Na3Bi} ARPES measurements of the low-energy band structure of Na$_3$Bi (001) single crystals. (A) shows a stacking plot of constant energy contours at different binding energies in the $k_x-k_y$ plane. Red dotted lines are guides to the eye. (B-D) show schematics of the dispersion obtained by ARPES measurements slicing through the three-dimensional Dirac cones at different momentum locations (red planes), displaying conceptually the origin of linear and hyperbolic spectrums. (B-C)(i-iii) measure the dispersion at $k_{x,y} = 0, 0.06$ and 0.11 \AA$^{-1}$, respectively, whereas D(i-iii) shows the dispersion at $k_z = 0, 0.13$ and 0.24 \AA$^{-1}$. Red dotted lines shows the fits obtained using only one set of velocity parameters ($v_x = 2.75$~eV$\cdot$\AA, $v_y = 2.39$~eV$\cdot$\AA, and $v_z = 0.7$~eV$\cdot$\AA).
(From  Z.~K.~Liu {\it et al.}, Science 343, 6173 (2014) \cite{Liu13Na3Bi}. Reprinted with permission from AAAS.)
}
\end{figure}

Measurements on Na$_3$Bi (001) single crystals have shown a pair of three-dimensional Dirac points near the $\Gamma$-point \cite{Liu13Na3Bi}. As seen in Figure~\ref{fig:Na3Bi}, for measurement cuts through the Dirac point the dispersion is linear along all three directions $k_x,k_y$ and $k_z$, whereas for measurement cuts that miss the Dirac point, the dispersion becomes hyperbolic, exactly as expected for a three-dimensional Dirac spectrum. 
The Dirac nature of the spectrum is further confirmed by the need for only the Fermi velocities along the three coordinate axes, $v_x = 2.75$~eV$\cdot$\AA, $v_y = 2.39$~eV$\cdot$\AA, and $v_z = 0.7$~eV$\cdot$\AA, in order to provide excellent fits of the ARPES data. These values display a significant out-of-plane anisotropy of the Dirac spectrum. It was also shown that  additional {\it in-situ} K-atom surface doping destroys the surface spectrum while still preserving the bulk Dirac spectrum, explicitly demonstrating that bulk crystal symmetries protect the three-dimensional Dirac semimetal state. The surface K-atom doping also provided $n$-doping of the bulk Dirac cones, shifting the Fermi level from the Dirac point to the conduction band.

For Cd$_3$As$_2$ two almost simultaneous early ARPES studies both showed three-dimensional Dirac points with the accompanied linear dispersion in all three directions \cite{Neupane13, Borisenko13}. One of the studies reported a linear Dirac cone spectrum with an apex at 0.2~eV binding energy located at the Brillouin zone center when projected onto the (001) surface \cite{Neupane13}. Despite the small doping of the Dirac spectrum, only the Dirac conduction band was found to be present at the Fermi level, while the Dirac valence band was found to co-exist with a parabolic bulk valence band. Moreover, a surprisingly high Fermi velocity of about 10.2~eV$\cdot$\AA\ was reported, which is in agreement with much older mobility measurements \cite{Rosenberg59}. The other study found two Dirac points located slightly displaced from $\Gamma$ along the $\Gamma-Z$ line and also at finite binding energy, $\sim 75$~meV and with a Fermi velocity $5 \pm 2$~eV$\cdot$\AA~\cite{Borisenko13}.

Although several theoretical proposals exist for candidate Weyl semimetal materials, no clear experimental data is yet available. It is also worth keeping in mind that ARPES by itself can only measure the linear spectrum around the nodal points but can often not dissolve the degeneracy of the bands at these points. This can make it hard to separate Weyl semimetals from Dirac semimetals using only ARPES without any additional information, which could determine the Weyl or Dirac nature of the nodal points. One possible key detail is that Weyl semimetal has true Fermi arcs on any surface, whereas in Dirac semimetals two arc segments are connected.

%% file: Subsection_QPI.tex
\subsection{Scanning tunneling spectroscopy}
In contrast to ARPES, STS is a real space surface measurement technique.
However, the spatial dependence of impurity- or boundary-induced states encodes information about the quasiparticle dispersion and inherent quasiparticle symmetries. Therefore, so-called Fourier transformed scanning tunneling spectroscopy (FT-STS) can be used to also experimentally probe energy dispersion relations. In addition, valuable information can also be obtained about the chirality of the quasiparticles from FT-STS, even when the pseudospin vector is not necessarily associated with the electron spin. Thus FT-STS provide highly complementary information to ARPES. In this section we will review key FT-STS measurements on graphene, topological insulators, and $d$-wave cuprate superconductors, establishing both linear Dirac energy dispersion relations and the chiral properties of the Dirac fermions.

In STS, the tunneling current between a sharp metallic tip and a conducting sample is measured across a vacuum gap of a few {\AA}ngstroms as a function of the position $r$ and bias voltage $V$. This yields a map of the local density of states (LDOS) $N(\br,E)$ for each energy $E=eV$ \cite{Tersoff_Hamann_PRB1985}. In absence of disorder $N(\br,E)$ is spatially uniform. Impurities, grain boundaries or other imperfections, however, lead to elastic scattering of quasiparticles. That is, states with different wave vectors $\bk$ and $\bk'$ from the constant energy contour (CEC) $\epsilon_\bk=\epsilon_{\bk'}=E$ can be mixed and interference patterns in the LDOS with characteristic wave vectors $\bq=\bk-\bk'$ can occur. A Fourier transform of the LDOS map at a given bias therefore shows intensity at wave vectors connecting different points of the CEC. More quantitatively, the FT-STS amplitude depends on the number of available initial and final states on a given CEC separated by $\bq$. The FT-STS amplitude is, hence, proportional to the joint density of states $g(\bq,\omega)=\int\diff k N_\bk(\omega)N_{\bk+\bq}(\omega)$, where $N_\bk(\omega)$ is the momentum dependent DOS \cite{Balatsky_RMP}. Thus, maxima are expected for "nesting vectors" $\bq$, which connect parallel segments of the CEC.

\subsubsection{Graphene}
\begin{figure}%
\centering
\includegraphics[width=0.7\columnwidth]{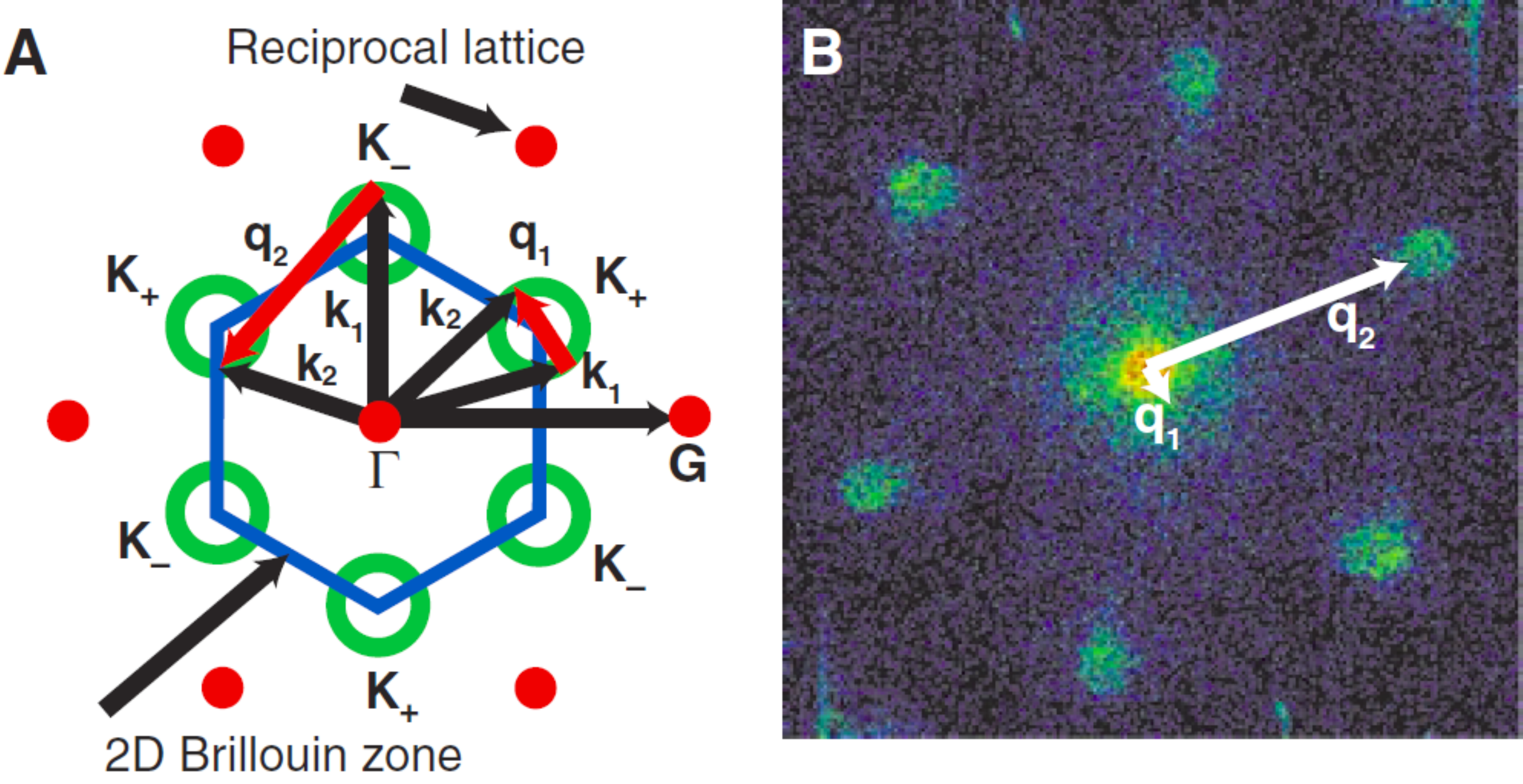}%
\caption{Quasiparticle interference in graphene. (A) Schematic of the two-dimensional
Brillouin zone (blue line), constant energy contours (green rings) at the
$K_\pm$ points, and the two dominant classes of scattering vectors which
create the interference patterns. $\bk_1$ and $\bk_2$ denote the wave vectors of incident and scattered carriers.
Scattering wave vectors $\bq_1$ (short red arrow) are seen to connect points on a single constant energy circle,
while wave vectors $\bq_2$ (long red arrow) connect points on constant energy circles between adjacent $K_+$ and $K_-$ points. Red circles indicate graphene reciprocal lattice points with origin $\bG$.
(B) Fourier transformed STS image of an $100\,{\rm \AA}\times 100\,{\rm \AA}$ area of defective bilayer graphene on SiC. $\bq_1$ scattering forms the small ring at $\bq\approx 0$, whereas $\bq_2$ events create the six circular disks at the $K_\pm$ points. (From  G.~M.~Rutter {\it et al.}, Science 317, 5837 (2007) \cite{Rutter_Science_2007}. Reprinted with permission from AAAS.)
}%
\label{fig:Stroscio_QPI}%
\end{figure}
For doped graphene, the CECs consist of rings around $K$ and $K'$, as illustrated in Figure \ref{fig:Stroscio_QPI}(a). Possible scattering wave vectors can thus connect points on the same CEC (intravalley scattering, $\bq_1$) or reach from $K$ to $K'$ (intravalley scattering, $\bq_2$). According to the nesting condition, FT-STS should exhibit rings of pronounced intensity with radius $q=2k_F$ (i.e.~twice the radius of the Fermi wave vector or the CEC radius $k_F$) around $\bq=0$ and $\bq=K-K'$ (and symmetry equivalent points). The dispersion of the ring radii with bias voltage thus reflects the quasiparticle dispersion $\epsilon_k$.
Indeed, several experiments have revealed quasiparticle interference maps for mono- and bilayer graphene on different substrates \cite{Rutter_Science_2007, Brihuega_PRL08, Crommie_NPhys_08, LeRoy_PRB_09, Mallet_PRB12} and demonstrated the expected linear dispersion. Moreover, the measured Fermi velocities have been found to be of the order of $v_F\approx 1.2$ - $1.5\cdot 10 ^6$\,m/s \cite{Crommie_NPhys_08,Mallet_PRB12}, which are roughly in agreement with ARPES \cite{Sprinkle09, Bostwick07} and magnetotransport measurements \cite{Geim2005, Zhang2005}. Quasiparticle interference resulting from inter- \cite{Brihuega_PRL08, LeRoy_PRB_09, Mallet_PRB12} and intra-valley scattering \cite{Crommie_NPhys_08} has also been detected.

\begin{figure}%
\includegraphics[width=\columnwidth]{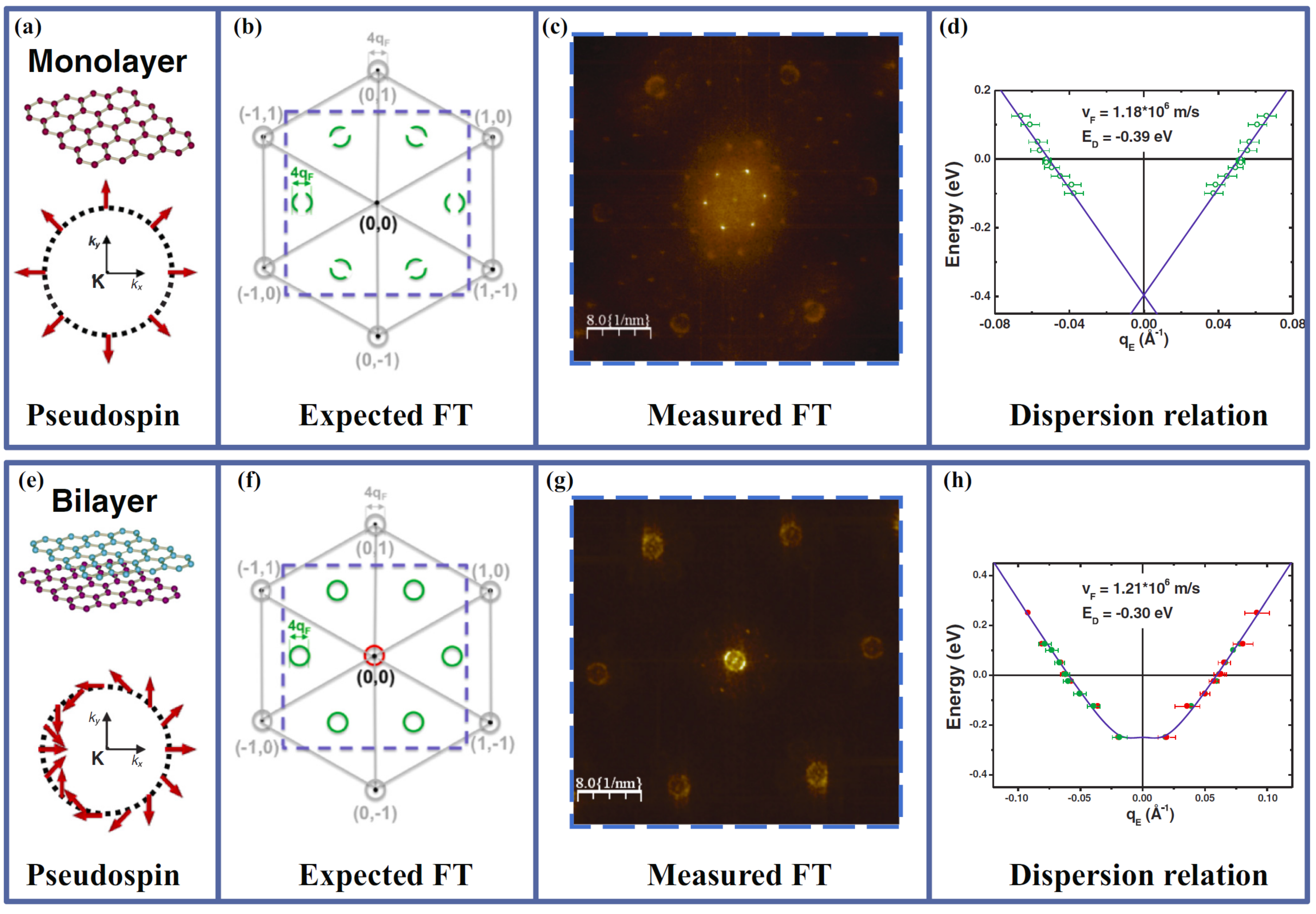}%
\caption{(a) Pseudospin texture of monolayer graphene. (b) Theoretically expected FT-STS map taking into account the Fermi surface topology and the pseudospin of the monolayer. (c) FT-STS map obtained from STS measurements on monolayer graphene on SiC(0001). Note the good agreement between (b) and (c). (d) Dispersion relation derived from the FT-STS for monolayer graphene. (e)-(h) are the same as (a)-(d) but now for bilayer graphene. Note the correspondence between (f) and (g): the pseudospin has no significant impact for the bilayer, contrary to the monolayer case. 
(Reprinted figure with permission from P.~Mallet {\it et al.}, Phys.~Rev.~B~86, 045444 (2012) \cite{Mallet_PRB12}. \href{http://link.aps.org/abstract/PRB/v86/p045444}{Copyright \copyright~(2012) by the American Physical Society.})
}%
\label{fig_QPI_Gr_Mallet}%
\end{figure}

A more detailed analysis of FT-STS images (c.f.~Figure \ref{fig_QPI_Gr_Mallet}) gives access to the chiral properties of the Dirac fermions \cite{Brihuega_PRL08,Peres_NJP2009,Bena_QPI_review2011,Mallet_PRB12}. Regardless of the nature of the scattering centers, the Dirac fermion pseudospin can suppress certain interference spots if the FT-STS procedure involves pseudospin averaging. FT-STS spots near the center of the first Brillouin zone naturally average out LDOS modulations inside a unit cell, which exactly corresponds to pseudospin (i.e.~sublattice) averaging in the case of graphene. Since intra valley back scattering in graphene results in pseudospin reversal, i.e.~orthogonal pseudospin states, the $2k_F$ interference rings are suppressed in the center of the Brillouin zone, as seen in Figure \ref{fig_QPI_Gr_Mallet}(b) and (c). In contrast, bilayer graphene exhibits a different pseudospin texture, which leaves the FT-STS maps rather unaffected in Brillouin zone center. FT-STS has been able to resolve this distinction between mono- and bilayer graphene \cite{Brihuega_PRL08,Mallet_PRB12}. In the case of intervalley scattering, the Dirac pseudospin textures are expected to impose characteristic angular modulations on FT-STS rings near the Brillouin zone corners, which were also observed in the same experiments \cite{Brihuega_PRL08,Mallet_PRB12}. For graphene, quasiparticle interference thus provides a solid experimental verification of both the linear dispersion and the quasiparticle chirality.

\subsubsection{Topological insulators}
\begin{figure}%
\centering
\includegraphics[width=\columnwidth]{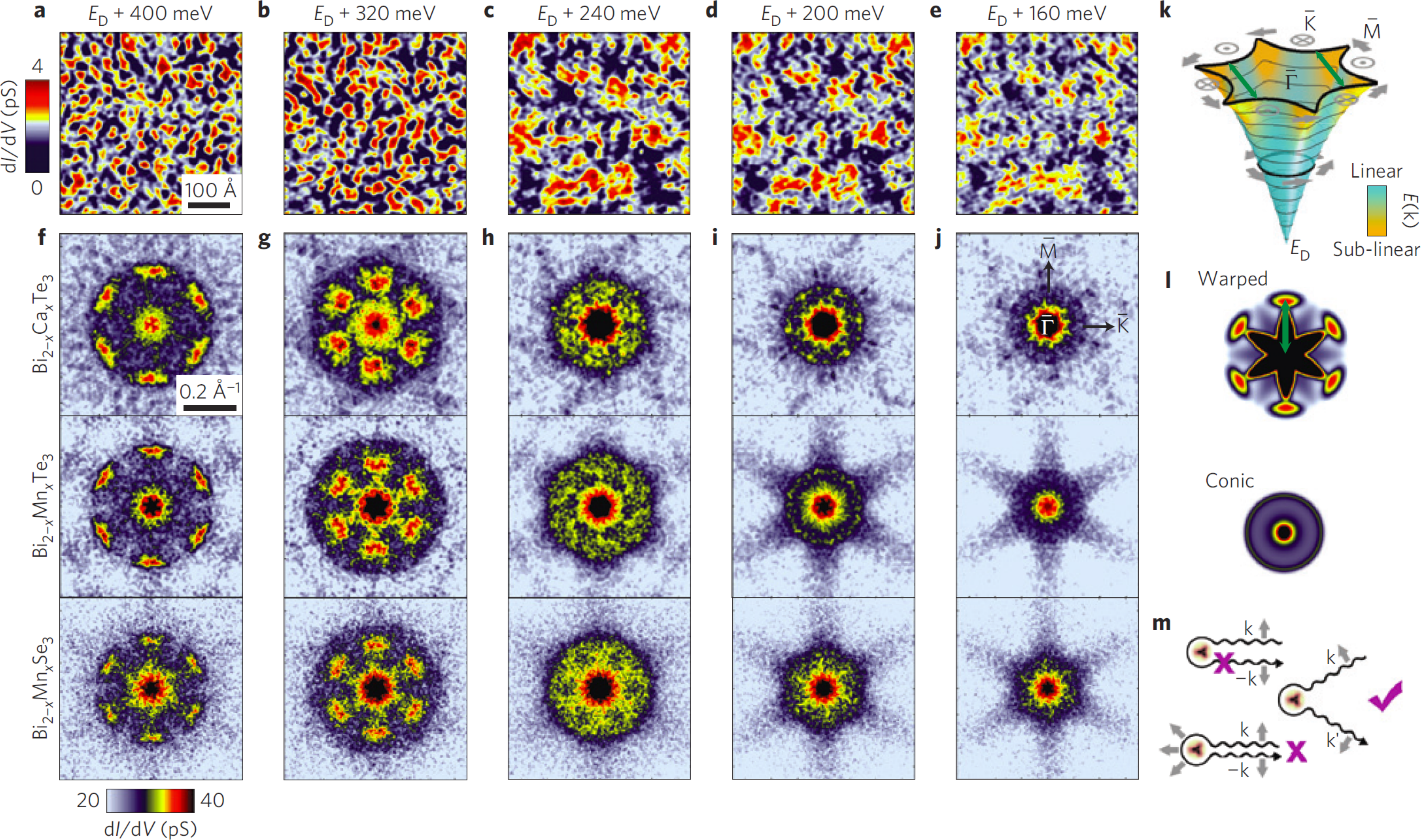}%
\caption{Quasiparticle interference in topological insulators. (a--e) Real-space quasiparticle interference patterns on the surface of Ca-doped Bi$_2$Te$_3$ at different energies. (f--j) Fourier transforms of the quasiparticle interference patterns from Ca- and Mn-doped Bi$_2$Te$_3$ and Mn-doped Bi$_2$Se$_3$ given in top, middle and bottom panels, respectively. All compounds show similar patterns in $\bq$-space, consisting of six strong peaks along the $\bar\Gamma$~--~$\bar{\rm M}$ directions at high energies and circular patterns at lower energies. (k) Schematic surface band structure of Bi$_2$Te$_3$ and the associated spin texture. Warping at high energies supports approximate-nesting conditions, those along $\bar\Gamma$~--~$\bar{\rm M}$ indicated by green arrows. (l) Calculated spin-dependent scattering probability captures the Fourier space quasiparticle interference patterns for the high-energy warped surface and the low-energy conical surface, respectively. (m) Illustration of processes and their contribution to the quasiparticle interference pattern: helicity forbids back scattering off non-magnetic impurities (top) but allows oblique scattering and interference that have a finite overlap between initial and final spin states (middle). A magnetic impurity allows spin-flip back scattering but not interference of the initial and final spin states which remain orthogonal. 
(Adapted by permission from Macmillan Publishers Ltd: Nature Physics, H. Beidenkopf {\it et al.}, Nature~Phys.~7, 939 (2011) \cite{Yazdani_NPhys2011}, copyright \copyright ~(2011).)
}%
\label{fig:Yazdani_TI_QPI}%
\end{figure}
Recently, considerable efforts have also been made to probe the Dirac nature and the robustness of the surface states in three-dimensional topological insulators by means of quasiparticle interference maps. Expected features, such as a linear dispersion and the absence of certain quasiparticle interference spots due to the spin textures of the surface states have been established for several topological insulators, including Bi$_{1-x}$Sb$_x$ \cite{Yazdani_Nature09}, Bi$_2$Te$_3$ \cite{Xue_TI_QPI_PRL_2009,KapitulnikPRL,Madhavan_PRL_2011,Yazdani_NPhys2011}, as well as Bi$_2$Se$_3$ \cite{Yazdani_NPhys2011,Kimura_PRB2012,Stroscio_TI_QPI_PRB2013}. In topological insulators, the effect of (pseudo)-spin orthogonality upon back scattering on the quasiparticle interference pattern turns out to be at least as drastic as in the case of graphene. For Bi$_2$Te$_3$ and Bi$_2$Se$_3$ there is just a single Dirac cone (no valley degeneracy) and pronounced FT-STS signatures occur only at higher energies where there is significant warping of the CECs \cite{KapitulnikPRL, Madhavan_PRL_2011, Yazdani_NPhys2011, Kimura_PRB2012, Stroscio_TI_QPI_PRB2013, Kimura_PRB2012, Stroscio_TI_QPI_PRB2013}, as illustrated in Figure \ref{fig:Yazdani_TI_QPI}. At energies where the Dirac cone is spherical, direct back scattering would provide nesting and characteristic rings of radius $2k_F$ would be expected. Due to pseudospin orthogonality, however, the detection of interference patterns due to directly back-scattered Dirac fermions is suppressed in non-spin-polarized STS, see Figure \ref{fig:Yazdani_TI_QPI}(m). In analogy with graphene, no $2k_F$-rings are seen here. At higher energies, hexagonal warping leads to various nesting vectors, see Figure \ref{fig:Yazdani_TI_QPI}(k), which connect parts of the CECs with nonorthogonal pseudospins and give rise to pronounced FT-STS spots. These FT-STS results clearly played pivotal role in establishing the chiral properties of the quasiparticles in the topologically protected surface surface. Naturally, spin-polarized STS can directly probe the Dirac particle (pseudo)-spin in topological insulators and related experiments are currently on the way.

\subsubsection{High-temperature cuprate superconductors}
\begin{figure}%
\centering
\includegraphics[width=0.8\columnwidth]{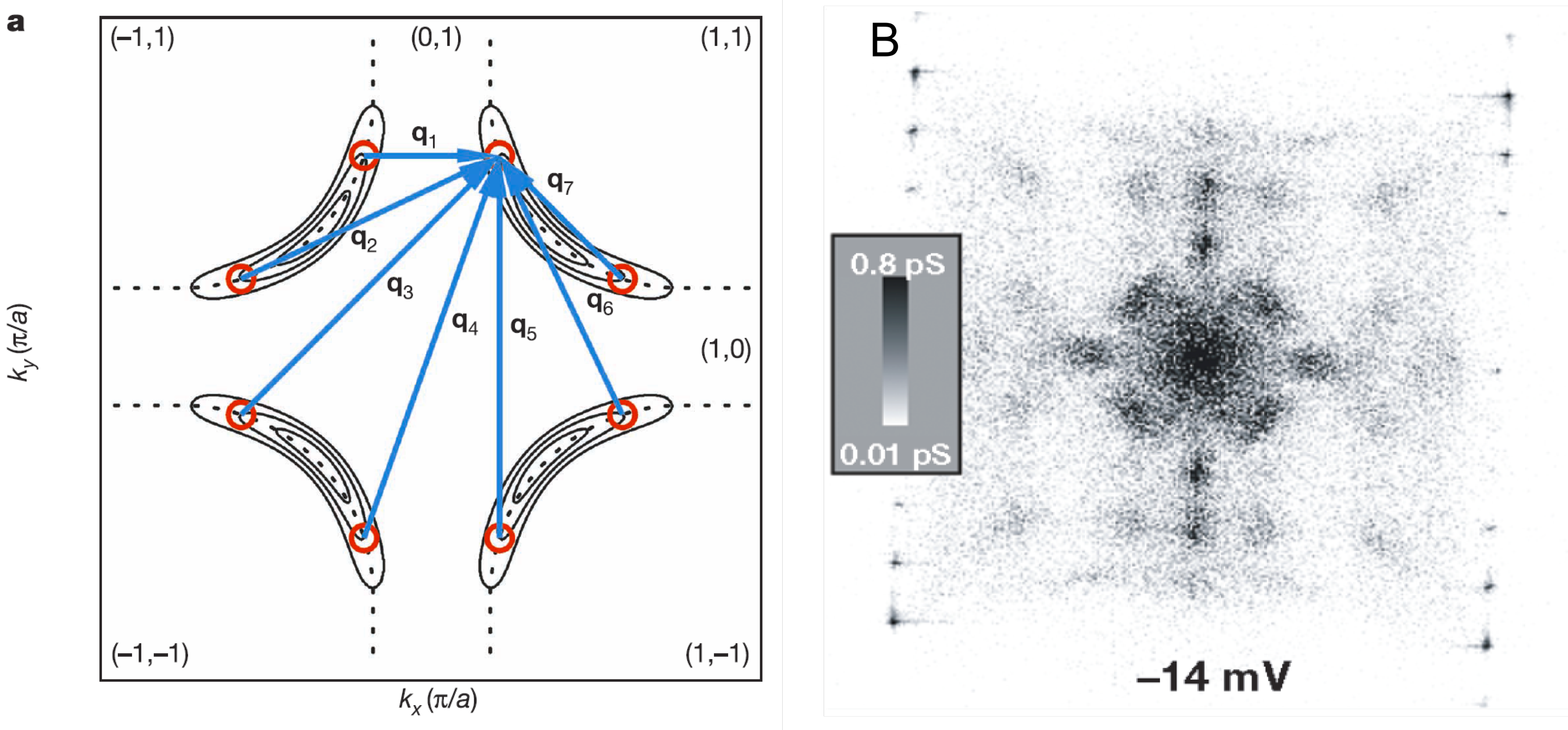}%
\caption{(a) $\bk$-space electronic states in a high-temperature cuprate superconductor. Normal-state Fermi surface (dashed) and constant energy contours for the Bogoliubov quasiparticles (solid banana shape curves). Arrows denote scattering $\bq_i$ vectors responsible for quasiparticle interference patterns. (b) FT-STS map of a Bi$_2$Sr$_2$CaCu$_2$O$8+\delta$ surface, revealing (most) of the LDOS modulations indicated by the arrows in (a).
(Adapted by permission from Macmillan Publishers Ltd: Nature, K. McElroy {\it et al.}, Nature 422, 592 (2003) \cite{McElroy_Nature_2003}, copyright \copyright ~(2003).)
}%
\label{fig:d-wave_QPI}%
\end{figure}
While arguably the most complex electronic Dirac materials, historically $d$-wave cuprate superconductors were the first class of Dirac materials to be studied by means of FT-STS \cite{Hoffman_Hudson_Science2002,Hoffman_McElroy_Science2002,McElroy_Nature_2003,Howald_PRB_2003}. In $d$-wave cuprate superconductors, the Bogoliubov quasiparticles are the native excitations and exhibit banana-shaped constant energy contours at energies below the maximum gap, as displayed in Figure \ref{fig:d-wave_QPI}(a). This shape directly affects quasiparticle interference patterns. As the normal state Fermi velocity in general exceeds the nodal velocity by far (c.f.~Section \ref{subsec:d-wave}), the largest contribution to the $\bk$-resolved density of states $N_\bk(\omega)\sim |\nabla_k E_\bk|^{-1}$ arise from the eight tips of the CECs. Consequently, the largest FT-STS intensity is expected at wave vectors $\bq$ which connect different CEC tips, simply due to the large density of the initial and final states. The experimental data in Refs.~ \cite{Hoffman_McElroy_Science2002,McElroy_Nature_2003,Hanaguri_Science_2009} were found to be in good agreement with this model, as illustrated in Figure \ref{fig:d-wave_QPI}(b). Energy dependence of the FT-STS spots have also given access to the nodal velocity and the normal state Fermi velocity, which turned out to be in accordance with ARPES measurements \cite{McElroy_Nature_2003}.

For graphene and topological insulators, the pseudospin averaging of the STS probe turned out to suppress certain FT-STS spots. For $d$-wave superconductors, the situation is, however, different. The Bogoliubov quasiparticles are coherent superpositions of electrons and holes and the Nambu pseudospin entering the $d$-wave superconducting Dirac Hamiltonian simply refers to electrons and holes (rather than e.g.~the sublattice in the case of graphene). Bogoliubov states where the electron component exceeds the hole component, yield an enhanced probability for electron tunneling into a superconductor but suppressed hole tunneling. Hence the tunneling intensity at a positive sample bias is large but reduced at negative bias \cite{Balatsky_RMP}. In other words: at any finite bias, one component of the Nambu pseudospin is preferentially probed by STS and there is no (complete) pseudospin averaging.

The field of FT-STS on high-temperature cuprate superconductors has further evolved in recent years and allowed also for phase sensitive investigations of the order parameter \cite{Hanaguri_Science_2009}.
The classical problem of competing charge or spin orders in cuprates can be naturally addressed in FT-STS, where (static) modulations manifest as non-dispersing spots. Corresponding features have been found in various experiments and have been reviewed in Ref. \cite{Balatsky_RMP}.

%% file: Section_ManyBodyEffects.tex
\section{Many-body interactions}
\label{sec:MP}
Having established the microscopic origin of Dirac materials, as well as reviewed multiple experimental results measuring the Dirac spectrum in these materials, we now turn to reviewing common properties among Dirac materials. First we focus on the properties of many-body interactions in Dirac materials.
Many-body effects, and in particular electron-electron interactions, in condensed matter physics are a very complex issue as it involves understanding the (collective) behavior of a macroscopic number of particles. 
It is also an unusually rich and constantly developing field. 
We will here not even attempt to give a comprehensive review of many-body effects in all different Dirac materials, but only try to point out a few effects present in Dirac materials because of their peculiar low-energy spectrum, which are qualitatively different from normal metals. These effects include an unusual electron-hole excitation spectrum, lack of electric screening, electron-electron interactions causing velocity renormalization of the slope of the Dirac spectrum and generation of mass gaps, as well as the possibility of achieving superconductivity in these systems.

For more information on many-body effects in Dirac materials, we mention Ref.~\cite{Kotov12} which contains a recent, very comprehensive review of electron-electron interactions in graphene. Other many-body effects in graphene, including electron-phonon interactions, were also recently reviewed in Refs.~\cite{AHC_RMP,Katsnelson_book}.
For the more recently discovered topological insulators, and also three-dimensional Dirac materials, the effects of electron-electron interactions are only started to be investigated. 
In fact, for topological materials there is currently a strong activity trying to extend topological classifications beyond simple band insulators, where per definition the electronic interactions are weak, to interacting systems, see for example a recent review in Ref.~\cite{Turner13}. 
Before proceeding, we also note that the situation for $d$-wave superconductors is quite different.
Here the Dirac fermions are Bogoliubov quasiparticles, which have a mixture of electron and hole properties, plus there is a superconducting condensate background which effectively screens the Coulomb repulsion. Thus, most many-body effects reviewed in this section 
do not apply to $d$-wave superconductors, and we will thus primarily focus the discussion on Dirac materials with electron (or hole) quasiparticles.

\subsection{Electronic excitations}
\begin{figure}[htb]
\begin{center}
\includegraphics[width=0.95\columnwidth]{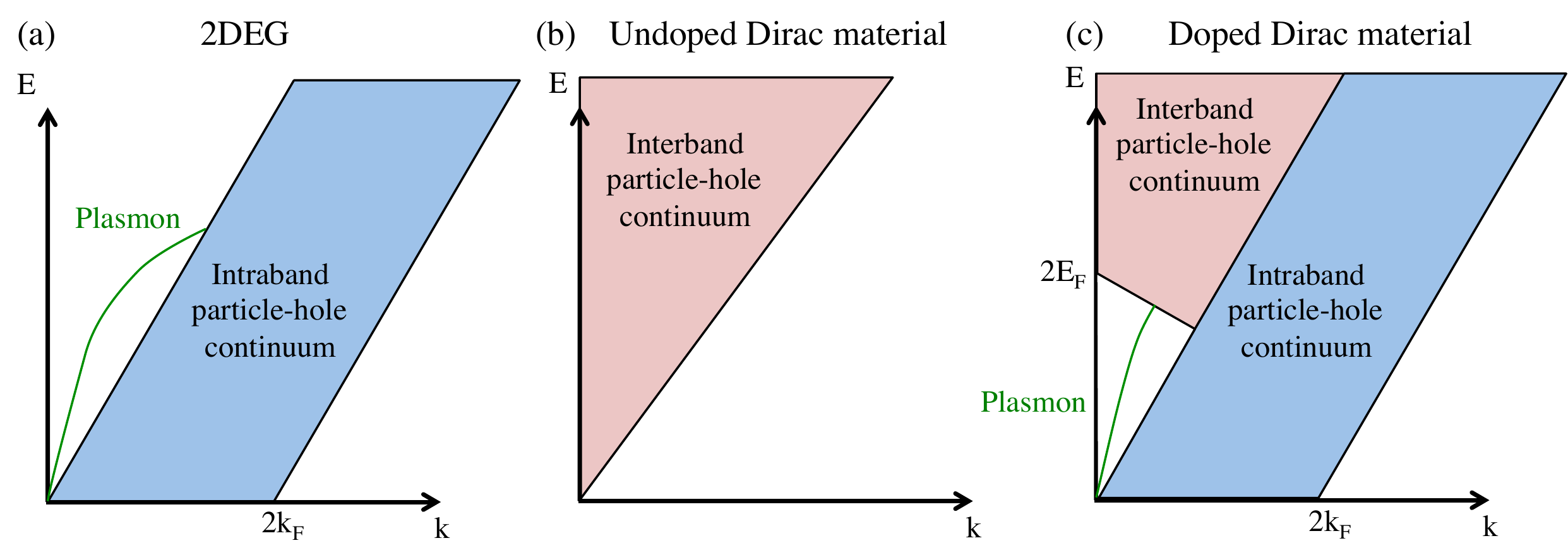}
\end{center}
\caption{\label{fig:eh_excit} Energy as function of momentum plot for the simplest electron-hole excitations in a (a) 2DEG, (b) two-dimensional undoped Dirac material, i.e.~with Fermi level at the Dirac point, and (c) doped Dirac material with the intraband (blue) and interband (red) electron-hole continuum and the collective plasmon mode (green line). 
}
\end{figure}

Before proceeding to look in detail on several specific electronic interaction effects in Dirac materials, it is informative to take a very elementarily viewpoint and briefly look at (semi)metallic systems in terms of their simplest excitations. The main excitations present in metallic systems are electron-hole excitations and the associated collective modes, such as plasmons, which are the quanta of the plasma oscillations in an electron gas.
Electron-hole pairs are excitations of the Fermi sea where an electron with momentum $\bk$ is excited above the Fermi level to a new state with momentum $\bk+\bq$, leaving a hole behind in the Fermi sea. For states close to the Fermi level the energy of this excitation is $\omega = \varepsilon_{\bk +\bq} - \varepsilon_{\bk} \approx v_F q$, with $q = |\bq|$. For a normal metal or doped semiconductor with a parabolic dispersion, the low-energy electron-hole excitations are always intraband transitions. These transitions exist down to zero energy, since it is always possible to produce electron-hole pairs with arbitrarily low energies. This is illustrated by the electron-hole continuum occupying the full energy region for momenta less than $2 k_F$ (the maximum momentum difference) in Figure \ref{fig:eh_excit}(a). For a two-dimensional electron gas (2DEG) it is also possible to have a collective plasmon excitation with energy $\omega \propto \sqrt{q}$ \cite{Shung86}. Since it is located outside the electron-hole continuum for small $q$ it will in principle have an infinite life-time. Now, compare this with a two-dimensional Dirac material with the Fermi level at the Dirac point. It is then not possible to have intraband electron-hole excitations, but any excitation needs to be an interband transition. Thus, there is a finite energy cost for producing an electron-hole pair at finite momentum transfer and the electron-hole continuum instead occupies the upper part of the energy vs.~momentum diagram, as shown in Figure \ref{fig:eh_excit}(b). This also means there can be no long-lived plasmon or other collective modes in undoped Dirac materials, since their energies fall within the electron-hole continuum making them significantly damped. For a doped Dirac material, intraband electron-hole transitions are again permitted, leading to a mixture of normal metal and (undoped) Dirac material properties, as shown in Figure \ref{fig:eh_excit}(c). For example, collective excitations are now again sustainable at long wavelengths. Similar qualitative behaviors can also be found for three-dimensional metals and Dirac materials, respectively. 

\subsection{Electric screening}
The unusual electron-hole excitation spectrum is related to the lack of screening in undoped Dirac materials. The polarization and thus the dielectric function are notably different in a two-dimensional Dirac material compared to a 2DEG. Within the random-phase approximation (RPA) the polarization function can be calculated analytically in undoped graphene (see for example Refs.~\cite{Shung86, Gonzalez94, Hwang07}):
\begin{align}
\label{eq:polarization}
\Pi(q,\omega) = \frac{q^2}{4\sqrt{v_F^2 q^2 - \omega^2}}.
\end{align}
For $\omega > v_F q$, the polarization function is imaginary indicating damping, which coincides with the electron-hole pair excitation region. Furthermore, the static polarization at $\omega = 0$ vanishes linearly with $q$. This means that there is no complete (long-range) electric screening in a Dirac material with the Fermi level at the Dirac point. For finite doping there is a crossover to normal metallic behavior.

The non-existent screening can also be understood from a renormalization group argument for the long-range Coulomb interaction $e^2/(4\pi \epsilon r)$, where $\epsilon$ is the dielectric constant.
If we integrate out all high-energy electronic modes outside a certain interval around the Dirac point, where the Fermi level is located, then no gapless modes have been integrated out. Thus, no non-analytic terms can be generated during this integration. Now, the original electron-electron interaction has a $1/r$ distance dependence and thus has a non-analytic form in momentum space. But since no non-analytic terms were generated during the renormalization group procedure, the long-distance behavior of the Coulomb interaction is unchanged and there is thus no screening of the long-distance electron repulsion in the system \cite{YeSachdev98, Herbut06}.  Formulated slightly differently this means that there is no charge ($e$) renormalization in two-dimensional Dirac materials.

For a two-dimensional material there is also the additional issue of the nature of the dielectric constant, since a two-dimensional  material is always embedded in a three-dimensional medium. Consider for example graphene embedded in a three-dimensional medium with a dielectric constant $\epsilon_d$. At long distances the background dielectric constant applicable in the graphene sheet is, in fact, independent on the screening within the graphene sheet itself. This can be seen by solving a simple electrostatic problem for a point charge inserted in the middle of a thin dielectric slab in a three-dimensional medium with dielectric constant $\epsilon_d$. At distances large compared to the slab thickness the screening is entirely determined by $\epsilon_d$ (see e.g.~Ref.~\cite{Emelyanenko08}). Thus the dielectric constant for two-dimensional materials is set by their three-dimensional host. 
Such substrate screening effects have been measured experimentally in graphene \cite{Jang08, Ponomarenko09}. There are also measurements on graphene within a graphite host suggesting that graphene might have an intrinsic dielectric function amplified by excitonic effects \cite{Reed10}.

Above we established that the long-range $1/r$ tail of the Coulomb interaction prevails in a two-dimensional Dirac material. 
A related question is what happens to short-range electron interactions, or contact interactions? This question can be answered in an renormalization group sense by a simple dimensional analysis. In two dimensions the Dirac fields scale as inverse length. Then any short distance (contact) coupling constant $g$, which multiplies four Dirac fields, must have the dimension of length. In a perturbative expansion $g$ must therefore be multiplied by a power of length, but in a critical theory only the thermal length or the wavelength of the external perturbations is available. Since both of these length scales become very long, the perturbative expansion terms in $g$ will become small. Short-range electron-electron interactions are thus perturbatively irrelevant at low energies around the non-interacting fix point \cite{Gonzalez01,Herbut06}.

\subsection{Velocity renormalization}
\begin{figure}[htb]
\begin{center}
\includegraphics[width=0.7\columnwidth]{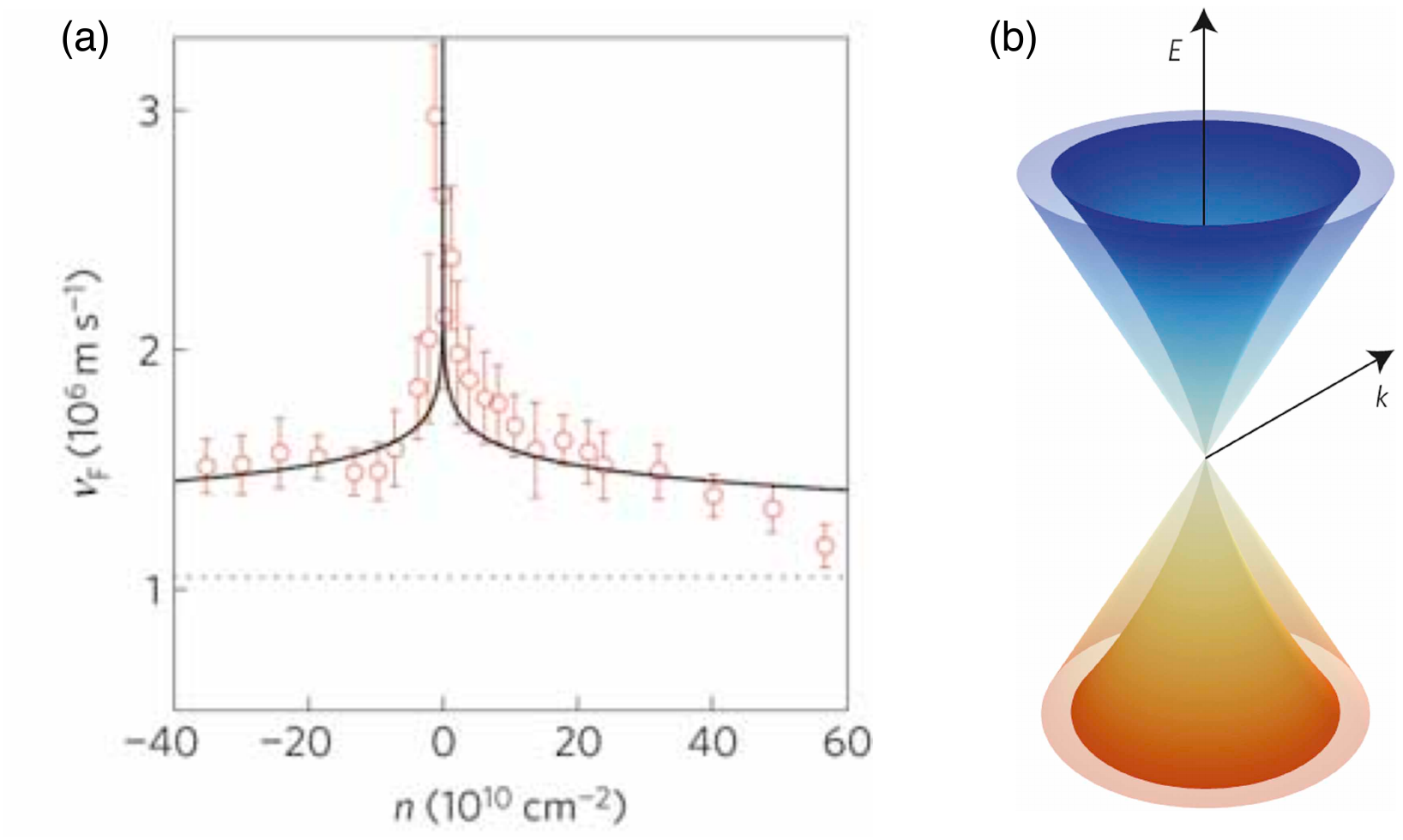}
\end{center}
\caption{\label{fig:vel_renorm} (a) Fermi velocity $v_F$ dependence on the carrier density $n$ in suspended graphene measured by extracting the cyclotron mass from Shubnikov-de Haas oscillations. Solid line is a fit to renormalization group results within RPA, which displays the logarithmic enhancement of $v_F$ close to the Dirac point located at $n = 0$.
(b) Sketch of the electronic spectrum of graphene without (outer transparent cone) and with electron-electron interactions (inner cone). The outer cone is the single particle spectrum $E= \hbar v_F k$, the inner cone illustrates the renormalization group flow of the logarithmically divergent $v_F$ as the quasiparticle energy $E$ approaches zero.
(Adapted by permission from Macmillan Publishers Ltd: Nature Physics, D.~C.~Elias {\it et al.}, Nature Phys.~7, 701 (2011) \cite{Elias11}, copyright \copyright~(2011).)
}
\end{figure}

In the view of the lack of screening, the question is what are the consequences of the Coulomb repulsion? We just reviewed that short-range interactions are irrelevant in the weak coupling regime, so left is only the long-range Coulomb tail. 
For this, the ratio of the Coulomb repulsion to the kinetic energy prefactors, i.e.~the fine-structure constant $\alpha = e^2/(4\pi \epsilon \hbar v_F)$ is important. 
A dimensional analysis shows that $\alpha$ is dimensionless, and it is thus a marginally relevant perturbation. However, despite the similarity with the $3+1$D QED fine-structure constant, there are notable differences. As we have seen above, the charge is not renormalized in Dirac materials, due to the instantaneous nature of the Coulomb repulsion. Any renormalization group flow must therefore originate in the Fermi velocity $v_F$. Since Lorentz invariance is broken by the instantaneous Coulomb repulsion there is no formal problem with a renormalized $v_F$, compared to in QED, where the speed of light $c$ is unchanged. Perturbative calculations in the weak-coupling limit have shown that $v_F$ grows logarithmically towards infinity during a renormalization group flow to lower energies \cite{Gonzalez94}, thus shrinking the fine structure constant.
Also the tilted and anisotropic Dirac spectrum in the high-pressure phase of the organic semiconductor $\alpha$-(BEDT-TTF)$_2$I$_3$ has been shown to have a logarithmically increasing velocity with decreasing energies \cite{Isobe12}. Since titling parameter at the same time was found to not be renormalized, velocity renormalization is the only consequence of (weak) long-range Coulomb repulsion in a two-dimensional Dirac material.
However, $v_F$ can obviously never exceed the speed of light. This limit is reinstated if one properly includes a retarded form of the Coulomb interaction via the exchange of a photon, then $v_F$ will saturate at $c$  \cite{Gonzalez94}. 
In practice this is not a relevant concern, since $v_F$ only grows logarithmically and therefore the physics in a real Dirac material is realistically given in a regime where $v_F$ is increased compared to its bare value, but never really close to $c$. Evidence for a logarithmic velocity increase at low energies has recently been reported by measuring the cyclotron mass in suspended graphene \cite{Elias11}, as seen in Figure \ref{fig:vel_renorm}.

The increased $v_F$ due to Coulomb interactions means that the excitations close to the Dirac point have a larger velocity, but also that there will be fewer of them, i.e.~the density of states is reduced close to the Dirac point, as also clearly seen in Figure \ref{fig:vel_renorm}(b). The latter is in line with an intuitive understanding of electron-electron interactions, which usually reduce the density of states or even open a gap in the excitation spectrum. Physical consequences of the reduced density of states include, among other effects \cite{Sheehy07}, a suppression of the specific heat below its non-interacting value \cite{Vafek07PRL}. Interestingly, the increased particle velocities compensate for the reduced density of states in the conductivity, which is not suppressed at low frequencies \cite{Herbut08}. 

Going beyond lowest order perturbation theory, by for example preforming an RPA treatment for many electron flavors (valley + spin in the case of graphene), also leads to a renormalization of the quasiparticle weight $Z$ \cite{Gonzalez99, Son07,Foster08, Kotov12} . If the initial value of $\alpha$ is large, then $Z$ decreases fairly fast. But eventually, when $\alpha$ has decreased sufficiently, $Z$ will stop decreasing and level off. This means that the quasiparticle weight is not fully lost and the system continues to have well-defined quasiparticles even at the Dirac point. 

\subsection{Mass gap}
Very generally speaking, the reduced density of states at and near the Dirac point in Dirac materials make electron-electron effects less effective in causing symmetry breaking and generation of an energy gap in Dirac materials compared to normal metals. For example, it is much harder to reach the density of states needed to satisfy the Stoner criterion for magnetism within the Hartree-Fock approximation or for a measurable BCS superconducting transition temperature in a Dirac material than in a normal metal. Clearly, the velocity renormalization from long-range Coulomb interactions discussed in the previously, does not help in this regard, nor does the fact that short-range interactions are perturbatively irrelevant \cite{Gonzalez01,Herbut06}.
This is notably different from systems with higher order band crossing points, which give rise to a finite density of states even for crossings at the the Fermi level. For example, short-range interactions have been shown to be marginally relevant in the renormalization group sense for quadratic band crossings \cite{Sun09}. This explains the remarkable difference between graphene and bilayer graphene. In bilayer graphene the simplest interlayer coupling gives rise to a quadratic band crossing and multiple recent experiments have reported interaction-driven gapped states in suspended and undoped graphene bilayers \cite{Feldman09, Weitz10, Mayorov11, Bao12, Velasco12}.
For Dirac materials the question, however, still remains if it is possible to cause a quantum phase transition into a symmetry broken state with a finite mass for (relatively) strong electron-electron interactions? Unfortunately, going beyond the weak-coupling regime makes theory hard to control and it is difficult to reliably determine the full phase diagram. In fact, the phase diagram is dependent on the details of the electron-electron interactions. 

\subsubsection{Excitonic state}
One much discussed possibility for a spontaneously generated mass in two-dimensional Dirac materials is an excitonic state generated by the long-range Coulomb potential. The study of this phenomenon dates back to the discussion of chiral symmetry breaking in QED in 2+1 dimensions \cite{Pisarski84, Appelquist86}, and is still continuing, see e.g.~references in \cite{Kotov12}. However, due to both the instantaneous and three-dimensional nature of the Coulomb repulsion in a two-dimensional material, the results are somewhat different in a Dirac materials compared to in pure 2+1 dimensional QED.

A gap equation for the excitonic state has been obtained from a self-consistent solution for the self-energy within the RPA, being exact for large number of fermion flavors $N$ \cite{Khevshchenko01, Gorbar02, Khveshchenko04, Khveshchenko09, Liu_Li09, Gamayun10}. It turns out that the gap is strongly momentum dependent, due to the long-range tail of the Coulomb interaction, and it reaches a maximum value at small momenta. The instability has also been shown to be enhanced by the application of a perpendicular magnetic field, due to the formation of Landau levels \cite{Khveshchenko01b, Gorbar02}.

In graphene, where $N = 4$ (spin $+$ valley degeneracy), some calculated values of the critical fine-structure constant for generating a finite mass is $\alpha_c = 0.92$ \cite{Gamayun10} and $\alpha_c = 1.13$ \cite{Khveshchenko09}. 
Numerical Monte Carlo calculations of the corresponding lattice field theory have also recently been preformed \cite{Drut09, Drut09b, Drut09c,Brower12,Ulybyshev13}. These calculations have not only been able to strongly support the actual existence of an excitonic state but have also estimated $\alpha_c = 1.1$  in graphene \cite{Drut09b}.
These values of $\alpha_c$ in graphene should be compared to $\alpha = 2.2$ for graphene in vacuum. However, the presence of a substrate in general always significantly lowers $\alpha$. For example, graphene on SiO$_2$ generates $\alpha = 0.79$, which is below the critical coupling and indicates that only suspended graphene can host an excitonic gap, if it even exists. 
Very recent Monte Carlo calculations, also including the screening from the $\sigma$-electrons in the graphene sheet \cite{Wehling_Coulomb_PRL11}, have found that freestanding graphene does not exhibit chiral symmetry breaking, but the transition point is quite close if the interaction strength can only be slightly increased \cite{Ulybyshev13}. 
These theoretical results should also be compared with experimental results on graphene. For example, $\alpha \approx 0.14$ has been estimated by imaging the dynamical screening of charge in a freestanding graphene sheet within a graphite host, i.e.~well below the critical value for an excitonic insulator state \cite{Reed10}. So far, there is no experimental evidence for a gap in freestanding monolayer graphene, even down to 1~K \cite{Elias11, Mayorov12}.

The physical structure of the excitonic state in graphene has been shown to come in several varieties depending on the details of the Coulomb repulsion and the nature of the excitonic pairing between Dirac cone valleys. Proposals range from charge density wave states modulating the electronic density between the two sublattices \cite{Khevshchenko01} to Kekule dimerization, where the unit cell is tripled \cite{Hou_Chamon07}. If strong short-range repulsion is also included, there is in addition a strong tendency for time-reversal symmetry breaking and formation of an antiferromagnetic state \cite{Herbut06, Herbut09}.

An excitonic gapped phase has recently also been studied for the three-dimensional inversion-symmetric Weyl semimetals \cite{Wei13Weyl}. For interactions exceeding a critical strength, a number of excitonic phases can be found in a Weyl semimetal, most of which are gapless. However, one of these, a chiral intranodal excitonic state with the order parameter having different signs for two Weyl points, has been found to be gapped \cite{Wei13Weyl}. This mass term is allowed since it is an effective hybridization between the two Weyl points.   

\subsubsection{Short-range repulsion}
The possibility that (strong) short-range repulsion can generate a mass gap in the Dirac spectrum has also been frequently discussed. Most work has concentrated on the Hubbard model, which only includes a (strong) on-site repulsion term, on the half-filled honeycomb lattice, with extensions also including repulsion between nearest neighbor sites. This would model undoped graphene in the limit where all long-range Coulomb repulsion is sufficiently screened to be unimportant. Indeed, the $1/r$ Coulomb repulsion tail has been found to be marginally irrelevant in a renormalization group sense at the interacting fixed point for short-range repulsion \cite{Herbut06,Herbut09}. This together with strong screening by any substrate provides motivation for only studying the short-range part of the electronic interactions in the strong coupling limit.

The Hubbard model on the half-filled honeycomb lattice does not suffer from the fermion sign problem and thus quantum Monte Carlo methods provide an accurate numerical solution. 
Early Monte Carlo studies found a Mott insulating state at $U/t \gtrsim 5$, where $U$ is the on-site repulsion and $t\approx 2.7$~eV is the nearest neighbor hopping in graphene, such that the band width is $W = 6t$ \cite{Sorella92,Paiva05}. This should be compared to the Hartree-Fock mean-field result $U/t = 2.23$, which indicates that quantum fluctuations are very important. 
More recent Monte Carlo results have found a gapped Mott antiferromagnetic state for $U/t > 4.3$, with a possible spin-liquid state in the interval $3.5 < U/t < 4.3$, formed by short-range resonance valence bonds \cite{Meng10}. These results were obtained on clusters containing up to 648 sites. However, later results, using clusters as big as 2592 sites, found only a direct transition between the Dirac semi-metal state and the antiferromagnetic insulator \cite{Sorella12}, thus disfavoring the formation of a spin-liquid state in the transition region.

\begin{figure}[htb]
\begin{center}
\includegraphics[width=0.7\columnwidth]{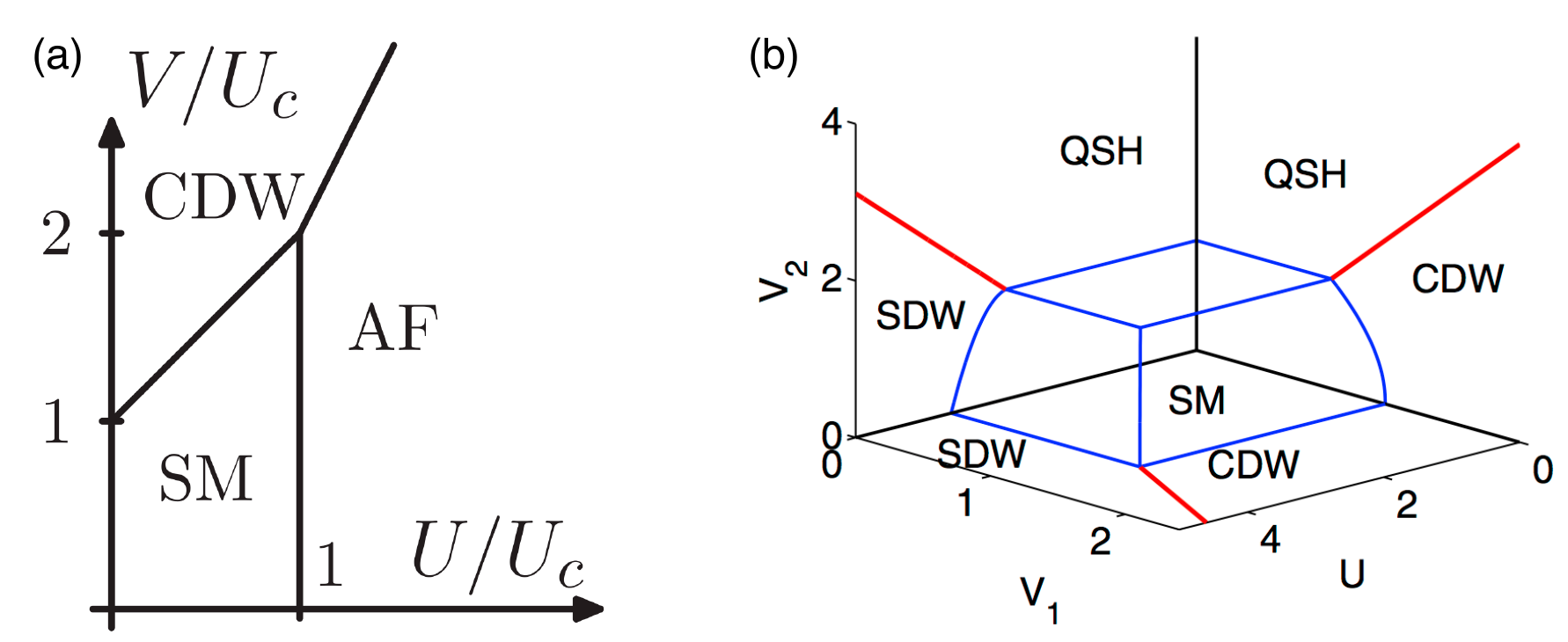}
\end{center}
\caption{\label{fig:SM_AF_CDW} Dirac semimetal to insulator transitions predicted by renormalization group calculations of the extended Hubbard model on the half-filled honeycomb lattice, modeling undoped graphene.
(a) Phase diagram from renormalization group large-$N$ expansion calculations for on-site repulsion $U$ and nearest neighbor repulsion $V$. The semimetal (SM) state is found below the critical couplings $U_c$ and $V_c$, where antiferromagnetic (AF) or a charge density wave (CDW) insulating states are reached, respectively. 
(Reprinted figure permission from I.~F.~Herbut, Phys.~Rev.~Lett.~97, 146401 (2006) \cite{Herbut06}. \href{http://link.aps.org/abstract/PRL/v97/p146401}{Copyright \copyright~(2006) by the American Physical Society.})
(b) Phase diagram from mean-field and functional renormalization group calculations for on-site $U$, nearest $V_1$, and next-nearest $V_2$ neighbor repulsion. The spin-density wave (SDW) phase is the antiferromagnetic insulator state, whereas a topological quantum spin Hall (QSH) state is now also possible for strong next-nearest neighbor repulsion. 
(Reprinted by permission from S.~Raghu {\it et al.}, Phys.~Rev.~Lett.~100, 156401 (2008) \cite{Raghu08}. \href{http://link.aps.org/abstract/PRL/v100/p156401}{Copyright \copyright~(2008) by the American Physical Society.})
}
\end{figure}

The antiferromagnetic state found for a strong Hubbard-$U$ term breaks the sublattice (chiral) symmetry by assigning opposite magnetic moments on the two sublattices. 
In the presence of longer-range repulsion, this state competes with the excitonic charge density wave instabilities \cite{Khevshchenko01, Gorbar02, Khveshchenko04, Khveshchenko09, Liu_Li09, Gamayun10} discussed in above, which also break the chiral symmetry. This competition has been studied in some detail by including a nearest neighbor repulsion $V$ together with the on-site $U$ term into an extended Hubbard model. Renormalization group results have found both antiferromagnetic and charge density wave fixed points, depending on the relative strength of $U$ and $V$ for this model \cite{Herbut06,Herbut09}, as illustrated in Figure~\ref{fig:SM_AF_CDW}(a). These fixed points are unstable relative to the non-interacting fixed point at weak coupling but have runaway directions for strong coupling, giving an ordered phase at strong interactions. 
Additional mean-field and functional renormalization group results on the half-filled honeycomb lattice supports the existence of these ordered phases \cite{Honerkamp08,Raghu08}, but also proposes the existence of a quantum spin Hall state if the next-next nearest neighbor repulsion is sufficiently strong \cite{Raghu08}, as displayed in Figure \ref{fig:SM_AF_CDW}(b). The topologically non-trivial quantum spin Hall state is an insulating state, which breaks the $SU(2)$ symmetry associated with choosing the spin projection axis but preserves time-reversal symmetry. First principles calculations find sizable local and non-local Coulomb interactions in graphene ($U=3.3t$, $V=2.0t$). Further analysis shows that the Dirac fermion phase of graphene is stabilized against antiferromagnetic strong coupling phases by the non-local interaction terms \cite{Wehling_Coulomb_PRL13}.

\subsection{Superconductivity}
An energy gap can also be generated in the Dirac spectrum by a superconducting order parameter. According to BCS theory the superconducting critical temperature $T_c \propto \exp[-1/[N(0)g)]$, where $N(0)$ is the density of states at the Fermi level and $g$ is the attractive pair potential. Dirac materials with their low, or for undoped materials even vanishing, density of states at the Fermi level are thus quite resilient against forming an intrinsic superconducting state. In undoped Dirac materials there is even a quantum critical point, where superconductivity only appears above a certain critical attractive coupling \cite{CastroNeto01, Kopnin08, Zhao06}, which is in stark contrast to the Cooper instability in a normal Fermi liquid. 
For a two-dimensional Dirac material the quantum critical point for $s$-wave superconductivity is at $g_c = \pi v_F/D$, where $D$ is an ultraviolet cutoff \cite{CastroNeto01, Uchoa07} and the superconducting gap grows with the coupling $g>g_c$ as $\Delta = D(1 - g_c/g)$ \cite{CastroNeto01}.
Away from the Dirac point, when the density of states at the Fermi level is finite, there is no longer a quantum critical point and the gap dependence on the coupling crosses over to the weak-coupling BCS result with increasing doping levels \cite{Uchoa05,Kopnin08}. 
Interestingly, adding disorder to a two-dimensional Dirac material with an attractive interaction has been found to destroy the semimetal state and instead favor superconductivity \cite{Nandkishore13}. Thus near the Dirac point disorder seems to possibly even assist superconductivity.
Of course, if the system is strictly two-dimensional these BCS mean-field results only give the temperature where the Cooper pair amplitude disappears. Phase coherence on the other hand is reached at the lower Kosterlitz-Thouless temperature. This is due to the increasing importance of thermal fluctuations in reduced dimensions. However, for non-suspended graphene or the surface state in a topological insulator there should be no such dimensional concerns.

Based on the above, superconductivity is thus perhaps not the first property that comes to mind when one thinks about Dirac materials, apart of course from the $d$-wave superconductors, which are Dirac materials just because of superconductivity. Nevertheless, superconducting Dirac materials have and are continuing to render a lot of attention and we will here review both results on proximity-induced superconductivity and searches for intrinsic superconducting states in several different Dirac materials. We will also discuss superconducting vortex states in Dirac materials, which can host unusual zero-energy states.

\subsubsection{Proximity-induced superconductivity}
Independent on the strength of possible intrinsic attractive couplings, it is in general possible to generate a superconducting state in a material by proximity to an external superconductor. Cooper pairs and superconductivity can in this way be proximity-induced into a Dirac material. 

Proximity-induced superconductivity was first demonstrated in graphene in a Josephson junction consisting of two superconducting Ti/Al bilayer contacts on top of a graphene sheet \cite{Heersche07}. Dirac fermions have experimentally been shown to give rise to ballistic transport even on the micrometer scale, which means that graphene can sustain supercurrents even in long Josephson junctions \cite{Heersche07,Miao07,Shailos07,Du08}. 
The Josephson current has been calculated in this types of junctions using both continuum and lattice models \cite{Titov06,Moghaddam06, Maiti07, Black-Schaffer08}. Notably, a significant skewing of the current-phase relation has been found both theoretically \cite{Black-Schaffer08, Hagymasi10,Black-Schaffer10Tdep} and experimentally \cite{Chialvo10,English13}, and attributed to both inverse proximity effect in the superconducting regions and current depairing \cite{Black-Schaffer10Tdep}.
 
\begin{figure}[htb]
\begin{center}
\includegraphics[width=0.5\columnwidth]{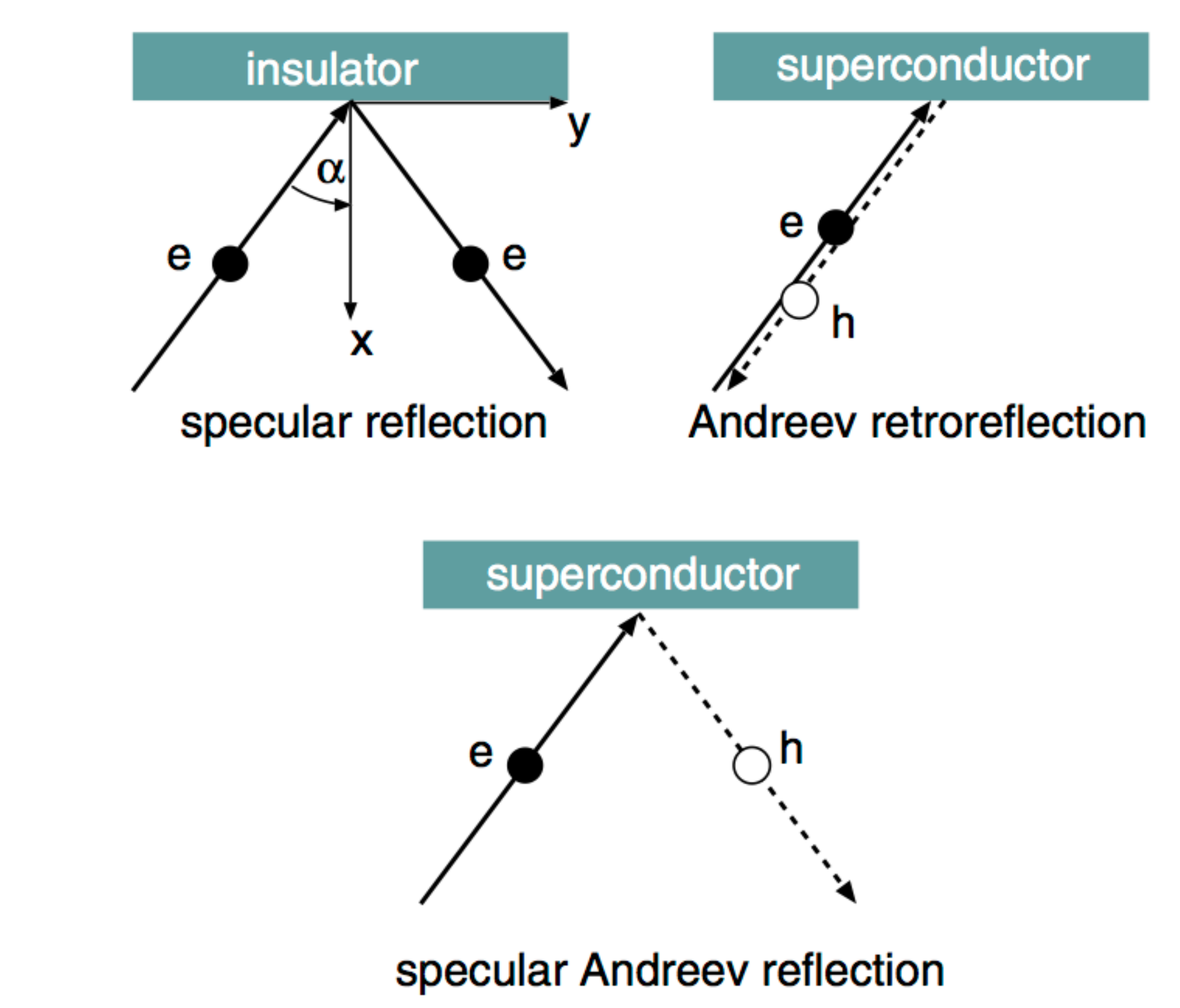}
\end{center}
\caption{\label{fig:Andreev}  Reflection processes in which an electron (solid lines) undergoes specular reflection at a metal-insulator interface (top left) or is converted to a hole (dashed lines) in an Andreev process. The hole is retroreflected for a normal metal-superconductor interface (top right), while there is instead a specular Andreev reflection process for an undoped Dirac material-superconductor interface (bottom).
(Reprinted figure by permission from C.~W.~J.~Beenakker, Phys.~Rev.~Lett.~97, 067007 (2006) \cite{Beenakker06}. \href{http://link.aps.org/abstract/PRL/v97/p067007}{Copyright \copyright~(2006) by the American Physical Society.})
}
\end{figure}

Another very interesting phenomenon due to the Dirac spectrum in a superconductor-normal state system is specular Andreev reflection. The Andreev scattering process converts an incoming electron from the normal side into an outgoing hole on the normal side, while at the same time creating a Cooper pair on the superconducting side of the interface. For a normal metal N region, the hole is retroreflected, but in Dirac materials there is instead specular reflection \cite{Beenakker06}, as schematically shown in Figure \ref{fig:Andreev}. This is because an electron in the conduction band (above the Dirac point) is converted into a hole in the valence band (below the Dirac point), if the Fermi level is located at the Dirac point. This property has also been shown to give rise to Andreev modes propagating along graphene-superconductor interfaces \cite{Titov07}. 
This specular Andreev reflection process, first established in graphene \cite{Beenakker06}, is naturally also present in the Dirac surface state of a three-dimensional topological insulator \cite{Linder10PRB}.
Very recently, specular Andreev reflection was also established at a Weyl semimetal-superconductor interface \cite{Chen13Weyl}. A review on Andreev processes and Josephson junctions in graphene is available in Ref.~\cite{Beenakker08RMP}.

In topological insulators superconductivity has also recently been proximity-induced from both conventional \cite{Sacepe11, Veldhorst12, Wang12SC,Yang12SC} and high-temperature $d$-wave superconductors \cite{Zareapour12, Wang13TISC} Most experiments have so far been focusing on transport properties such as Josephson supercurrents, including measurement of the Josephson supercurrent in the insulating bulk regime, relevant for Dirac material properties \cite{Cho13TISC}.
Theoretically, an induced $s$-wave order parameter added to the topological insulator surface state has been shown to behave at low energies like an effectively spinless $p+ip$-wave state \cite{Fu08TISC}. This leads to zero-energy states, found for example at superconductor-ferromagnet interfaces or in vortex cores, being Majorana modes \cite{Read00,Fu08TISC}. Majorana modes will be discussed in some detail below.
Explicit calculations have verified that an $s$-wave superconducting state induces a spin-triplet $p$-wave amplitude on the surface of the topological insulator, due to the spin-momentum locking in the Dirac cone \cite{Stanescu10,Black-Schaffer11QSHI, Lababidi11}. A full classification of all induced superconducting amplitudes, depending on the symmetry of the external superconductor, has also been preformed \cite{Black-Schaffer13Bi2Se3SC}. Interestingly, for external spin-triplet $p$-wave superconductors, induced spin-singlet $s$- or $d$-wave amplitudes dramatically affects the gap structure of the surface state.

There also exist some theoretical proposals for proximity-induced superconductivity in the three-dimensional Weyl semimetals \cite{Meng12Weyl, Das13Weyl}. For example, superlattices of alternating conventional superconductors and topological insulators can create a Weyl superconductor \cite{Meng12Weyl}. Superconductivity here splits each Weyl node in two and the resulting Bogoliubov-Weyl nodes can be pairwise independently controlled, allowing for a number of different phases \cite{Meng12Weyl}.

\subsubsection{Intrinsic superconductivity}
The low density of states at the Fermi level for undoped and lightly doped Dirac materials make it harder to achieve intrinsic superconductivity in these materials, at least from a weak-coupling perspective. In addition, the superconducting mechanism and the symmetry of the superconducting order parameter are most often very material specific, or at least dependent on crystal structure and the nature of the Dirac fermion excitations. There are therefore less apparent common characteristics for intrinsic superconductivity in Dirac materials. Nevertheless, we will here review some existing works on several different Dirac materials.

Of the known Dirac materials, intrinsic superconductivity has so far been most studied in graphene. The closely related intercalated graphites, such as CaC$_6$ and KC$_8$, where alkali metals are positioned in-between single layer graphene planes, have been known low-temperature conventional $s$-wave superconductors for a long time \cite{Hannay65}. They have recently also received renewed attention, trying to understand the details of the phonon-driven mechanism \cite{Csanyi05, Weller05, Calandra05, Emery05}. Phonon-mediated superconductivity has also been shown to exist in the thinnest limits of these intercalated graphite, i.e.~that of intercalated bi-layer graphene \cite{Mazin10, Kanetani12}, and it has also been proposed to exist when alkali atoms are deposited directly on a single layer of graphene \cite{Profeta12}. In the latter case, a plasmon-mediated mechanism has also been proposed \cite{Uchoa07}. However, note that alkali atoms produce a large $n$-type doping of the Dirac cone, and they superconducting gap does not appear at the Dirac point.

In contrast to the $s$-wave superconducting state caused by phonons in graphene-like materials, it was recently suggested that phonon-induced superconductivity in a three-dimensional inversion-symmetric Weyl semimetal instead favors a finite-momentum, or so-called Fulde-Ferrell-Larkin-Ovchinnikov (FFLO), pairing state \cite{FuldeFerrell64,Larkin65} over the conventional $s$-wave BCS state \cite{Cho12Weyl}. The center of momentum for the FFLO pairs is fixed by the momentum of the Weyl nodes.

Superconductivity due to strong electron correlations has also been investigated, especially in graphene, using either a (extended) Hubbard model or the $t$-$J$ model. These models have been shown to give rise to a spin-singlet $d$-wave superconducting state, in much the same manner as for the high-temperature cuprate superconductors. However, the sixfold symmetry of the honeycomb lattice forces the two $d$-wave solutions, $d_{x^2-y^2}$ and $d_{xy}$, to become degenerate, and therefore the time-reversal symmetry breaking $d_{x^2-y^2}+id_{xy}$-wave (d+id'-wave) symmetry combination has been shown to emerge \cite{Black-Schaffer07, Honerkamp08, Pathak10, Ma11, Wei13Weyl, Gu13}. The $d+id'$-wave symmetry notation refers to the symmetry of the order parameter in the full Brillouin zone and when the kinetic energy is diagonal, i.e.~in the band structure basis. Projected down to the individual valleys the $d_{x^2-y^2}+id_{xy}$ state has $p_y+ip_x$ symmetry in both valleys \cite{Black-Schaffer07,Linder09b}. The order parameter thus only has nodes at the $K,K'$ points and is thus a fully gapped state at finite doping. The $d+id'$ state is a topologically non-trivial state with two chiral edge states carrying a finite, but not quantized, spontaneous edge current \cite{Black-Schaffer12PRL}.
The $d+id'$ symmetry of the superconducting state is a direct consequence of the sixfold lattice symmetry in graphene. The order parameter symmetry thus persists even when the doping level is significantly increased towards the van Hove singularity, where the two Dirac cones in graphene merge to produce a single Fermi surface centered around the $\Gamma$ point \cite{Nandkishore11, Wang11, Kiesel12}. Therefore the $d+id'$ symmetry is not a consequence of the Dirac physics in graphene, but of the lattice symmetry. In fact, a large doping is in general required to avoid the quantum critical point present in undoped graphene. Interestingly, many topological insulators also have a hexagonal lattice structure, thus making the same time-reversal symmetry breaking $d+id'$-wave state favored in topological insulators for superconductivity stemming from strong electron-electron interactions \cite{Black-Schaffer13Bi2Se3SC}. 

Beyond spin-singet superconductivity, some spin-triplet superconducting states have also been discussed for graphene \cite{Bergman09}. For example, a spin-triplet $p$-wave Kekule state, which breaks translation symmetry, has been proposed for attractive nearest neighbor interaction \cite{Roy10}, whereas an $f$-wave state might be favored in vertical graphene heterostructures with graphene layers separated by BN spacing layers \cite{Guinea12}.

Intrinsic superconductivity has experimentally also been found in several doped topological insulators. The currently most prominent example is Cu-doped Bi$_2$Se$_3$, where Cu is intercalated between the Se layers. Cu$_x$Bi$_2$Se$_3$ is a low-temperature superconductor with a transition temperature of $2.2-3.8$~K for $0.1< x < 0.6$ \cite{Hor10}. It stirred a lot of attention when it was proposed to be a topological superconductor with an unconventional odd-parity order parameter and topologically protected surface states \cite{Fu&Berg10}, but, so far, experimental results are seemingly contradictory with regards to the symmetry of the order parameter \cite{Sasaki11, Kriener11, Levy13}. The Cu intercalation causes the Fermi level to be well within the bulk conduction band, and thus Cu$_x$Bi$_2$Se$_3$ is not a bulk insulator in the normal state. APRES measurement have, however, shown that the surface state is still preserved although the Fermi level is rather far from the Dirac point \cite{Wray10}, so superconductivity does not gap out the Dirac point.

Sn$_{1-x}$In$_x$Te is another doped topological insulator which is a low-temperature superconductor with possible topological surface states \cite{Sasaki12}. In this case the parent structure SnTe is a topological crystalline insulator, where crystal symmetries protect the Dirac surface states. Recent ARPES measurements have found both a small bulk hole-like Fermi pocket, as well as a topological surface state in the normal state, suggesting the possibility of topological superconductivity \cite{Sato13SnInTe}.

Intrinsic low-temperature superconductivity has also been found in the undoped topological insulators Bi$_2$Se$_3$ \cite{Kirshenbaum13} and Bi$_2$Te$_3$ \cite{Zhang12Bi2Te3} with the application of high pressure. In both cases superconductivity has been attributed to a metallization of the bulk due to pressure, but with the surface state still possibly surviving \cite{Zhang12Bi2Te3}.

\subsubsection{Supeconducting vortices}
In a superconducting vortex the phase of the superconducting order parameter winds around a core, such that $\Delta = |\Delta(\rho)| e^{in\phi}$, where $(\rho,\phi)$ are cylindrical coordinates and $n$ is the so-called vorticity, which is an integer number. Vortices are formed in the superconducting state in response to an external magnetic field, and they allow the magnetic field lines to penetrate the superconductor at the vortex cores, where the superconducting order parameter necessarily goes to zero.

A vortex in even the simplest (spin-singlet $s$-wave) superconducting state in a two-dimensional undoped Dirac material has been shown to have very intriguing properties, as it allow zero-energy states in the core of the vortex \cite{Jackiw81}. These zero-energy states are protected by an index theorem stating that there are $n$ zero-energy states in the vortex for every Dirac cone \cite{Weinberg81}. At finite doping, it has been shown that one zero-energy state is still present for odd vorticity (odd number $n$), whereas for even vorticity there are only non-zero energy states \cite{Bergman09,Khaymovich09}. 

In graphene there is a fourfold Dirac cone degeneracy, due to the valley and spin degrees of freedom, and there should thus be four zero-energy modes per single ($n=\pm 1$) vortex core. However, the index theorem only applies to a slightly modified Dirac Hamiltonian compared to that in real graphene. For example, including the Zeeman spin-splitting term, which is always present in a magnetic field in a real material, shifts the zero-energy states to a small but finite energy \cite{Ghaemi12}. This effect is quantitatively small, since the Zeeman term is in general small, and thus still leaves near zero-energy states. Moreover, the index theorem works only in a continuum theory, while the honeycomb lattice of graphene has been shown to split the zero energy states to finite energies \cite{Bergman09}. 

In topological insulators there is only one single Dirac cone on the surface, and there will thus be one single zero-energy state in the vortex core. A zero-energy Bogoliubov quasiparticle state has equal hole and electron components as well as equal spin-up and spin-down character. The zero-energy state is therefore its own antiparticle, which is the definition of a  Majorana fermion, see Ref.~\cite{Wilczek09} for a recent short review. The conclusion that vortices on the surface of a topological insulator host Majorana fermions was originally reached by a slightly different line of reasoning than the above, which evoked the existence of a spinless $p+ip$ superconducting state \cite{Fu08TISC}, but the results are nonetheless the same. Using the index theorem to establish the number of vortex zero-energy states directly differentiates between the multiple zero-energy vortex states in graphene and the single zero-energy Majorana mode on the topological insulator surface, simply due to the number of Dirac cones.
A Majorana mode has also been shown to exist at the interface between a superconducting and a ferromagnetic region on the topological insulator surface \cite{Fu08TISC, Fu09TISC}. Intimately related to the occurrence of this interface Majorana mode in topological insulators is an identification of zero modes bound to vortex cores and at sample edges in graphene \cite{Bergman09}.

%% file: Subsection_magneticfield.tex
\subsection{Magnetic field dependence}
\subsubsection{Landau level quantization}
Another fundamental example of a similar response for two-dimensional Dirac materials to external probes is the characteristic form of the Landau quantization in magnetic fields. For Dirac particles with well-defined charge, like electrons and holes in graphene or topological insulators, a magnetic field enters the Dirac Hamiltonian through the minimal substitution $\bp\to\bp+e\bA$, where $\bA$ is the vector potential. A perpendicular magnetic field $B$ can be represented in the Landau gauge $\bA=(yB,0,0)$. Due to the remaining translation symmetry in $x$-direction, the ansatz $\psi(x,y)=e^{ikx}\phi(y)$ can be made for the eigenstates of the resulting Dirac equation, which leads to \cite{AHC_RMP,Katsnelson_book}:
\begin{equation}
\hbar v_f\left[\begin{array}{cc} 0 & \partial_y-k+eyB/\hbar \\
	- \partial_y-k+eyB/\hbar & 0
\end{array}\right]\phi(y)=E \phi(y).
\label{eq:Dirac_mag_field}
\end{equation}
The magnetic field sets a length scale $l_B=\sqrt{\hbar/eB}$, the so-called magnetic length, which can be used to obtain a dimensionless form of Eq.~(\ref{eq:Dirac_mag_field}):
\begin{equation}
\hbar\omega_c\left[\begin{array}{cc} 0 & a \\
	a^\dagger & 0
\end{array}\right]\phi(\xi)=E \phi(\xi),
\label{eq:Dirac_mag_field_II}
\end{equation}
where $\xi=y/l_b-k l_b$ is the dimensionless length, $\omega_c=\sqrt{2}v_F/l_b$ is the cyclotron frequency, and $a=\frac{1}{\sqrt{2}}(\xi+\partial_\xi)$ ($a^\dagger=\frac{1}{\sqrt{2}}(\xi-\partial_\xi)$) are the ladder operators of a one-dimensional harmonic oscillator. With the well-known eigenstates $\ket{n}$ of the one-dimensional harmonic oscillator ($a^\dagger a\ket{n}=n\ket{n}$), the solutions to Eq.~(\ref{eq:Dirac_mag_field_II}) are easily constructed. There is clearly \textit{one} zero-energy mode
\begin{equation}
\ket{\phi_0}=\ket{0}\otimes\left(\begin{array}{c}
	0 \\ 1
\end{array}\right),
\label{eq:WF_zero_LL}
\end{equation}
as well as eigenstates ($n>0$)
\begin{equation}
\ket{\phi_{n,\pm}}=\ket{n}\otimes\left(\begin{array}{c}
	0 \\ 1
\end{array}\right)\pm\ket{n-1}\otimes\left(\begin{array}{c}
	1 \\ 0
\end{array}\right),
\label{eq:WF_n_LL}
\end{equation}
at finite energies
\begin{equation}
E_{n,\pm}(B) = \hbar \omega_c\sqrt{|n|} =\hbar v_F \sqrt {2eB|n|/\hbar}.
\label{EQ:DiracB}
\end{equation}
There are two features which make the Landau level quantization of massless Dirac fermions distinct from two-dimensional Schr{\"o}dinger fermions. First, there is a square-root dependence of the Landau level energies on the magnetic field and level index $n$, in contrast to the linear dependencies on $B$ and $n$ for Schr{\"o}dinger fermions. The experimental demonstration of these square-root dependencies in fact played pivotal roles in establishing experimental evidence for Dirac fermion excitations in graphene \cite{Geim2005, Zhang2005, JiangStormer07, LiAndrei07, Andrei_PRL09} and on the surface of three-dimensional topological insulators \cite{Zhang_LL_TI_PRL10,Hanaguri_LL_TI_PRB09}. 

The square-root magnetic field dependence leads to much larger Landau level spacings and cyclotron energies than in typical two-dimensional electron gases or metals. For the free two-dimensional electron gas, a typical laboratory magnetic field on the order of $10$~T leads to $\hbar\omega_c\approx 1\,{\rm meV}\approx k_B 10$~K, to be contrasted with $\hbar\omega_c\approx 100\,{\rm meV}\approx k_B 1000$~K for graphene. This is the reason why quantum Hall physics can be observed in Dirac materials at unusually high temperatures, in graphene even at room temperature \cite{Novoselov_Science07}. 

The second distinct feature of the Landau level sequence for Dirac fermions is the the zeroth Landau level, Eq. (\ref{eq:WF_zero_LL}), which is independent on field and always located at zero energy. This level is shared by electrons and holes to equal amounts. Its existence also gives rise to a quantum Hall effect with quantization of the Hall conductance $\sigma_{xy}$ at half-integer values: $\sigma_{xy}=(n+\frac{1}{2})e^2/h$ with integer $n$. Measurements of this quantum Hall effect sequence have thus been put forward as hallmark of magnetotransport dominated by Dirac carriers. Graphene has four Dirac cones and the experimental demonstration of $\sigma_{xy}=(4n+2)e^2/h$ Hall conductance plateaus \cite{Geim2005, Zhang2005}, literary ignited the enormous amount of research activities dedicated to this material. In three-dimensional topological insulators, electron transport by topological surface states is often obscured by defect-related bulk contributions. Recently, progress in increasing sample qualities \cite{YAndo_PRB10,Xiong_PhysicaE,YAndo_PRL11} in Bi-based topological insulator compounds has been reported. Moreover, strained HgTe has been shown to be a promising material for achieving topological surface state dominated transport and signatures of the unusual half-integer Hall quantization have been measured in this material \cite{Molenkamp_PRL11}.

Generally, the zeroth Landau level is special in that it is fully pseudospin polarized, as can be seen from Eq.~(\ref{eq:WF_zero_LL}). For graphene, a given perpendicular magnetic field leads to a zeroth Landau level which is entirely localized in sublattices A or B for the valleys at K and K', respectively. Reversing the magnetic field reverses this valley dependent sublattice polarization. The same pseudospin polarization translates also to topological insulators, where the real electron spin coincides with the Dirac pseudospin and the zeroth Landau level is thus fully spin-polarized.

Any vector field $\mathbf{A}$ which couples to the Dirac pseudospin in terms of $\bm\sigma\cdot\mathbf{A}$ affects the Dirac fermions like a gauge field and effectively produces pseudomagnetic fields $\mathbf{B}=\nabla\times \mathbf{A}$. In graphene, strain modulates the hopping integrals between adjacent atoms and causes pseudomagnetic fields \cite{Kane_Mele_PRL97, Lammert_Crespi_PRB00, Wehling_EPL08, Vozmediano_PhysRep2010}. Random pseudomagnetic fields due to ripples cause electron scattering and are considered to be an important scattering mechanism in freely suspended graphene samples, as reviewed in Ref.~\cite{Katsnelson_book}. In three-dimensional topological insulators, exchange coupling to magnetic impurities acts on the topological surface state electrons like an effective gauge field if the impurity moment is oriented in-plane \cite{Honolka_PRL12}. Random in-plane oriented magnetic impurities thus mimic random pseudomagnetic fields and cause electron scattering very similar to ripples in graphene. Dilute ensembles of Fe atoms adsorbed on the Bi$_2$Se$_3$ surface were demonstrated to exhibit an in-plane magnetic easy axis \cite{Honolka_PRL12} and exchange coupling likely realizes random pseudomagnetic fields there.

For sufficiently strong pseudomagnetic fields, pseudo Landau levels become observable. In strained graphene bubbles on a Pt surface, pseudomagnetic fields exceeding $300$~T have been reported and signatures of pseudo Landau levels in scanning tunneling spectra were then revealed \cite{Crommie_Science2010}. As strain does not break time reversal symmetry, the pseudomagnetic fields in valleys $K$ and  $K'$ must have opposite sign. Thus, the sublattice polarization of the zeroth \textit{pseudo} Landau level is the same in the $K$ and $K'$ valley and detectable in local probe experiments with sublattice resolution. For graphene, this has been demonstrated by {\it ab-initio} calculations \cite{Wehling_EPL08}, but not yet experimentally verified. Artificial graphene allows even more freedom in engineering strain, resulting in pseudomagnetic fields and pseudo Landau levels by means of atomic manipulation. Here, the sublattice polarization of the zeroth Landau level has been directly observed in STM experiments \cite{Manoharan_Nature2012}.

While the above described magnetic field response and Landau level quantization of charged Dirac particles is indeed universal in two dimensions (at least in the low-field limit), Dirac materials without charge conservation such as $d$-wave superconductors have to be considered more carefully. The Bogoliubov quasiparticles in $d$-wave superconductors do not have a well-defined charge and the response of the superconducting condensate to an external magnetic field has to be taken into account \cite{Franz_Tesanovic_PRL00}.

%% file: Subsection_chiralanomaly.tex
\subsubsection{Chiral anomaly}
Moving beyond two dimensions to the three-dimensional Weyl semimetals each Landau level is instead a one-dimensional dispersing mode along the magnetic field direction. Also here the zero-energy level is very interesting: $\epsilon_0 = \chi \hbar v_F \bk \cdot \hat{B}$, i.e.~the zero-energy level is a chiral mode whose dispersing direction is set by the chirality $\chi$ of the Weyl point. At low temperatures it is only this zero-energy level that is relevant and it thus determines the low-energy physics, usually referred to as the quantum limit. If also applying an electric field ${\bf E}$ in the same direction as the magnetic field $\bf B$ all states will move along the electric field according to $\hbar \dot{\bk} = -e {\bf E}$. For the chiral zero-energy level this means that electrons are either disappearing or appearing at the Weyl point depending on the chirality of the Weyl point. Charge is thus not conserved around an individual Weyl point and we get a modified charge continuity equation:
\begin{align}
\label{eqn:chargeeq}
\frac{\partial n}{\partial t} + \nabla \cdot {\bf J} = \pm \frac{e^2}{4\pi^2\hbar^2} {\bf E}\cdot {\bf B}.
\end{align}
This non-conservation of charged particles for a single Weyl point is known as the chiral anomaly, or the Adler-Bell-Jackiw anomaly in high-energy field theory \cite{Bell69, Adler69}. 
Physically, the anomaly provides yet another explanation to why Weyl points always have to appear in pairs with opposite chirality in materials, since overall there needs to be charge conservation in a real material. 

Despite the overall charge conservation in the full unit cell, the chiral anomaly can still give rise to interesting effects. For a pair of Weyl points separated in momentum space and with opposite chirality, the chiral anomaly gives a charge imbalance, or equivalently valley polarization, between the two nodes according to
\begin{align}
\label{eqn:valleypol}
\frac{\partial (n_+ - n_-)}{\partial t}  = \frac{e^2}{2\pi^2\hbar^2} {\bf E}\cdot {\bf B}.
\end{align}
Since the Weyl nodes are separated in momentum space, this charge imbalance can only be relaxed by large momentum scattering and can thus have a very long relaxation time in clean systems. One direct consequence is that the longitudinal conductivity along the applied magnetic field, which is proportional to the relaxation time, can be very large \cite{Nielsen83}. The longitudinal conductivity is also proportional to the magnetic field, since a Weyl semimetal in a magnetic field effectively consist of decoupled one-dimensional chains along the field direction. Thus  the resistivity decreases with magnetic field, resulting in negative magnetoresistance \cite{Hosur13}. Negative magnetoresistance has recently been measured in the topological insulator Bi$_{1-x}$Sb$_x$ in a magnetic field near its topological phase transition, which could be a system with both Weyl points and Landau levels \cite{KimLi13}. This experiment thus provides indications of the chiral anomaly through transport measurements.

Another related consequence of the chiral anomaly is a giant anomalous Hall effect, where the Hall conductance \cite{Yang11}
\begin{align}
\label{eqn:GH}
{\bf G}_H = \frac{e^2}{2\pi h} \sum_i \chi_i {\bk}_i,
\end{align}
with $\bk_i$ and $\chi_i$ being the momenta and chirality of the Weyl points, respectively. Thus the Hall conductance is directly proportional to the momentum space separation between the two Weyl points. If thus the Weyl point locations are determined by a separate experiment, for example by ARPES measurements, then it is possible to consider this to be a quantized Hall response.
This anomalous Hall effect can also be understood when considering that the region in the Brillouin zone between two oppositely charged Weyl points can be seen as a collection of two-dimensional planes with a quantized Hall effect because of a finite Chern number, as previously discussed in Section~\ref{subsec:Weyl}. The total Hall response should thus be directly proportional to the distance between the two Weyl points.
More details on the chiral anomaly in Weyl semimetals, and especially its consequences for transport properties, can be found in a recent review in Ref.~\cite{Hosur13}.